\documentclass[modern]{aastex61}
\usepackage{CJK}
\usepackage{amsmath}
\usepackage{graphics}
\newcommand{\s}{\,{\rm s}}      
\newcommand{\yr}{\,{\rm yr}}    
\newcommand{\cm}{\,{\rm cm}}    
       
\newcommand{\parsec}{\,{\rm pc}}
\newcommand{\erg}{\,{\rm erg}}        \newcommand{\K}{\,{\rm K}}
\newcommand{\eV}{\,{\rm eV}}    \newcommand{\keV}{\,{\rm keV}}

  \newcommand{\ks}{\,{\rm ksec}}

\newcommand{\cmmthree}{\,{\rm cm}^{-3}}
\newcommand{\g}{\,{\rm g}} \newcommand{\Ba}{\, \rm g\,cm^{-1}\,s^{-2}}

\newcommand{\dd}[1]{\mathrm{d}#1}

\newcommand{\xray}{{\rm X-ray}}

\newcommand*{\rom}[1]{\uppercase\expandafter{\romannumeral #1\relax}}  

\begin{document}
\begin{CJK*}{UTF8}{bkai}

\title{Non-equilibrium Ionization in Mixed-morphology Supernova Remnants}

\author[0000-0002-0747-0078]{Gao-Yuan Zhang (張高原)}
\affiliation{School of Space Science and Astronomy, Nanjing University, 163 Xianlin Avenue, Nanjing 210023, People's Republic of China}
\affiliation{Harvard-Smithsonian Center for Astrophysics, 60 Garden St, Cambridge, MA 02138}

\author[0000-0002-7597-6935]{Jonathan D. Slavin}
\affiliation{Harvard-Smithsonian Center for Astrophysics, 60 Garden St, Cambridge, MA 02138}

\author[0000-0003-3462-8886]{Adam Foster}
\affiliation{Harvard-Smithsonian Center for Astrophysics, 60 Garden St, Cambridge, MA 02138}

\author[0000-0003-4284-4167]{Randall K. Smith}
\affiliation{Harvard-Smithsonian Center for Astrophysics, 60 Garden St, Cambridge, MA 02138}

\author[0000-0003-3175-2347]{John A. ZuHone}
\affiliation{Harvard-Smithsonian Center for Astrophysics, 60 Garden St, Cambridge, MA 02138}

\author[0000-0002-5683-822X]{Ping Zhou (周平)}
\affiliation{School of Space Science and Astronomy, Nanjing University, 163 Xianlin Avenue, Nanjing 210023, People's Republic of China}
\affiliation{Anton Pannekoek Institute for Astronomy, University of Amsterdam, Science Park 904, 1098 XH Amsterdam, The Netherlands}

\author[0000-0002-4753-2798]{Yang Chen (陳陽)}
\affiliation{School of Space Science and Astronomy, Nanjing University, 163 Xianlin Avenue, Nanjing 210023, People's Republic of China}



\begin{abstract}
The mixed morphology class of supernova remnants (MMSNRs) comprises a substantial fraction of observed remnants and yet there is as yet no consensus on their origin. A clue to their nature is the presence of regions that show X-ray evidence of recombining plasmas. Recent calculations of remnant evolution in a cloudy interstellar medium (ISM) that included thermal conduction but not non-equilibrium ionization {(NEI)} showed promise in explaining observed surface brightness distributions but could not determine if recombining plasmas were present. In this paper we present numerical hydrodynamical models of MMSNRs in 2D and 3D including explicit calculation of NEI effects. 
Both the spatial ionization distribution and temperature-density diagrams show that 
recombination occurs inside the simulated MMSNR, and that both adiabatic expansion and thermal conduction cause recombination, albeit in different regions. Features created by the adiabatic expansion stand out in the spatial and temperature-density diagrams, but thermal conduction also plays a role.
Thus thermal conduction and adiabatic expansion both contribute significantly to the cooling of high temperature gas.
Realistic observational data are simulated
with both spatial
and spectral input from various regions. We also discuss the possibility of analyzing 
the sources of recombination and dominant hydrodynamical processes
in observations using temperature-density diagrams and spatial maps.
\end{abstract}

\nobreak
\nopagebreak
\section{Introduction} \label{sec:intro}

Mixed-morphology supernova remnants (MMSNRs)
are a class of SNR 
characterized by \citet{Rho1998},
who defined them as containing a radio shell with centrally brightened thermal \xray\ emission. This class is also known as thermal composite SNRs, named by \citealt{Jones1998}. 
The external radio shell is believed to be associated with the blastwave shock propagating into the interstellar medium (ISM), but 
the bright thermal \xray\ interior is {hard} to explain. MMSNRs 
are usually highly absorbed and almost always found in regions with 
molecular clouds (MCs). 
{
Observational evidence has confirmed that about half out of known 37 MMSNRs are interacted with MCs \citep{Zhang2015}.}
Given the complex environments the SNRs are evolving in, there have been several possible
mechanisms proposed for the brightness of the thermal \xray\ interior,
such as a radiatively cooled rim \citep{Harrus1997,Rho1998}, 
thermal conduction in the interior hot gas \citep{Cox1999, Shelton1999}, evaporation of gas from the shock-engulfed cloudlets \citep{White1991}, 
shock reflection \citep{Chen2008, Zhang2015}, and
even projection effects for some particular SNRs \citep{Petruk2001,Zhou2016}.
\citet{Slavin2017} investigated the influence of a cloudy surrounding medium
on the evolution of SNRs with numerical hydrodynamical simulations
that included thermal conduction. 
The results agree with those of \citet{White1991} in that centrally brightened
thermal emission is generated, though emission from shocked clouds, which
was not included by \citet{White1991}, was found to be important.
A limitation of the calculations in \citet{Slavin2017} was that the plasma was assumed to 
be {in} collisional ionization equilibrium (CIE). 

While it has long been known that shocks should produce plasma that is far from
CIE and under-ionized for its temperature, it was surprising that some MMSNRs 
(about 12 out of 37) were found to contain over-ionized (recombining) plasma  
\citep[See][and the references therein]{Suzuki2018}. Considering the modest 
effective area and the spectral resolution of currently available X-ray observatories our ability to diagnose such over-ionization is limited and so the fraction 
of such remnants could be larger. So far all SNRs found to have recombining plasma 
belong to the class of MMSNRs,
implying a possible correlation between the mechanisms for the overionization and the 
centrally peaked thermal emission.

In the standard Sedov-Taylor evolution of an SNR in a homogeneous ISM, material is heated by the shock and cools slowly and adiabatically because of expansion of the remnant. Thus to get a significant signature of recombination, some additional cooling mechanisms must be at work. Thermal conduction presents one possible cause of this cooling since when it is
operating freely, the hot central region of the remnant cools by conducting its heat to the cooler 
regions closer to the shock front, leading to flattening of the central temperature profile.
\citet{Zhou2011} performed {the first}
hydrodynamical simulation including a shock interacting a ring-like cloud
in 2D to investigate the over-ionization seen in a particular SNR, W49B, albeit without considering other cloud configurations or comparing other cooling sources in detail. 
Other possibilities have also been proposed, such as the transfer of energy from the thermal plasma to cosmic rays 
\citep{Suzuki2018} to explain the unexpected ionization.
In this paper, we explore the ionization effects of a cloudy ISM on SNR evolution, utilizing
hydrodynamic simulations that include thermal conduction and non-equilibrium ionization.


\section{Methods and models} \label{sec:method}
To study the effects of thermal conduction and the cloudiness of the ISM on the ionization inside evolving SNRs we performed
hydrodynamical simulations using the FLASH code v4.3\footnote{\url{http://flash.uchicago.edu/site/flashcode/}}
\citep{Fryxell2000}.
We used the same initialization as runs in \citet{Slavin2017} including using 2D cylindrical symmetry (except as noted below). The new aspect of this study is the
use of non-equilibrium ionization (NEI), which utilizes the new NEI unit developed by \citet{ZhangG2018}. That unit evolves the ionization simultaneously with the hydrodynamics.
The gas is assumed to be initially in CIE in the ambient medium at a temperature $T=10^4\K$
and a number density $n_{\rm H}=0.25\cmmthree$.
We do not include the effects of the magnetic field, radiative cooling or energy loss by cosmic ray acceleration in the simulations presented in this paper.


{
We use the \texttt{Diffuse} module of FLASH code for thermal conduction in
the simulation.
Both 
classical Spitzer conductivity \citep{Spitzer1956}, $\kappa\propto T^{5/2}$, and saturated conductivity
\citep{McKee1977}, $\kappa\propto\rho c^3$ are supported with a smooth transition from classical to saturated.}
(see the Appendix in \citealt{Slavin2017} for details). For the sake of comparison, 
we performed three different simulations (see Table~\ref{tab:models}): 1) a plane parallel shock encountering a single cloud, 2) a SNR in a homogeneous medium and 3) a SNR in a cloudy ISM (2D and 3D).



\subsection{Single cloud in a plane parallel shock}

We first simulated a single cloud in a shock tube (2D)
to investigate how the NEI state evolves around a dense cloud
(model ``A'' in Table~\ref{tab:models}). 
The shock propagates from left to right along the $x$-axis
in the domain where a single 
dense cloud is situated.
{
The plane parallel shock initialization is set up as a Sod shock tube with a higher
density and temperature on the left side and a lower density and temperature on the
right side. The single cloud is situated inside the right side region. 
The initial velocity of the left plasma is 300 km/s; and the right
plasma is stationary initially. The cloud in the right region is
100 times denser ($\rho_c=1\times10^{-22}\g\cmmthree$) than the surrounding plasma with 
a lower temperature ($T_c=1\times10^2\K$) to make sure the
pressure are equal in both of the cloud and the surrounding plasma. 
The geometrical and 
physical parameters can be found in Table~\ref{tab:param}.
}

\begin{deluxetable}{ccl}
  \tablecaption{Parameters for single cloud simulation\label{tab:param}}
  \tablehead{
    \colhead{Variable} & \colhead{Value} & \colhead{Description}\\
  }
  \startdata
  xmax    & $1\times 10^{18}\cm$ & The size of simulation box in x axis \\
  ymax    & $1\times 10^{18}\cm$ & The size of simulation box in y axis \\
  posn    & $0.2\times 10^{18}\cm$ & The position (in x axis) of the initial shock front\\
  cposx   & $0.5\times 10^{18}\cm$ & The position (x) of the cloud center \\
  cposy   & $0.5\times 10^{18}\cm$ & The position (y) of the cloud center \\
  crad    & $0.2\times 10^{18}\cm$ & Radius of the cloud \\
  rhoLeft & $1\times 10^{-23}\g\cmmthree$  & The initial density on the left of shock front\\
  rhoRight& $1\times 10^{-24}\g\cmmthree$ & The initial density on the right of shock front\\
  crho    & $1\times 10^{-22}\g\cmmthree$ & Density of the cloud \\
  tLeft   & $1\times 10^{6}\K$  & The initial temperature on the left of shock front\\
  tRight  & $1\times 10^{4}\K$  & The initial temperature on the right of shock front\\
  ct      & $1\times 10^{2}\K$ & Temperature of the cloud \\
  uLeft   & $3.0\times10^{7}\cm\s^{-1}$ & The initial velocity (along x-axis) on the left of the shock front \\
  uRight  & $0$ & The initial velocity (along x-axis) on the right of the shock front \\
  xl\_boundary\_type & diode & X-axis left boundary condition \\
  xr\_boundary\_type & outflow & X-axis right boundary condition \\
  yl\_boundary\_type & outflow & Y-axis left boundary condition \\
  yr\_boundary\_type & outflow & Y-axis right boundary condition \\
  \enddata
\end{deluxetable}

We use the boundary condition type \texttt{outflow}, which is 
a zero-gradient boundary 
condition that allows the simulated fluid flow out or into the domain,
for both boundaries in $y$-axis and the right boundary in $x$-axis.
The boundary condition type \texttt{diode} allows the fluid to flow out as well, 
but does not allow the fluid to return into the domain.
We use \texttt{diode} for the left boundary condition in $x$-axis,
so that we can see the gas stretching in the post-shock area.
The thermal conduction is enabled to compare
its effect on NEI with the dynamical processes, such as the
adiabatic expansion in the stretching area.

To show in which stage (ionizing or recombining) the plasma is, we use a variable 
that is the difference of average charge of the ions, 
\begin{equation}
\Delta \bar{c}=\bar{c}_{\rm eq}-\bar{c},
\label{eq:aver}
\end{equation}
where $\bar{c}=\sum\limits_{i=0}^{Z} f^{(i)} c_i$
is the average charge of the element with the atomic number $Z$ ($c_i=i$ is the
charge of the $i$th ion, with i=0 for the neutral atom; $f^{(i)}$ is the $i$th ion fraction satisfying 
$\sum\limits_{i=0}^{Z} f^{(i)} =1$), and 
$\bar{c}_{\rm eq}$ is the expected one in equilibrium at a given temperature 
\citep[See][]{ZhangG2018,Zhou2011}.
$\Delta \bar{c}>0$ implies an underionized plasma, while $\Delta\bar{c}<0$ an overionized
or recombining plasma.
We use this model to qualitatively describe the origins of the ionization evolution around 
the cloud. It is not identical to case of clouds 
shocked by a spherical shock in an SNR. However, it does show that
we can infer the dominant reasons
by investigating
the physical positions of different ionization states (See Fig.~\ref{fig:singlecloud} and \S~\ref{sec:results_1}).

\subsection{SNR explosion simulations}
\label{sec:cloudy}

Following the simulation initialization in \citet{Slavin2017}, 
we simulate SNRs exploding into different environments.
{
As shown in Table~\ref{tab:env}, a typical SNR kinetic energy ($E=10^{51}\erg$) is used within
the explosion center ($r_c\sim2.25\parsec$). The ambient plasma has a density of
$n_H=0.25\cmmthree$, a temperature of 1$\times10^4\K$. 
The 2D simulations use an axisymmetric cylindrical
geometry. In such a geometry, the clouds are tori like around the $z$-axis in 3D.
The explosion is initiated by thermal pressure in a spherical region of radius 2.25 pc. The density of the region is the same as the ambient medium and is not given any initial velocity.  Thus we are not including any ejecta component in these calculations.
After the initialization, the high pressure in the center pushes
the materials outward forming a shock front.
}

\begin{deluxetable}{ccl}
  \tablecaption{Simulation parameters of SNR explosion (2D)\label{tab:env}}
  \tablehead{
    \colhead{Variable} & \colhead{Value} & \colhead{Description}\\
  }
  \startdata
  geometry    & cylindrical & The geometry of the coordinates \\
  xmax    & 30 pc & The right edge in r axis \\
  xmin    & $0$ & The left edge in r axis \\
  ymax    & 30 pc & The right edge in z axis \\
  ymin    & $0$ & The left edge in z axis \\
  xl\_boundary\_type & axisymmetric & r-axis left boundary condition \\
  xr\_boundary\_type & reflect & r-axis right boundary condition \\
  yl\_boundary\_type & outflow & z-axis left boundary condition \\
  yr\_boundary\_type & outflow & z-axis right boundary condition \\
  expEnergy & 1$\times10^{51}\erg$ & The explosion energy \\
  rInit & 2.25 pc & The central ``ejecta'' mass radius \\
  rho\_ambient & 5.316$\times10^{-25}\g\cmmthree$ & The ambient density \\
  p\_ambient & 7.217$\times10^{-13}\Ba$ & The ambient pressure \\
  \enddata
\end{deluxetable}


To provide a baseline case,
we ran two simulation models in a homogeneous environment,
both with and without thermal conduction (models 
``B1'' and ``B2'' in Table~\ref{tab:models}). 
Model ``B1'' runs with thermal conduction to show the impact of the thermal 
conduction on the SNR evolution and the NEI states.
Without thermal conduction, model ``B2'' can be compared to the theoretical Sedov-Taylor solution in 1D.
Unlike the theoretical case, the initial explosion energy is not 
in a perfect point-like area in the hydrodynamic simulation.
So we use the position of the shock front ($r_s$) and 
the maximum density on the shock front ($n_{e,max}$) to normalize the results to compare to
the theory (See appendix \S~\ref{sec:theory} for the comparison).
To limit calculation time, only two-dimensional simulations were performed and 
only the NEI evolution of oxygen was calculated.


We simulate an SNR exploding into a cloudy environment with the NEI calculation,
(models ``C1'' and ``C2'' in Table~\ref{tab:models}).
The cloud distribution is randomly generated with the size following a 
power law (exponent of $-3$). 
The density ratio between the clouds and the 
inter clouds material is constant, $\chi=100$. 
We use the \citet{White1991} parameter, $C=\frac{f\chi}{1-f}$ ({WLC}), 
where $f$ is the filling factor {of the clouds}, 
to describe the cloud distribution. 
To investigate the impact of thermal conduction,
the model ``C1'' and ``C2'' runs were done with and without thermal conduction respectively.
The initial conditions are exactly
the same except for the thermal conduction.
The initial physical conditions of SNRs are also the same {as those of the} 
the homogeneous environment models ``B1'' and ``B2'' except for the clouds.

We perform several three-dimensional simulations (model ``D'')
of the SNR explosion in a cloudy environment as well. 
The physical parameters in the initial conditions are the
same with 2D simulations (with thermal conduction). 
{For our 3D simulations we use Cartesian coordinates in a box that is 60 pc on each side with the explosion occurring in the center. We use ``ouflow"  boundary conditions on all sides.}
{
The generation of 3D random cloud distributions follow the same method 
mentioned above.
The initialization files for FLASH code (all of the setups above) can be found in a Github repository (\url{https://github.com/TuahZh/MM-SNR-initializations}).
}


Because of the 
available computer power the physical resolution has to be decreased in the 3D simulation. 
The FLASH code uses an adaptive mesh refinement grid. In our simulations, 
the grid can be refined according to runtime variables of density and pressure.
All the 3D simulations are performed with a maximum resolution of 1024$^3$ and 512$^3$, and
2D simulations with a maximum resolution of 1024$^2$ for model ``A'' and
4096$^2$ for the rest. By comparing several density and temperature slice maps
in both 2D and 3D simulations, we confirm that the resolution does not considerably change the results.

\begin{deluxetable}{cccc}
  \tablecaption{SNR models\label{tab:models}}
  \tablehead{
    \colhead{Models} & \colhead{{WLC}\tablenotemark{$\dagger$}} & 
    \colhead{Thermal conduction} & 
    \colhead{Dimension}\\
  }
  \startdata
  A (single cloud) & - & Y & 2\\
  B1 (homogeneous environment) & 0  & Y & 2\\
  B2 (homogeneous environment) & 0  & N & 2 \\
  C1 (cloudy environment) & 10 & Y & 2\\
  C2 (cloudy environment) & 10 & N & 2 \\
  D (cloudy environment) & 10 & Y & 3\\
  \enddata
  \tablenotetext{\dagger}{C parameter defined by \citealt{White1991} (See \S~\ref{sec:cloudy}).}
\end{deluxetable}





\section{Results} \label{sec:results}

\subsection{Plane parallel shock interaction with a single cloud}
\label{sec:results_1}

\begin{figure}[htpb]
  \plotone{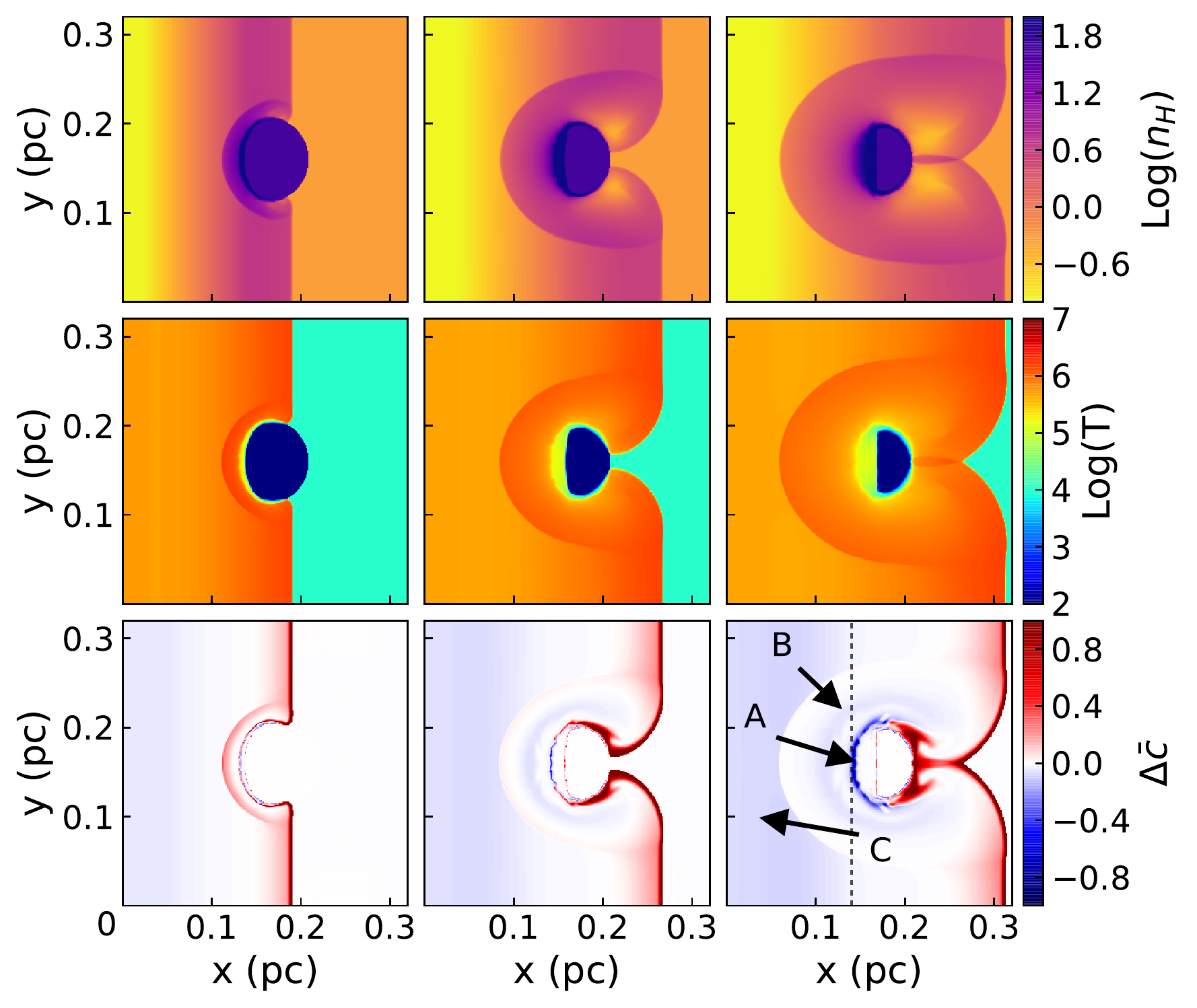}
  \caption{Three stages of the single cloud model. Time increase 
  from left to right panels. In each column, top, middle, and bottom
  panels show the density, temperature, average charge difference respectively.
  In the bottom right panel, 
  ``A'', ``B'', and ``C'' annotate three different recombining regions.
  The dotted vertical line depicts the separation of positive and negative
  $x$-axis component of velocity along the horizontal line through the cloud center. 
    \label{fig:singlecloud}}
\end{figure}

In Fig.~\ref{fig:singlecloud}, we show the evolution of the  
shock going through a cloud. As shown in the density and temperature figures,
the shock is distorted after impacting with the cloud. 
The distortion becomes more obvious after the engulfment, leaving a clearly
cooling region due to the expansion. It has lower density and temperature
than the adjacent regions in the post shock. The transmitted shock is also propagating inside
the cloud with a much smaller shock velocity (in green color in the temperature figures). 
Some ripple features that are
caused by the Rayleigh-Taylor instability occur in the compressed cloud region.
Additionally, a reflected shock propagates in the reverse direction.

By using the average charge difference, we can see the ionizing gas in red (positive) and
recombining gas in blue (negative) in the bottom panels.
As expected, the shock front is always ionizing and 
the expanding regions behind the shock are clearly recombining (annotated as ``C'' 
in the last figure). 
The reflected shock from the cloud reheats the post-shock gas.
A faint ring-like recombining feature appears between the reflected shock and the cloud 
(annotated as ``B''). Similar to the forward shock front, it is an expanding region
following the reflected shock front.
In the cloud, the gas 
is compressed as the shock propagates. Both ionization
and recombination can be seen at this region (annotated as ``A''). 
The recombination here 
is probably caused by the
thermal conduction in the conductive front or a mixture of the hot and cold
gas, especially considering the instability features shown in this region. 
As indicated in this single cloud simulation,
both thermal conduction and adiabatic expansion can contribute to the recombining
gas around the cloud. 
The mechanism for the recombination 
is probably dominated by thermal conduction 
inside the cloud or on the cloud rim. 
{The evaporated gas resulting from conduction consumes the thermal energy of the adjacent hot gas may also lead to rapid cooling \citep[as indicated in][]{Zhou2011} and hence recombination. }In the inter-cloud regions, the recombination
could probably caused by the adiabatic expansion.

\subsection{SNR in homogeneous medium}\label{sec:env}

\begin{figure}[htpb]
  \plotone{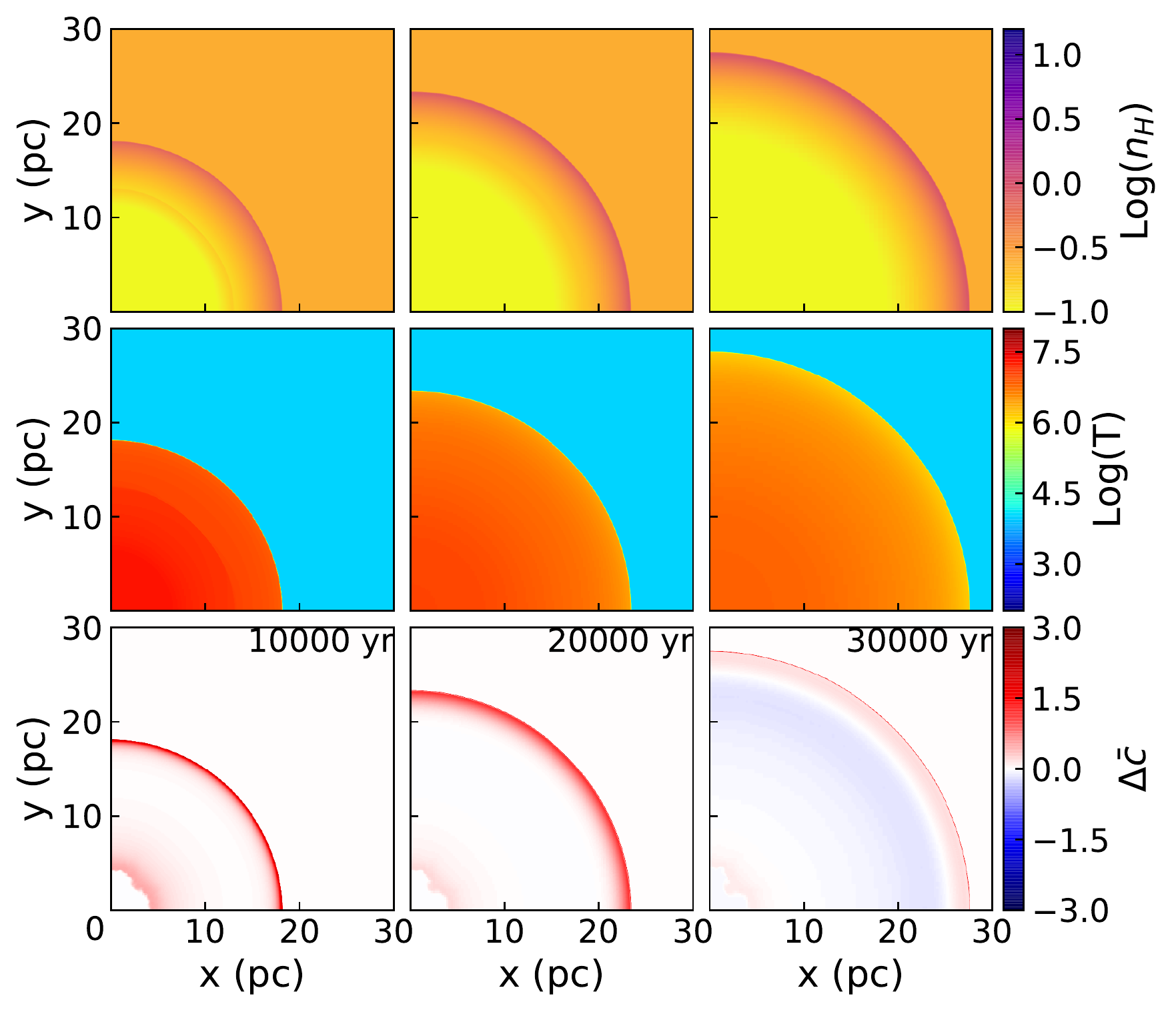}
  \caption{Three stages of the SNR model without clouds. 
  In each column, the top, middle, and bottom
  panels show the density, temperature, average charge difference 
  respectively with the same color scale. The innermost ionizing (red) features in
  the bottom panels are caused by the low density, which makes it harder to reach 
  equilibrium, but does not affect the results at larger radius.
    \label{fig:env}}
\end{figure}

This model has a homogeneous environment, resulting a similar result with
the self-similarity theoretical solution for a blast wave. Fig.~\ref{fig:env} shows the
model ``B1'' of a SNR in a homogeneous environment with thermal conduction. Thermal
conduction smooths the temperature distribution in the interior of the 
SNR. 
The density distribution is also affected by the thermal conduction, 
but this does not change the shell-like shape. 
The spectral emission measure
is proportional to the density-squared times the length along line of sight (LOS),
therefore
the \xray\ morphology will be also shell-like, unlike the center-filled morphology 
shown in MMSNR.
In the bottom row, the average
charge difference of oxygen is shown as the indication of ionization state
for this model. Before about 20000 yr, the plasma is strongly ionizing at the shock
front without any obvious recombining features. The interior becomes 
recombining at 30000 yr, caused by the adiabatic expansion as the 
shock front continue moving outwards.
The inner red rings at about 5 pc are the ionized
gas that is slow to reach equilibrium due to the extremely low density,
considering the ionization timescale $\tau=n_e t$. 
They are artifacts of the simulation which do not 
impact results at larger radius.

\subsection{SNR in a cloudy environment}

\begin{figure}[htpb]
  \plotone{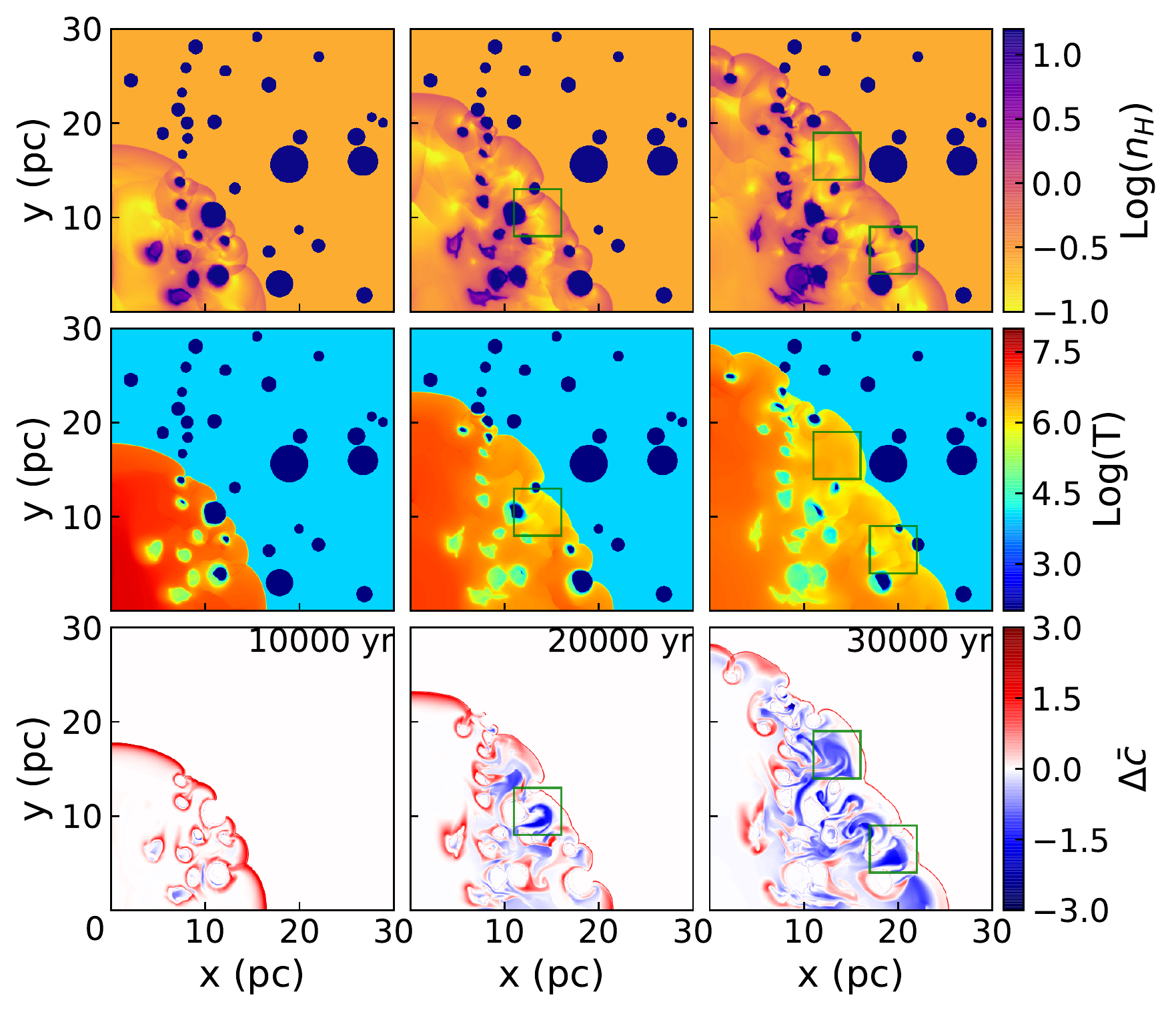}
  \caption{Three stages of the SNR in a cloudy environment
  model ``C1'' (with thermal conduction). Time increases 
  from left to right. In each column, the top, middle, and bottom
  panels show the density, temperature, average charge difference 
  respectively with the same color scale. The green boxes highlight selected
  recombining features that do not appear in the temperature
  and density maps.
    \label{fig:mmsnr}}
\end{figure}

\begin{figure}[htpb]
  \plotone{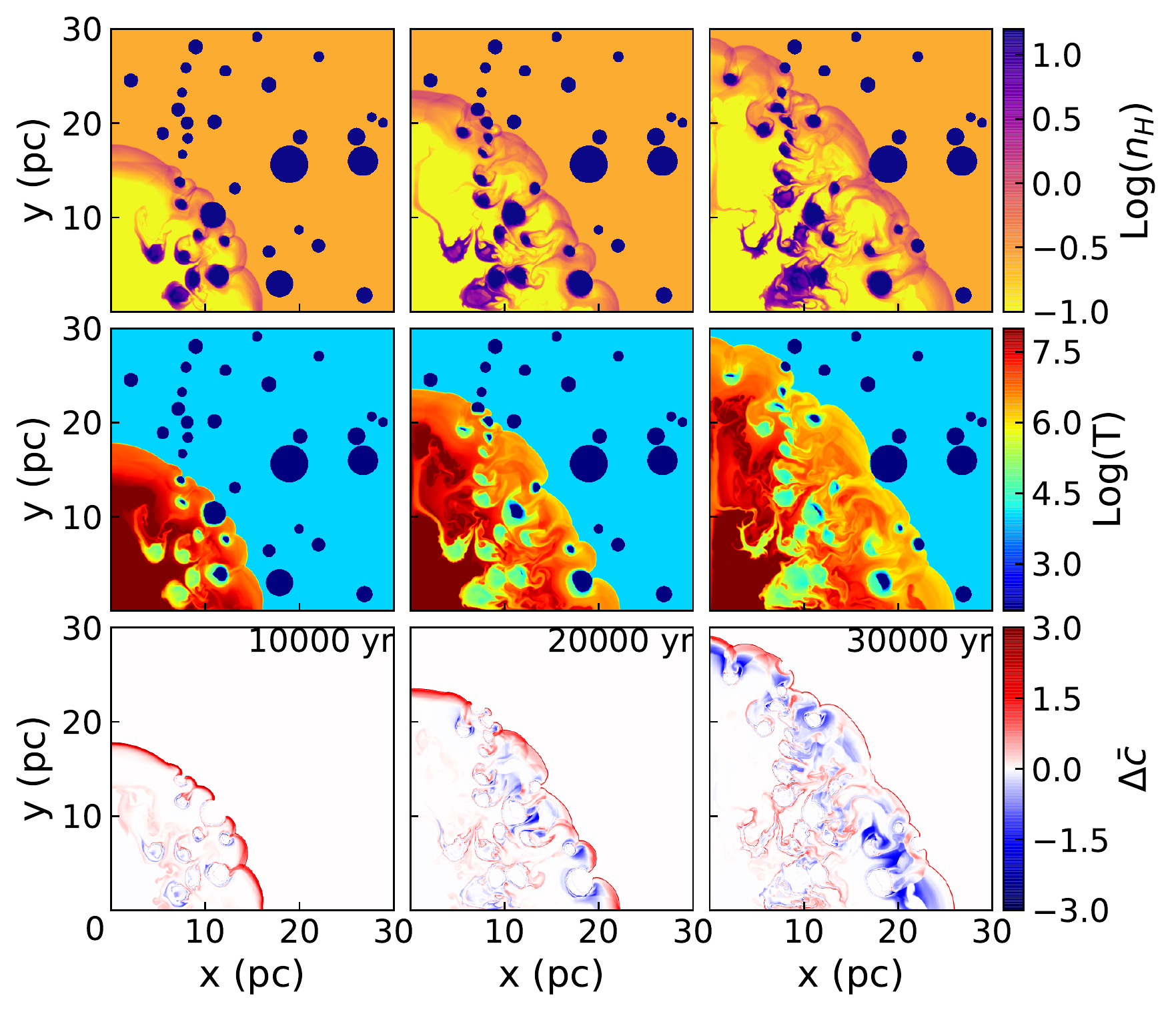}
  \caption{Three stages of the SNR model in a cloudy environment without 
  thermal conduction (model ``C2''). The layout is the same as Fig.~\ref{fig:mmsnr}. 
    \label{fig:nodiff}}
\end{figure}

Two SNR models in a cloudy environment, ``C1'' and ``C2'', are shown in 
Fig.~\ref{fig:mmsnr} and Fig.~\ref{fig:nodiff}. In model ``C1'', with
thermal conduction, the clouds are evaporated after being engulfed by
the SNR shock. The density in the interior then gets higher than the SNR
in a homogeneous environment. In the average charge difference figures, 
the shock front is again always ionizing as expected; but recombining features
show both in dense cloud regions and some low pressure (with low density
and low temperature) regions. Some interesting features are seen in the recombining
gas that do not appear in density or temperature figures (emphasized as 
green boxes in the figures). Comparing to the model
``B1'' (shown in Fig.~\ref{fig:env}), the recombination in both ``C'' 
models is more obvious,
and appears earlier in the SNR interior.
 

As shown in Fig.~\ref{fig:nodiff}, the clouds in model ``C2'' are not evaporated smoothly as in
model ``C1''. They are compressed and stripped with the forward shock and the reflected shock
passing by. 
The mixing of the hot and cold gas also increases the density profile, 
which is moving with the shock velocities, with
an almost-empty center left behind. In the average charge difference panel, recombining 
gas also appears in the post-shock area. However, it is in a smaller region 
than model ``C1'' with thermal conduction.

\subsection{Influence of thermal conduction on NEI}
\label{sec:tc}
\begin{figure}[htpb]
\plotone{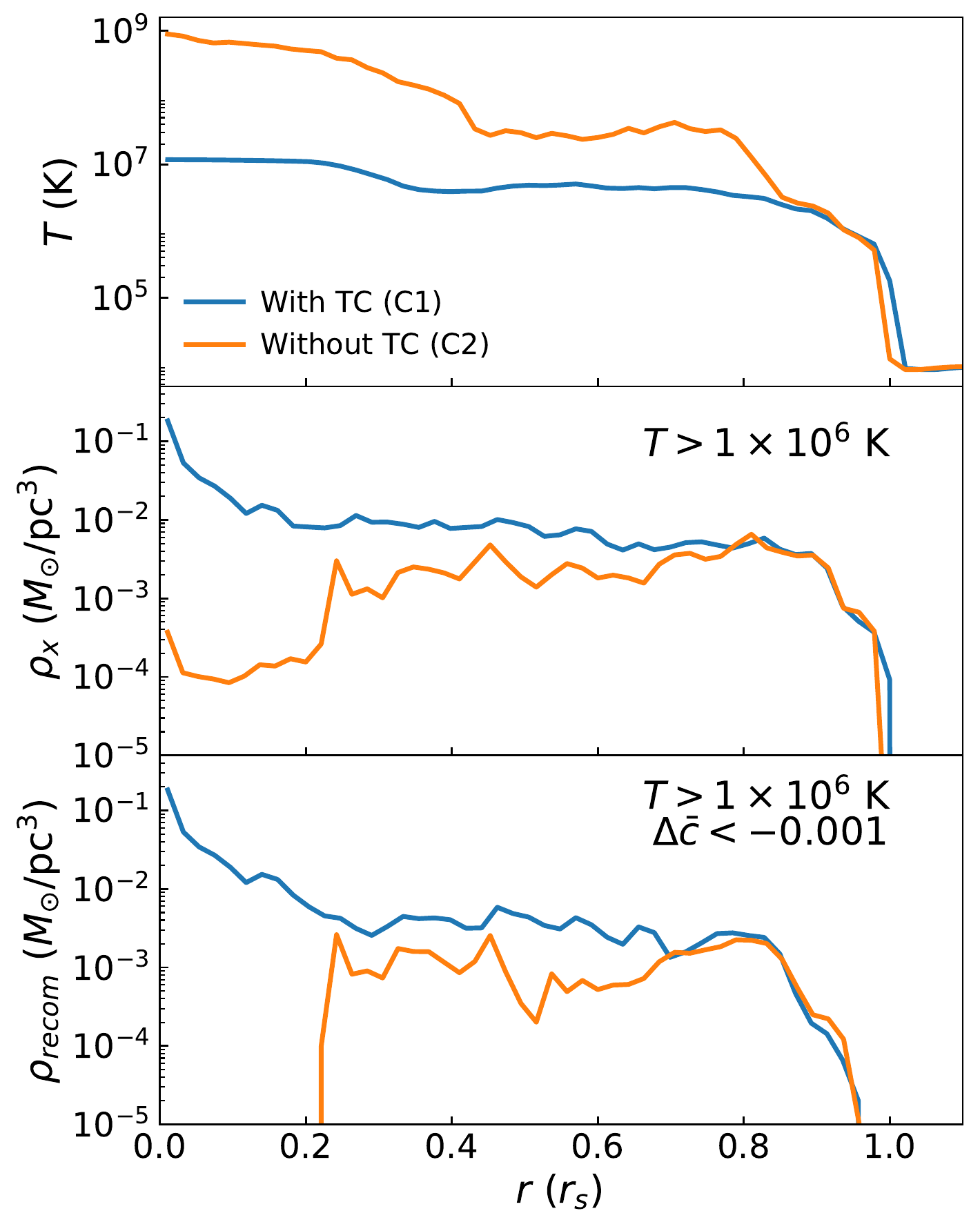}
  \caption{
  Average temperature (top), \xray\ emitting gas density (middle), and 
  recombining \xray\ emitting gas density (bottom)
  along the radial direction at 2$\times 10^{4}$ yr. 
  The blue lines are from the model ``C1'' (with thermal conduction; with TC),
  the orange lines are from the model ``C2'' (without thermal conduction; without TC).
    \label{fig:1d_2}}
\end{figure}

From the comparison between models ``C1'' and ``C2'', we conclude that 
the thermal conduction has significant influence on the NEI in SNR exploded into a cloudy 
environment. The top panel in Fig.~\ref{fig:1d_2} shows 
the average values of the temperature as a function of radius (in units of the shock front radius).
In model ``C1'', the temperature is smoothed along the 
radial direction, implying that the energy transfers between different
layers more efficiently due to thermal conduction. Then the recombination appears when
the inner hot plasma cools down rapidly. The \xray\ emitting gas ($T>1\times10^{6}\K$) 
is of primary interest. By accumulating the mass of \xray\ emitting gas in 
different shells with a thickness of 0.5 pc, the density can be calculated as a
function of radius (the middle panel in Fig.~\ref{fig:1d_2}).
Both models have a similar density profile in the shock front ($r>0.8 r_s$). 
However, 
the thermal conduction raises the density profile in the inner layers.
To investigate the 
recombining gas, we use $\Delta\bar{c}<-0.001$ to exclude the gas that
is ionizing or in equilibrium (See appendix \S~\ref{sec:charge2tau}). 
A radial density profile of the recombining \xray\
emitting gas is also generated with the same shell thickness (bottom panel in
Fig.~\ref{fig:1d_2}). The thermal conduction contributes to more over-ionized 
hot gas in the inner regions as well.


{The effect of thermal conduction can be separated to be the thermal conduction that 
is effective in large physical scales across the remnant \citep{Cox1999}, 
and the thermal conduction that  contributes to the evaporation of the dense clouds with smaller sizes
\citep{White1991, Cowie1981}. From our simulations, the large scale thermal conduction smooths
the temperature and density distribution; and the cloud evaporation 
is key to the generation of a thermal \xray\ emitting core and 
contributes to the fast cooling of the hot gas that shows over-ionized
features.}
Because of the magnetic field in or around the SNR, thermal conduction 
is probably overestimated by the {classical} Spitzer {and saturated} conductivity approximation.
The conductivity could be almost zero when the magnetic field is perpendicular to
{A high magnetic field amplification (and turbulence) is expected only when the shock velocity is high in very young SNRs that efficiently accelerate CRs \citep{Vink2003,Bell2004,Vink2012,Ji2016}.
Therefore, this effect does not play a major role after a few thousands years of evolution. 
In addition to shock generated turbulence, shear flow around clouds can lead to turbulence and complex field structures that can suppress thermal conduction and mass mixing \citep{Orlando2008}.
The actual effective thermal conductivity in a turbulent and magnetized environment such as SNRs evolving in a cloudy medium is uncertain. The reduction factor for conductivity relative to the unmagnetized value depends on such processes as turbulent transport and plasma instabilities 
\citep[e.g.\ ][]{Chandran2004, Komarov2016}
and is an area of active study. Studies of SNR evolution in a cloudy medium that include the magnetic field, anisotropic thermal conduction and NEI would be worth pursuing in the future.}


\subsection{Source of recombination}

\subsubsection{Thermal conduction and adiabatic expansion}

\begin{figure}[htpb]
  \plotone{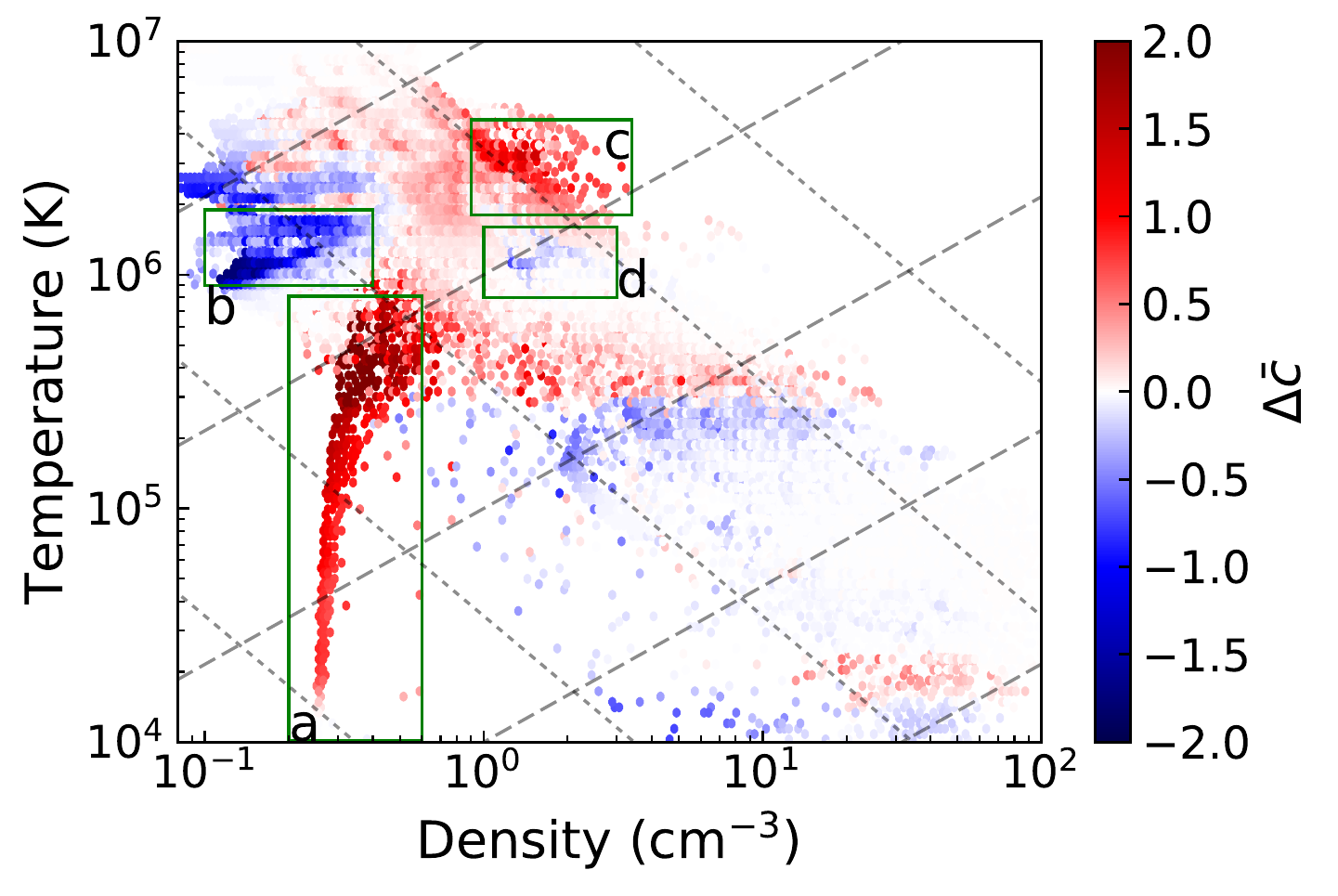}
  \caption{Phase plot of the average charge difference (for O; 
  at 20000 yr). The 
  dotted lines are the isobaric
  lines, and the dashed lines are adiabatic
  process lines in the phase map.
  Every pixel {shows} {the} average value {weighted by $n_H^2$}. Regions labeled ``a'', ``b'', ``c'',
  and ``d'' are shown in Fig.~\ref{fig:regions1}.
  \label{fig:dens_temp_delc_mass}}
\end{figure}

\begin{figure}[htpb]
\plottwo{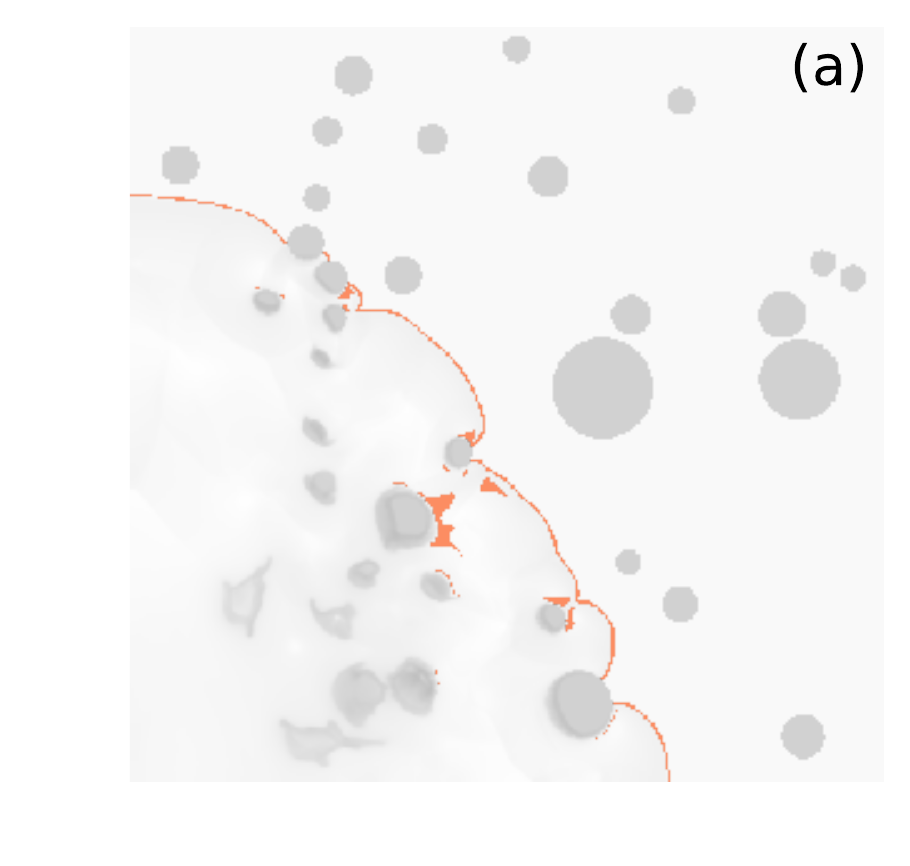}{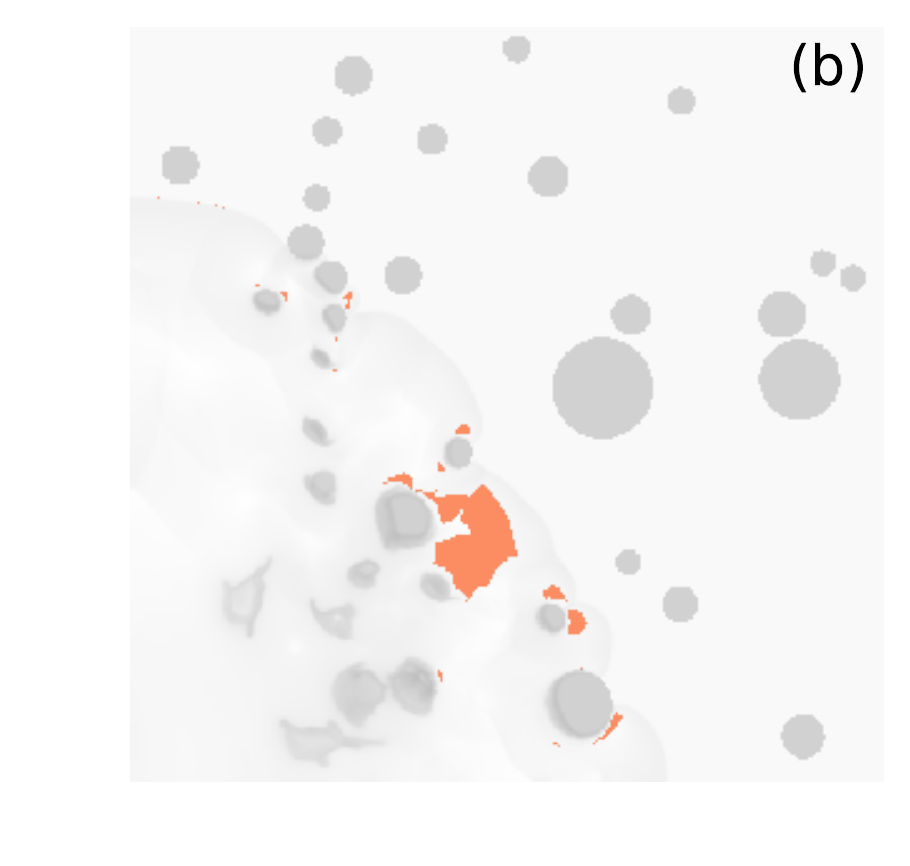}
\plottwo{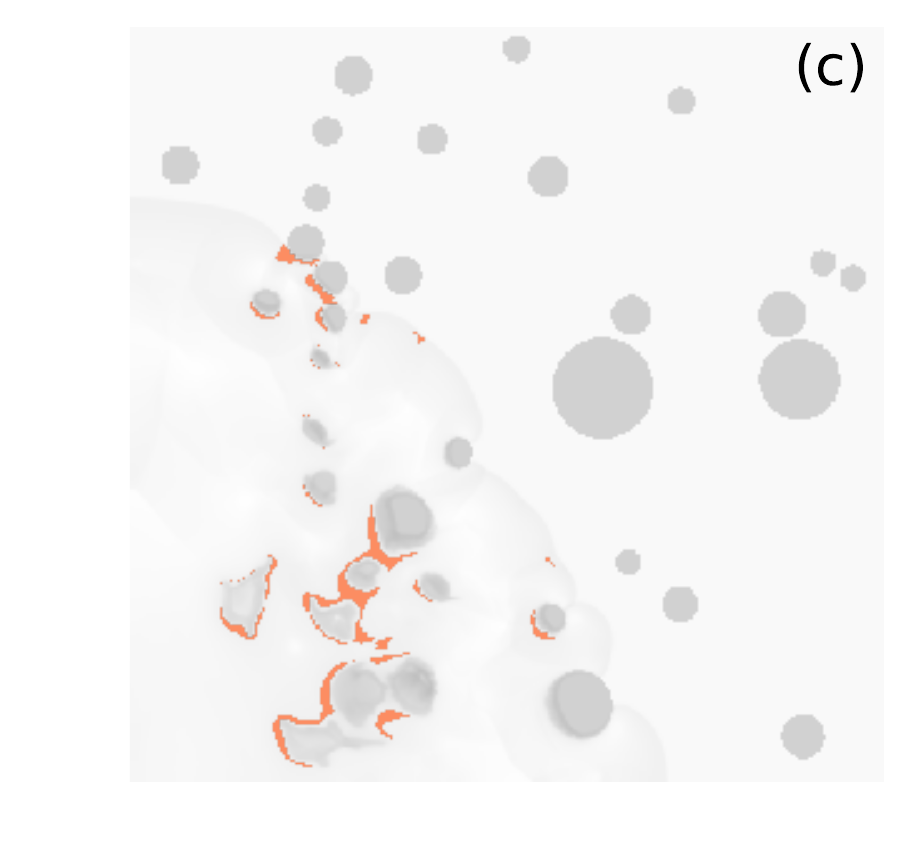}{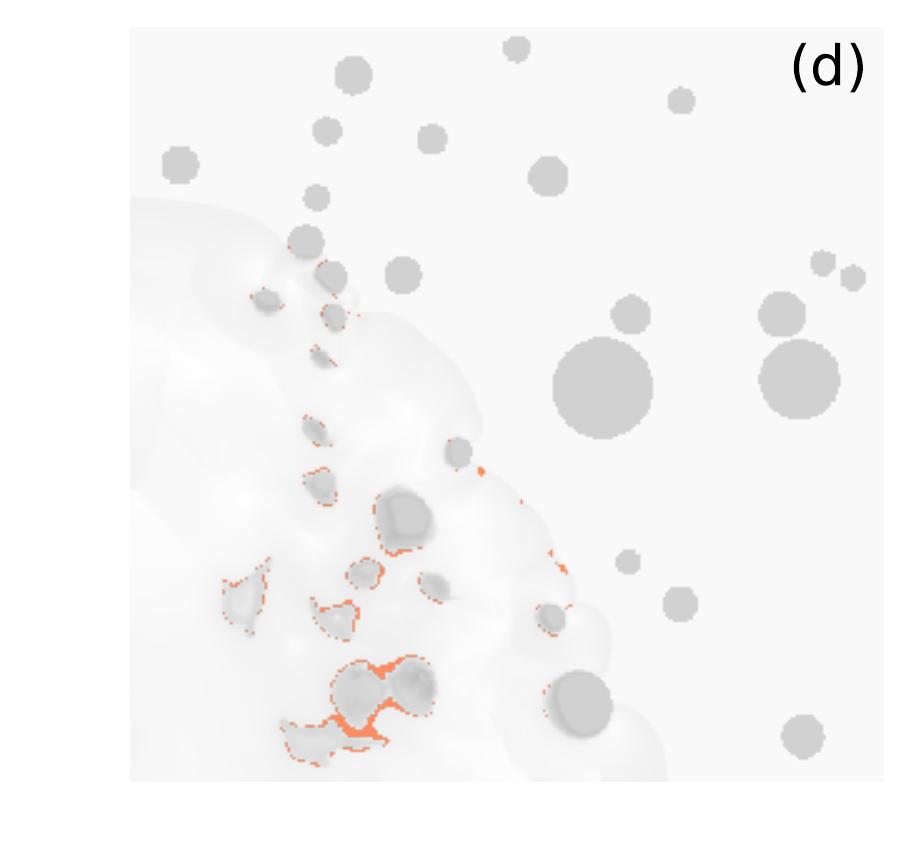}
\caption{Positions of regions in the temperature-density phase map 
(Fig.~\ref{fig:dens_temp_delc_mass}). Grey scale is the density distribution. 
\label{fig:regions1}}
\end{figure}

\textit{i.\ Evolution on the phase plots.}

As suggested in the single cloud simulation, there are two main
causes for recombination. One is the adiabatic expansion in the inter-cloud
area; the other is the energy transfer between interactions of hot and cold gas.
As suggested above (\S~\ref{sec:tc}), thermal conduction does influence the NEI process.
To show the different reasons for the NEI states, we scatter-plot all the 
pixels in model ``C1'' (refined to 512$\times$512) in a temperature-density phase
plot (Fig.~\ref{fig:dens_temp_delc_mass}). The 
average charge difference ($\Delta\bar{c}$) shown at each point is a  {density-squared-weighted} average
of the pixels in model ``C1'' with that same temperature and density.

Different physical processes can be shown in the phase plot.
Thermal conduction happens whenever hot and cold materials come into  
contact. 
Assuming that the gas is in rough kinetic balance, which would not cause abrupt expansion, the change of temperature and 
density should keep the pressure unchanged ($T\propto\rho^{-1}$) and 
follow an isobaric path in a phase diagram (dotted lines in 
Fig.~\ref{fig:dens_temp_delc_mass}).
When the pressure is much higher than the ambient gas, it performs a 
negative work, which decreases the internal energy and the temperature
very quickly. Assuming that there is no internal energy exchange, the adiabatic
process will follow $T\propto \rho^{\gamma-1}$ (dashed lines in 
Fig.~\ref{fig:dens_temp_delc_mass}), where $\gamma=\frac{5}{3}$ is 
the ratio of specific heats at constant pressure and at constant volume.
In Fig.~\ref{fig:dens_temp_delc_mass}, one immediately apparent feature is the ionizing
shock front at an almost constant density (region a).
In the \xray\ emitting gas, defined here as $T>10^6\K$, the ionizing and recombining plasmas seem to be separated
roughly by a certain value of pressure. The ionizing gas at a higher pressure has features that appear to follow isobaric lines (region c). 
Below region c, there is a recombining region at a lower temperature 
(region d). This is consistent with the influence of thermal conduction, which can
not only heat the cold gas but cool the hot gas. 
The recombining gas at a lower pressure (bottom left of the figure)
has some features that suggest it could be adiabatically expanding (region b). 
To check the physical positions of the gas in different ionization state,
the regions a, b, c, and d in Fig.~\ref{fig:dens_temp_delc_mass} are shown in
orange in Fig.~\ref{fig:regions1}, overlapping a density distribution to show
the cloud distribution.
The map of region a is consistent 
with the shock front and some ionizing regions behind clouds 
(See Fig.~\ref{fig:singlecloud} for a similar ionization distribution). 
Recombining region b is in the post-shock region behind
some clouds, which should be in expansion. Ionizing region c and recombining region
d seem to be next to the cloud regions. These are likely compressing regions,
where 
 thermal conduction dominates the cooling process.

\begin{figure}[htpb]
  \plottwo{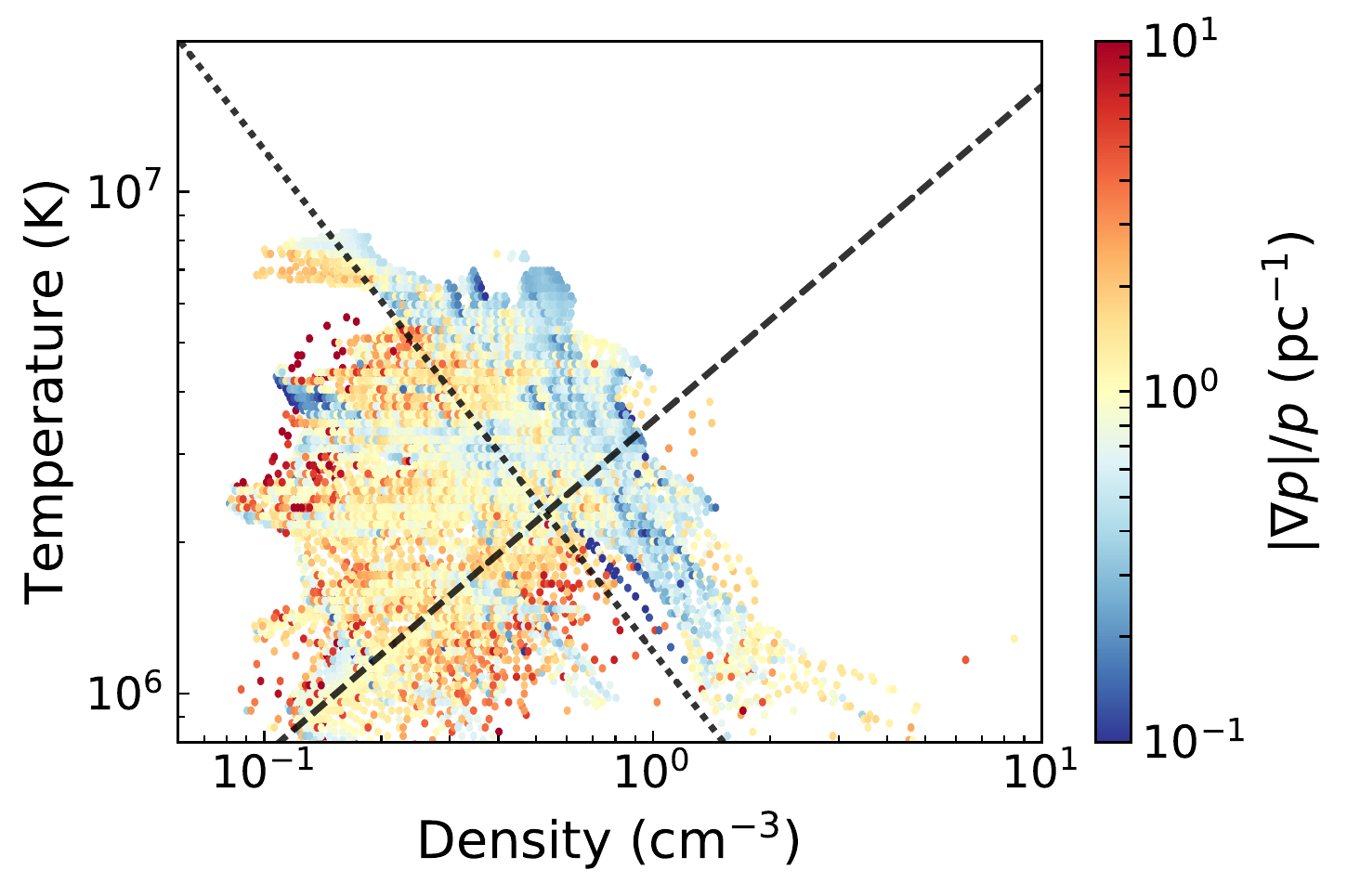}{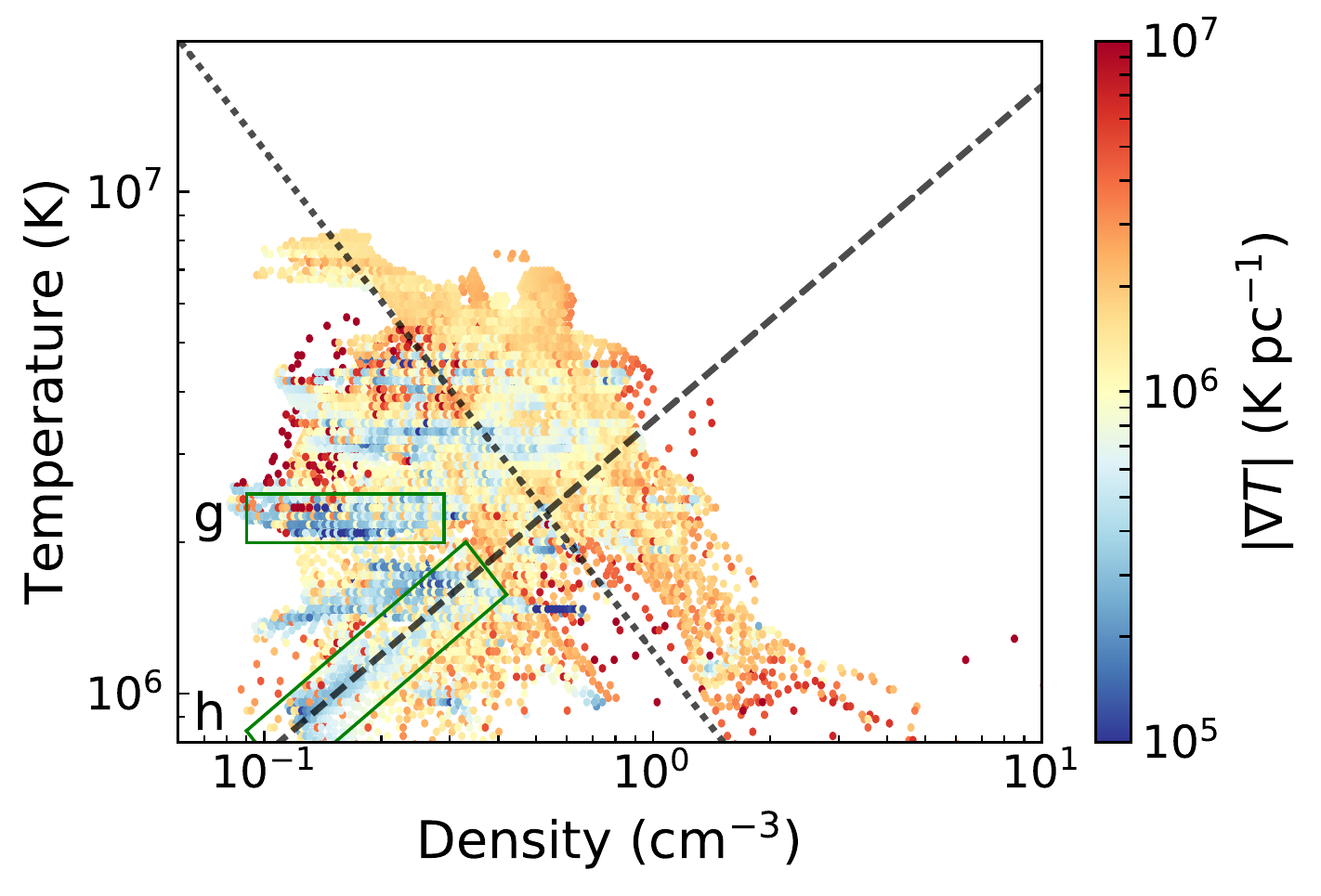}
  \plottwo{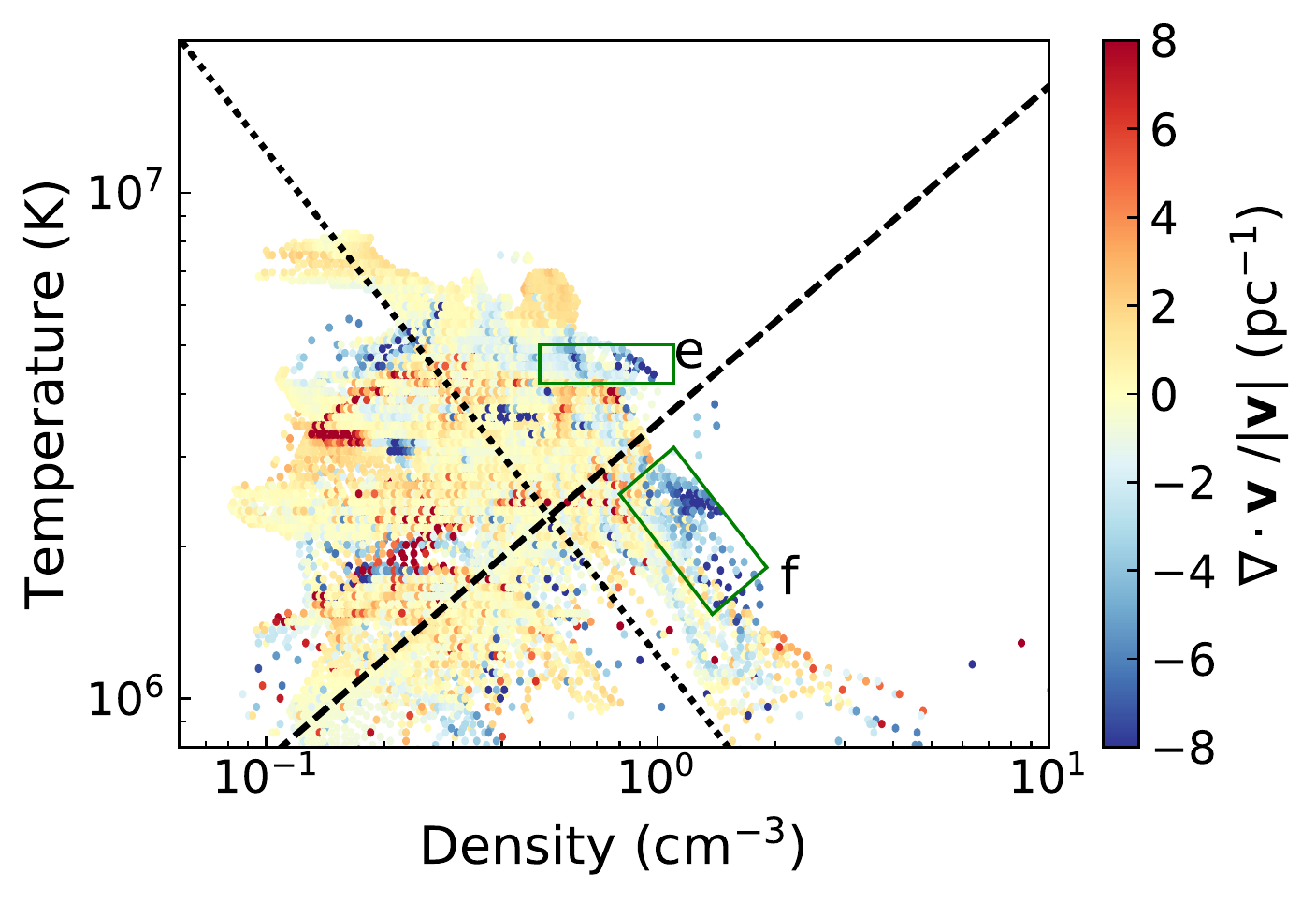}{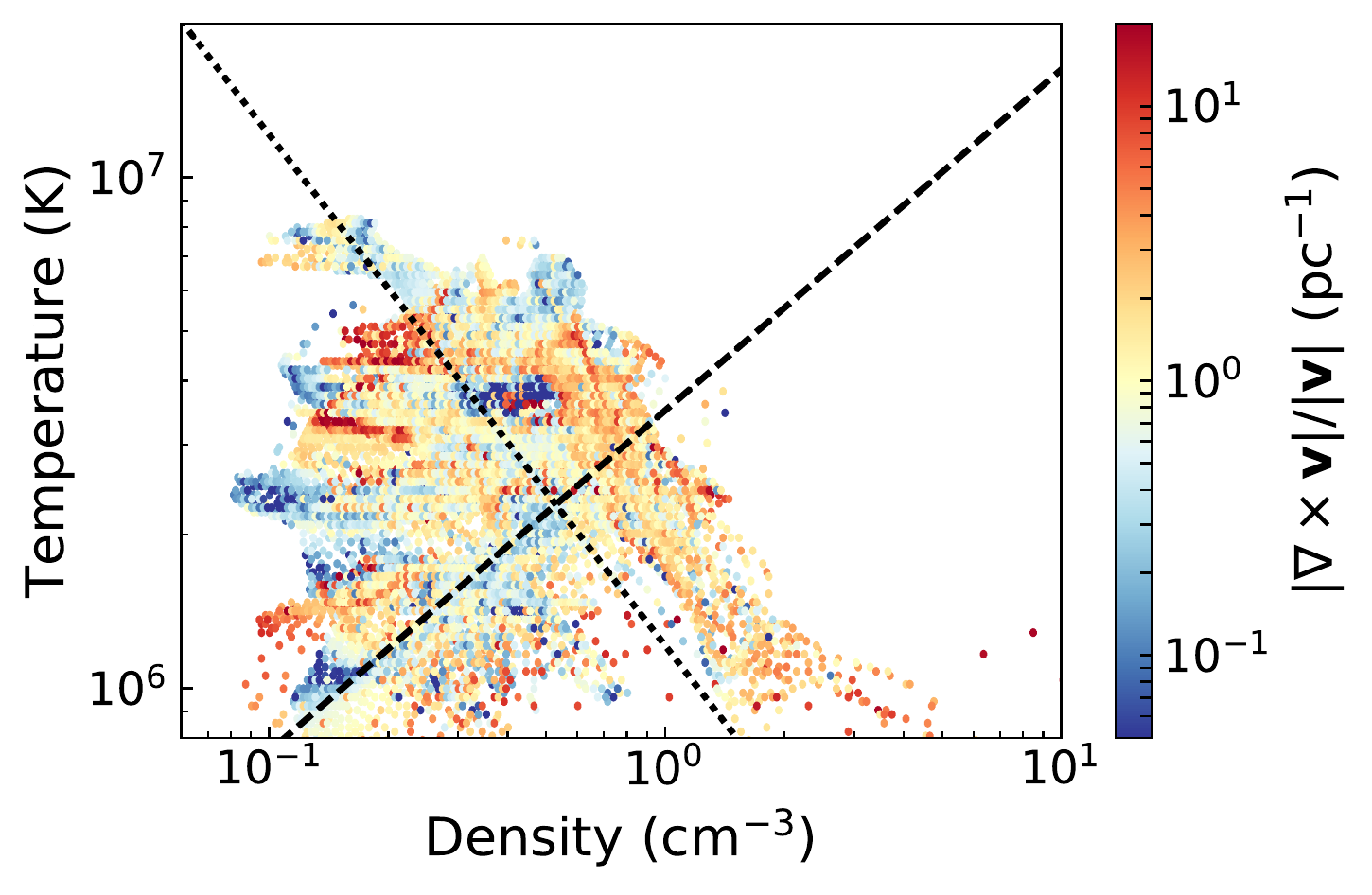}
  \caption{Phase plot (for O; at 20000 yr) for recombining plasmas
  with the magnitudes of relative gradient pressure vectors (top left),
  temperature gradient vectors (top right), relative velocity divergence, 
  and relative vorticity
  in color. The dotted line is the isobaric
  line for pressure 3.5$\times10^{-10}\Ba$, and the dashed line is an adiabatic
  process line. Four regions in the phase plot are also shown in the spatial distribution
  in Fig.~\ref{fig:regions2}. 
    \label{fig:dens_temp_gpop_gT}}
\end{figure}


\begin{figure}[htpb]
\plottwo{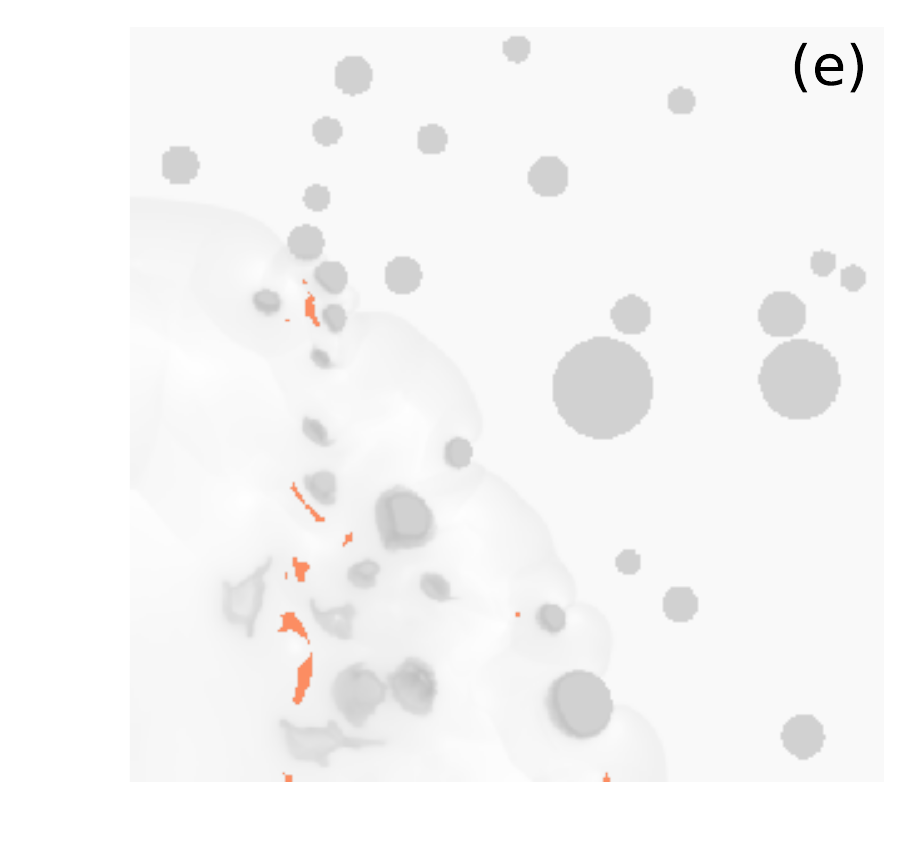}{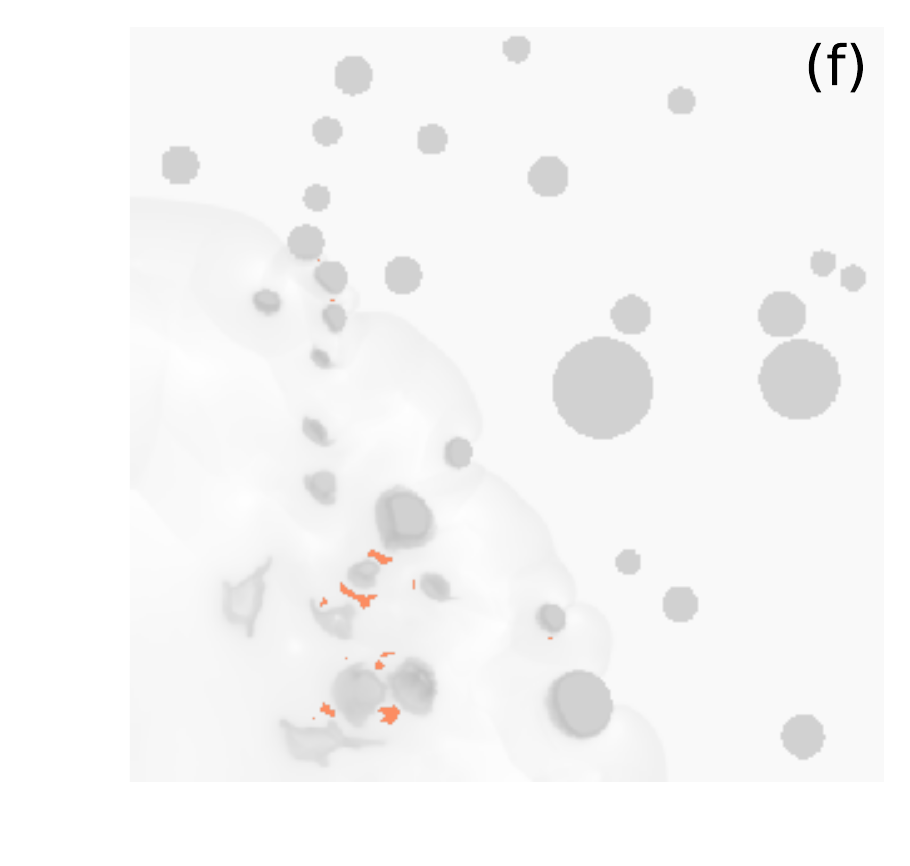}
\plottwo{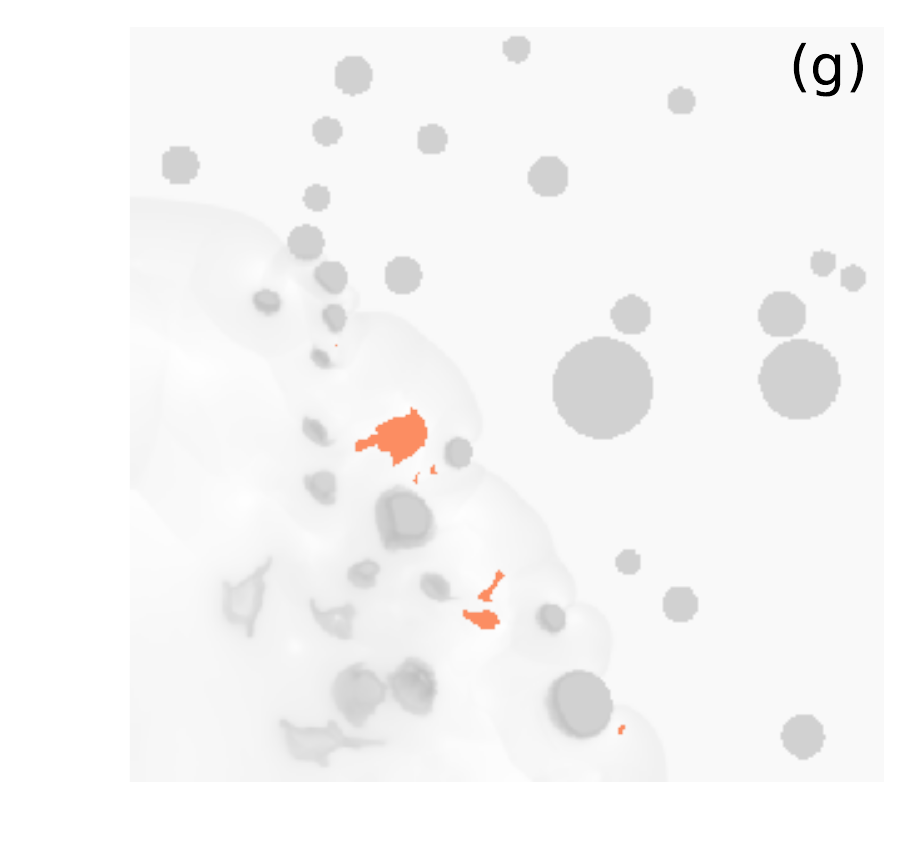}{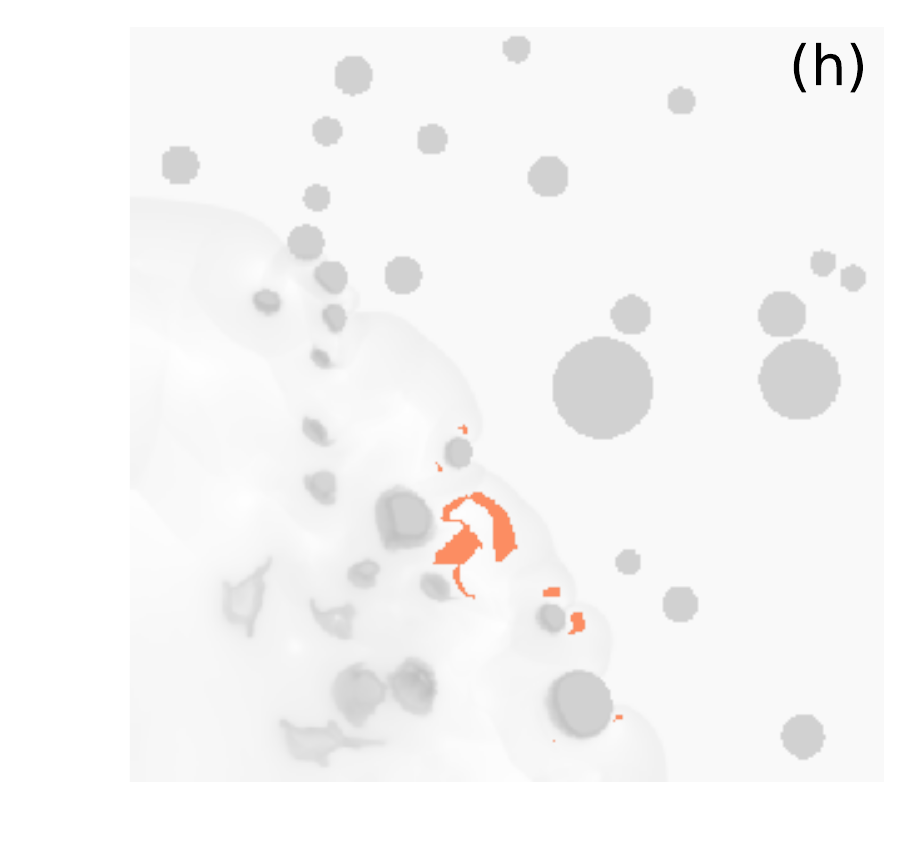}
\caption{Selected regions in the temperature-density phase map of recombining plasmsas.
(Fig.~\ref{fig:dens_temp_gpop_gT}). Grey scale is the density distribution. The recombination process in regions (e) and (f) seems to be driven by thermal conduction, while in regions (g) and (h) it is dominated by adiabatic expansion.
\label{fig:regions2}}
\end{figure}

Considering the ultimate goal of connecting the \xray\ observations to the underlying physical properties (see also appendix~\ref{sec:mimic}), Figures \ref{fig:dens_temp_delc_mass} and \ref{fig:regions1} imply 
the NEI amount ({\it e.g.} $|\Delta\bar{c}|$) is less important than the sign of $\Delta\bar{c}$. 
In the phase diagram (Fig.~\ref{fig:dens_temp_gpop_gT}) we show a range of quantities measured in the recombining plasma only (defined as $\Delta\bar{c}<-0.001$).
The partial differential (of gradient, divergence, and curl) calculations adopt
a central finite difference method.
The value of each point in this figure is the average value of the simulation
pixels that are recombining with the same temperature and density.
There are regions in this figure that can be analyzed to determine the
dominant physical reason for the recombination.
Thermal conduction, for example, needs a high temperature difference between the conducting
materials 
($\nabla T$), while adiabatic expansion has a positive velocity divergence 
($\nabla\cdot\vec{v}>0$). 
In the top left panel of Fig.~\ref{fig:dens_temp_gpop_gT}, 
the pressure gradient is relatively low 
at a higher pressure. When the pressure gradient is high in a low 
pressure environment, the plasmas are in an unstable state, and 
dynamical processes should dominate the temperature changes.
In the top right panel, there are some features in the bottom left
with a low temperature gradient,
corresponding to the higher relative pressure gradient areas in 
the top left panel (such as region g and h). In these regions expansion cooling could dominate, especially as some of them appear to follow
an adiabatic path (region h). In Fig.~\ref{fig:regions2}, 
regions g and h are all in inter-cloud regions that are also in between the 
shock front and cloud regions, also suggesting an adiabatic expansion cooling.
In the bottom left panel of Fig~\ref{fig:dens_temp_gpop_gT}, the velocity 
divergence shows whether the plasmas are expanding ($\nabla\cdot\vec{v}>0$, 
in reddish color) or compressing ($\nabla\cdot\vec{v}<0$, 
in blueish color). Regions with a negative velocity divergence, such as
region e and f, are not in expansion so the adiabatic expansion could not
contribute to the cooling in these regions. In the top right panel, thermal conduction could dominate with a high temperature gradient in the same areas. 
In the bottom right panel, the magnitude of vorticity depicts how the materials
are mixing together. The fluid instability that happens on the contacting 
surface of gas increases the vorticity. Expanding regions tend to have
a low vorticity on the contrary. Regions e and f have a high vorticity, also
suggesting the hot and cold plasmas are contacting with each other.
In Fig.~\ref{fig:regions2}, the map of region e seems to be in some complex 
cloud areas. Complex reflected shocks collide with each other. Region f seems
to be on the rims of clouds, where the thermal conduction should dominate the cooling.
It should be noted that both mechanisms can contribute the recombination all over
the phase map in previous time steps rather than the current time step. 
The recombination may not be consistent with cooling, because
there are some recombining gas parcels reheated without enough time to change the
ionization state.

\begin{figure}[htpb]
  \plottwo{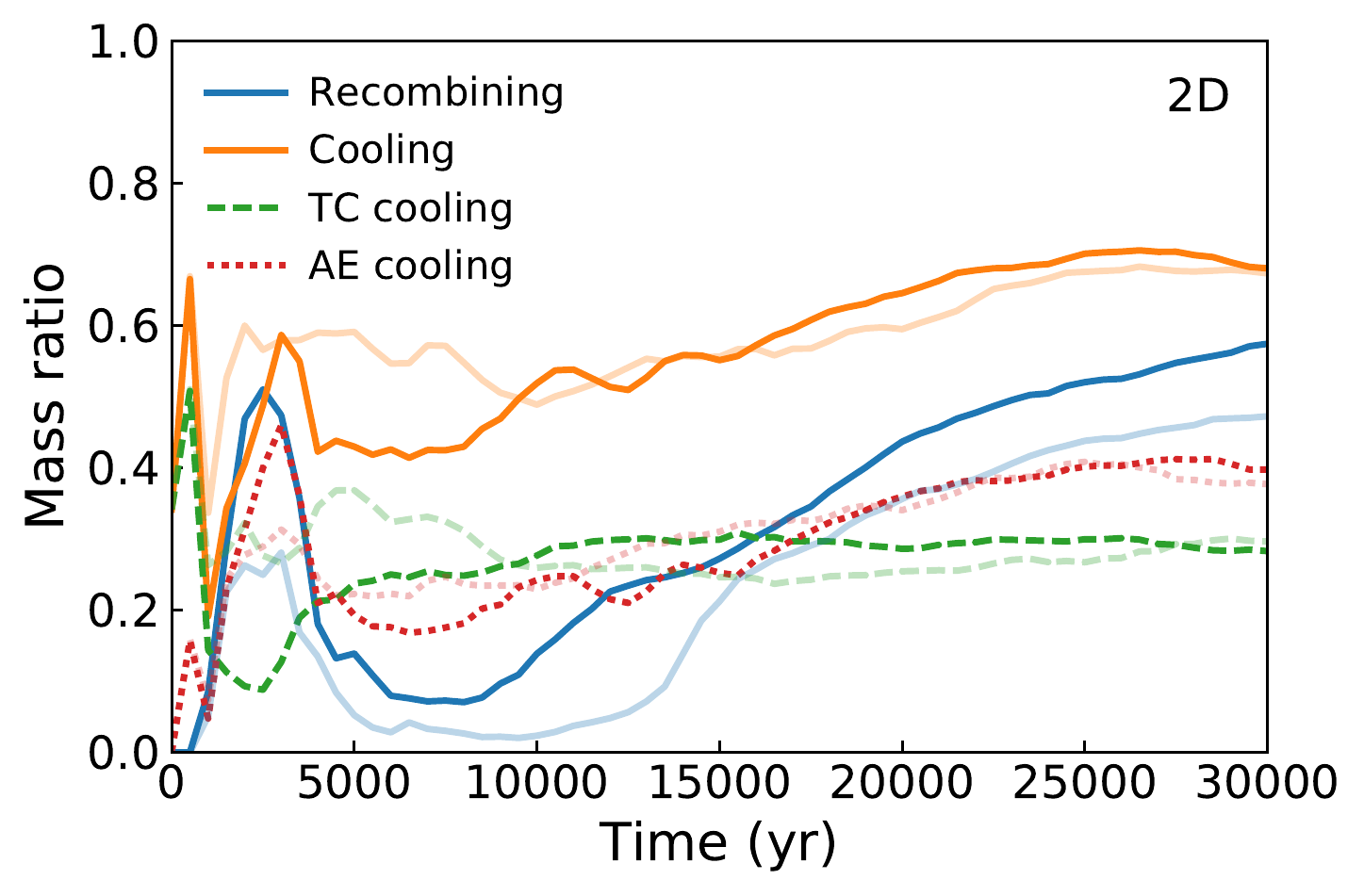}{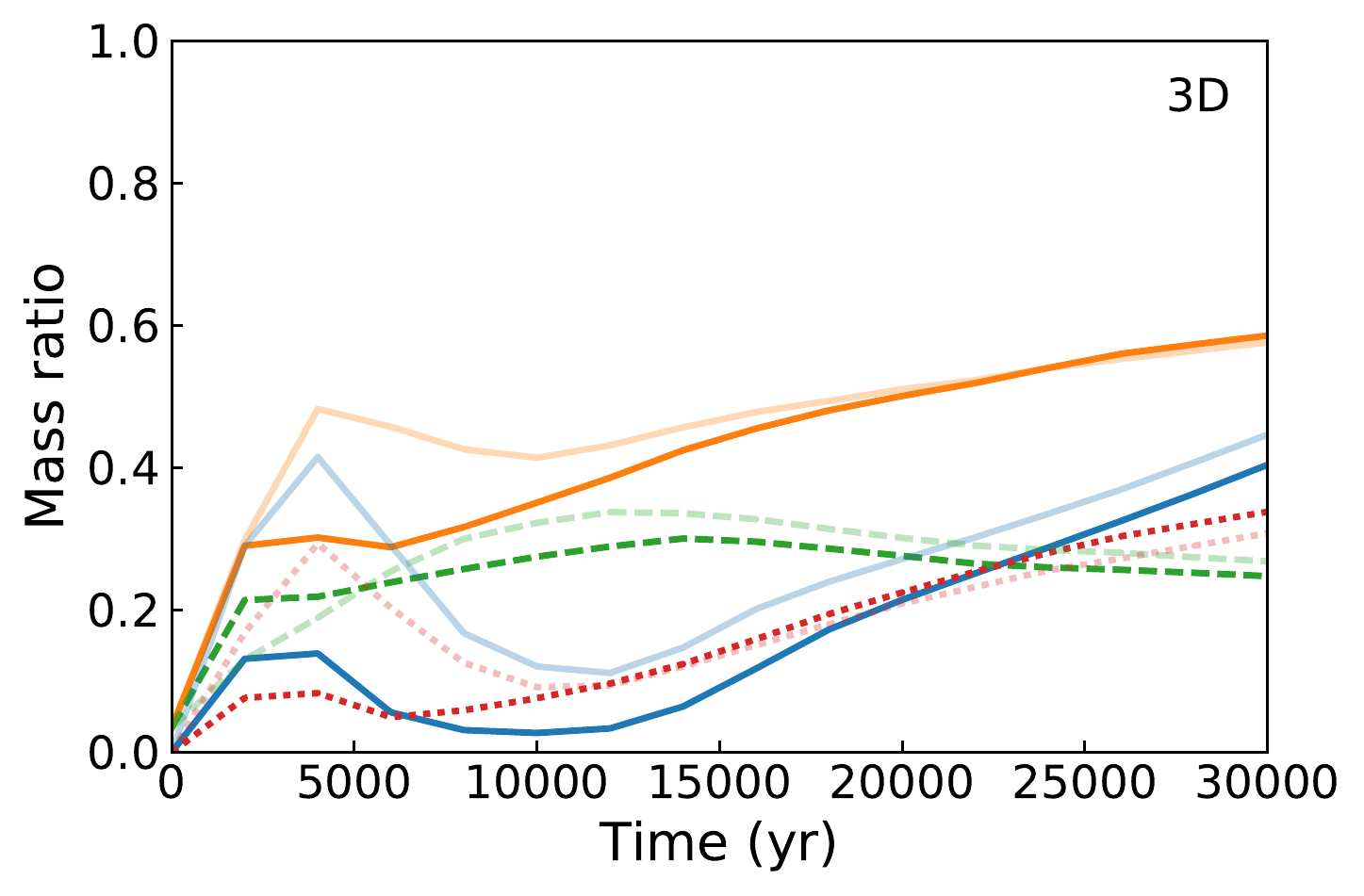}
  \caption{The mass ratios of recombining plasmas, cooling plasmas, thermal conduction 
  dominant cooling plasmas, and adiabatic expansion cooling plasmas in 2D simulations (left) 
  and 3D simulations (right)
  as a fraction of the total \xray\ emitting gas mass($T>1\times10^6\K$).
  A different
  random cloud distribution (with the same {WLC} parameter) is shown for both 2D and 3D 
  simulations in translucent lines.
    \label{fig:massratio}}
\end{figure}

\begin{figure}
    \centering
    \includegraphics[width=0.9\textwidth]{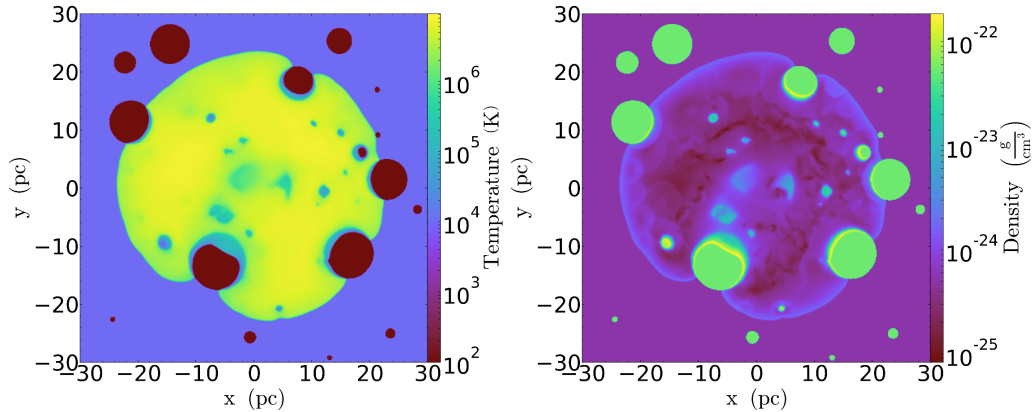}
    \caption{The temperature (left) and density (right) distribution of a slice plane at $z=0$ of the 3D simulation (at $2\times 10^4\yr$).}
    \label{fig:3d_slice}
\end{figure}

\begin{figure}
    \centering
    \includegraphics{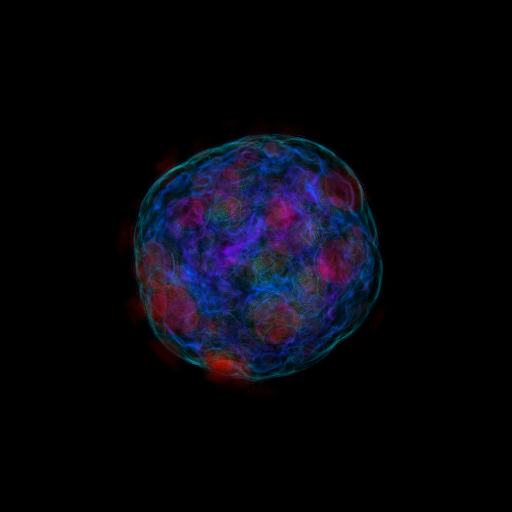}
    \includegraphics{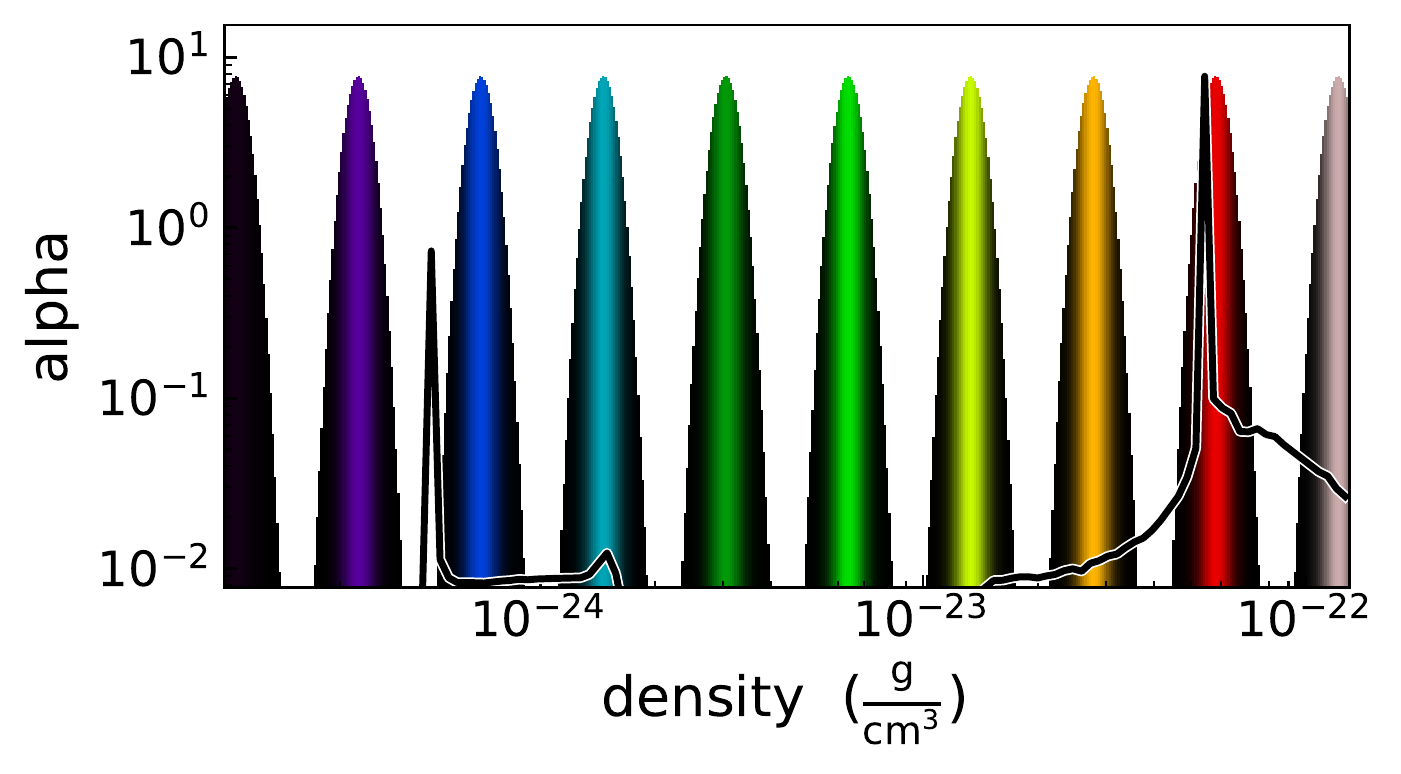}
    \caption{The density distribution of 3D simulation (at $2\times10^4\yr$) with a volume rendering method. The transfer function of the volume rendering is shown in the
    bottom panel, where the histogram of density is shown in black. From left to
    right, the first peak of the black line is at the initial density of the inter-cloud ISM, corresponding to a white region (the volumes with this density are left transparent). At the right wing of this peak, it is the density of shocked ISM, which is shown in blue and cyan. The second peak is the initial density of the dense clouds, corresponding to the red color in the top panel. All the clouds farther than $25\parsec$ are excluded.}
    \label{fig:3d_map}
\end{figure}

\begin{figure}
    \centering
    \includegraphics{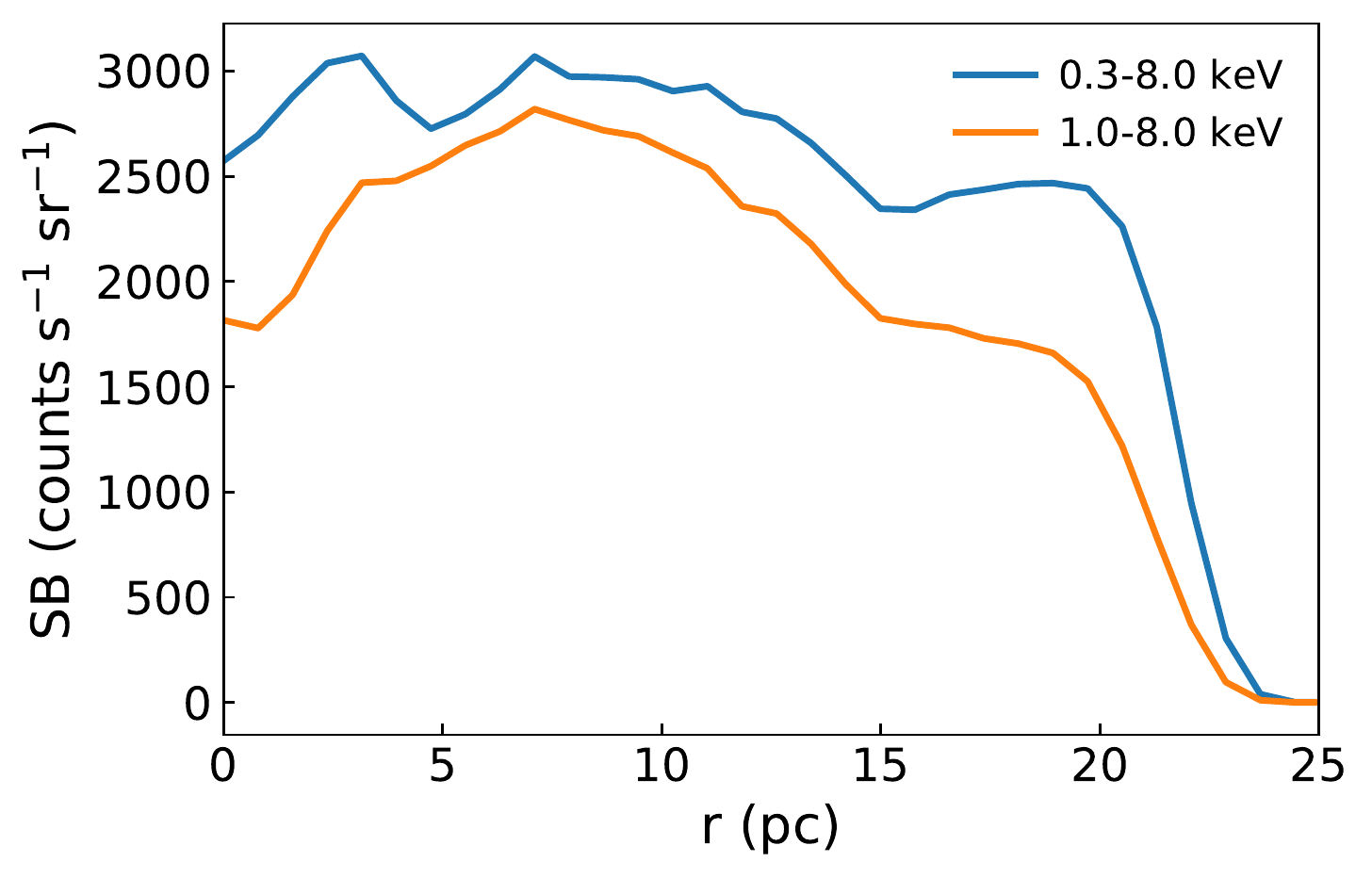}
    \caption{The radial surface brightness of the projected 3D simulation at 2$\times10^4\yr$ with a range of energy (0.3--8.0$\keV$ and 1.0--8.0$\keV$). The element abundances are assumed to be solar abundance and the emission model is assumed to be CIE (``{\em apec}'' model in {\em Xspec}).
    The photons are collected with the Athena instrument response. The harder band (1.0--8.0$\keV$) uses a higher exposure time (See the text).}
    \label{fig:sb_full}
\end{figure}

\textit{ii.\ Mass ratio of cooling components.}

In view of the hydrodynamic evolution, the reason for recombination can
be estimated by the temporal evolution of the cooling material. 
With the hydrodynamic equations in Lagrangian scheme, 
the internal energy change can be obtained with the equation,
\begin{equation}
\frac{\dd U}{\dd t}=-c^2\nabla\cdot\vec{v}+
\frac{1}{\rho}\nabla\cdot(\kappa\nabla T),
\end{equation}
where $U$ is the internal energy relating to $T$, $\rho$ is the density, $c$ is
the sound speed, $\vec{v}$ is the velocity vector, and $\kappa$ is the thermal 
conductivity. 
If only the cooling is considered, the first
term is the adiabatic expansion cooling, and the second term is the thermal conduction 
cooling. 
We can hence calculate both of the terms for a given parcel to determine whether
it is cooling or heating.
In cooling parcels, if the absolute value of the first term is greater than the one
of the second term, the adiabatic expansion is dominant (AE cooling); 
if the absolute value
of the first term is smaller than the one of the second term, the thermal conduction
is dominant (TC cooling). In the heating parcels, the dominant term can be determined with the
same method.
By summing up the mass in these categories, we can know how much of the gas
is cooling versus heating and ionizing versus recombining. As we are interested
in the \xray\ recombination phenomenon, 
Fig.~\ref{fig:massratio} plots the mass fraction of some categories, including 
cooling, TC cooling, AE cooling, and recombining
mass in the total hot gas ($T>1\times10^6\K$) mass.
In the 2D models, at an early stage, one or two of the clouds are engulfed, making the cooling mass
and recombining mass fraction increase abruptly. The AE cooling also seems
to exceed TC cooling. Later at around 10000 yr, 
the recombining mass proportion rises again. In the period from 5000 yr 
to about 15000 yr, TC cooling exceeds AE cooling. 
From Fig.~\ref{fig:mmsnr} (bottom left panel), 
the recombining plasmas are mainly in the clouds area, implying the thermal
conduction could possibly be the dominant term for the recombination during this time. 
The thermal conduction cooling mass remains nearly a constant, which is expected
considering the evaporation of clouds does not change dramatically \citep[as shown in][]{Slavin2017}. 
The adiabatic expansion proportion exceeds the thermal conduction at an even
later time. Some of the mushroom features in the late time recombining images 
(Fig.~\ref{fig:mmsnr}) also
suggest a contribution of adiabatic expansion. 

\subsubsection{The simulations in 3D}

In the 3D simulation, the total number of clouds increases over the 2D model.
{The clouds are spheres instead of rings in 2D cylindrical simulations.}
But the hydrodynamic behaviors are similar in a slice plot of the 3D simulation
(See {Fig.~\ref{fig:3d_slice} and} \citealt{Slavin2017}). 
{However, with the symmetry, 2D simulations will be
affected by the clouds on or near the $z$-axis and $x$-axis. 
In Fig.~\ref{fig:mmsnr}, 
we can see that
the clouds are distributed closer around $x$-axis than the one 
around $z$-axis. This effect causes a little higher temperature around
$z$-axis. A full volume 3D simulation can avoid this effect. In
Fig.~\ref{fig:3d_slice}, the slice plots in the 3D 
simulation show that the distribution of temperature and density are
affected by the distribution of the clouds.}
{With the Python module yt-project, a 3D volume rendering figure is 
shown in Fig.~\ref{fig:3d_map}. The color coding is tuned to show the shock front 
in cyan and the dense clouds in red. The shape of SNR that is distorted by
the distribution of the dense clouds is not symmetric. The cyan filaments
showing the shock front are formed by the projection effect. }

{
From the 3D simulation, we generated X-ray photons in each pixel according to 
the emission model and the physical properties in situ and projected them onto
an X-ray observation instrument (\texttt{XIFU} on Athena). By assuming that 
the plasma is in CIE (``{\em apec}'') and all the metal abundances are solar abundances
\citep{Anders1989}, Fig.~\ref{fig:sb_full} shows the average surface brightness
as a function of radius. In the X-ray band from 0.3$\keV$ to 8$\keV$, the
thermal emission is centrally brightened. Since MMSNRs are almost always 
heavily absorbed, the energy range from 1$\keV$ to 8$\keV$ has been shown in
this figure as well. The surface brightness in the center ($r<5\parsec$) is lower,
which might be caused by the empty center ($r<2.25\parsec$) in the initial condition.
Here, the smaller energy range ($1\keV$--$8\keV$) has an exposure time of 1000$\ks$; and the larger one ($0.3\keV$--$8\keV$) use a shorter exposure time
(100$\ks$).
In the appendix \S\ref{sec:mimic} we describe the details to generate 
simulated observations, and make use of the NEI information in the simulations.}

We have {also} produced the figures like Fig.~\ref{fig:mmsnr},
Fig.~\ref{fig:dens_temp_delc_mass}, and Fig.~\ref{fig:dens_temp_gpop_gT}
for 3D as well. They are similar to the figures of 2D, albeit with lower resolution.
The 2D simulations are shown here because the spatial distribution is easier to 
illustrate in 2D than 3D. In the right panel of Fig.~\ref{fig:massratio},
the mass ratio of recombining and cooling components is shown for 3D simulations.
The time resolution is sparser because of the considerable sizes of the files. It is
hard to investigate the early stages (about 2000 yr to 4000 yr) when the AE cooling
and TC cooling components change dramatically. The crossing of these two components
at the late stages is postponed from about 16000 yr to 23000 yr,
because of more fragmentary clouds.
{
In both 
2D and 3D simulation, the choice of distribution does not change the 
conclusion about the sources of recombination.}

We close this section by noting several key points.
\begin{enumerate}
\item We tested multiple different sets of random clouds distribution with the same {WLC}.  Although the results are not exactly the same, the overall trend is unchanged.
\item The cooling or heating components are shown as a function of their mass, not the total energy change. Thermal conduction can both heat and cool two groups of gas that are contacted 
simultaneously, leaving the total energy transfer as zero. 
\item With a lower limit (10$^6\K$) to the temperature, both thermal conduction and adiabatic expansion can cool an \xray\ emitting gas to be \xray-quiet. However, a cool gas can also be heated to an \xray\ emitting gas. So if the heating and ionizing components
are analyzed, the lower limit in temperature will underestimate the gas being heated. 
Since we focus on the recombining component in this paper, it does not impact our conclusions.
\item The times given here are for a specific ambient density. Different SNR observations should be compared to the simulation with corresponding initial conditions to make sure the evolution age is valid.
\end{enumerate}

\section{Discussion}
\subsection{Applications of the simulations to observations}

\begin{figure}[htpb]
\includegraphics[width=\textwidth]{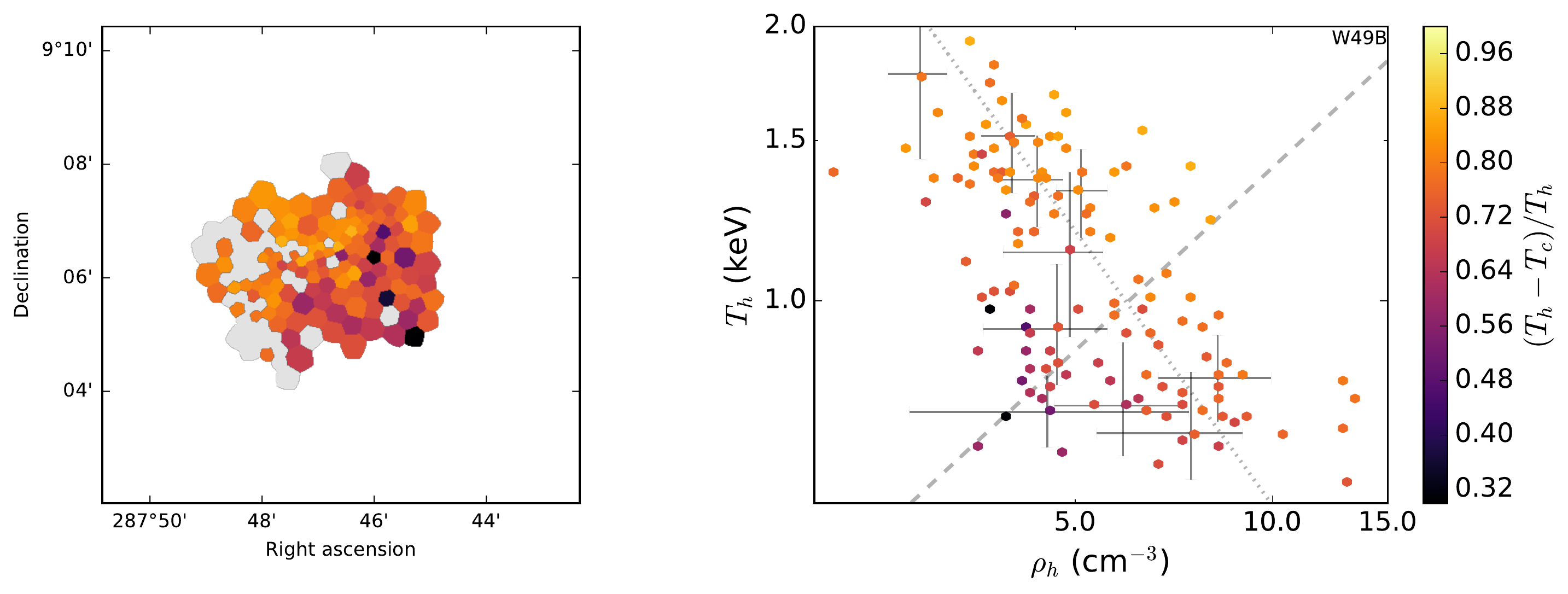}
  \caption{The relative temperature difference between the hot component and the cold component in a
  W49B \xray\ observation \citep{Zhou2017} is shown in the left panel with the 
  equilibrium points shown in gray. Recombining hot components in W49B (with the
  subscript of $h$) 
  are scattered into the temperature-density phase diagram (right panel).
The dashed line
  depicts an adiabatic evolution path and the dotted line depicts an isobaric line.
  Similar to Fig.~\ref{fig:dens_temp_gpop_gT}, bottom left points in the phase map
  tend to have low temperature difference, supporting an adiabatic expansion origin.
  Ten random points are chosen to show the error bar  (90\% confidence).
    \label{fig:w49b}}
\end{figure}

With the spatial resolution in \xray\ observations, a temperature-density diagram 
can show the relative position of each region in the phase map as the one
in Fig.~\ref{fig:dens_temp_gpop_gT}. Fig.~\ref{fig:w49b}
is an example for a W49B observation \citep{Zhou2017} with 
an adiabatic expansion line and an isobaric line shown as dashed and dotted lines. 
The error bars are from the 90\% confidence ranges of the hot component fitting,
where the errors from the cool component fitting 
are not included.
Only recombining (timescale 
$\tau<1\times10^{13}\s\cmmthree$) hot components are shown in the phase diagram (right). 
The colormap shows the relative
difference between the fitted temperature of hot components and the cooler components. 
It seems the low pressure regions tend to have a smaller relative temperature difference.
If the two components are assumed to be thermally conducting, the smaller the temperature difference the smaller the thermal conduction flux. Of course, this is not the same as a true thermal gradient; the two components may not be in actual contact but merely in the same line of sight.
{The points along the isobaric line with larger temperature differences show a probable thermal conduction contribution to the overionization.}
The points in the south west 
of the SNR with smaller temperature differences seem to be along an adiabatic expansion
line in the phase diagram, which could be a strong hint for an adiabatic expansion dominance. 
{With more \xray\ instruments coming on board, {the phase diagram} can be a useful tool to diagnose 
the NEI process in SNR plasmas.}

{
The distribution of the overionized plasma also depends on the SNRs' immediate environment. In the case of W49B, it was also suggested that both thermal conduction and adiabatic expansion resulted in the overionized plasma, but a density enhancement of the ambient medium is assumed near the SNR center.
In the numerical simulation of \citet{Zhou2011}, it was also suggested that both thermal conduction (mixing of hot and cold plasma) and adiabatic expansion resulted in the overionized plasma in W49B, but a density enhancement of the ambient medium is assumed in the SNR interior. This scenario is recently supported by \citep{Zhou2017}.}
{On the other hand, \citet{Miceli2010} proposed a scenario that W49B requires dense
clouds close to the explosion center.
When the blast wave and the 
ejecta break out into the surrounding low-density region, the plasma cools 
down rapidly and becomes overionized. It also supports the adiabatic 
expansion origin of the overionization. This scenario has been supported by
\citet{Lopez2013,Yamaguchi2018} and also been
proposed for SNR IC 443 \citep{Greco2018}.}

\subsubsection{Turbulent mixture}\label{sec:mix}

\begin{figure}[htpb]
  \plotone{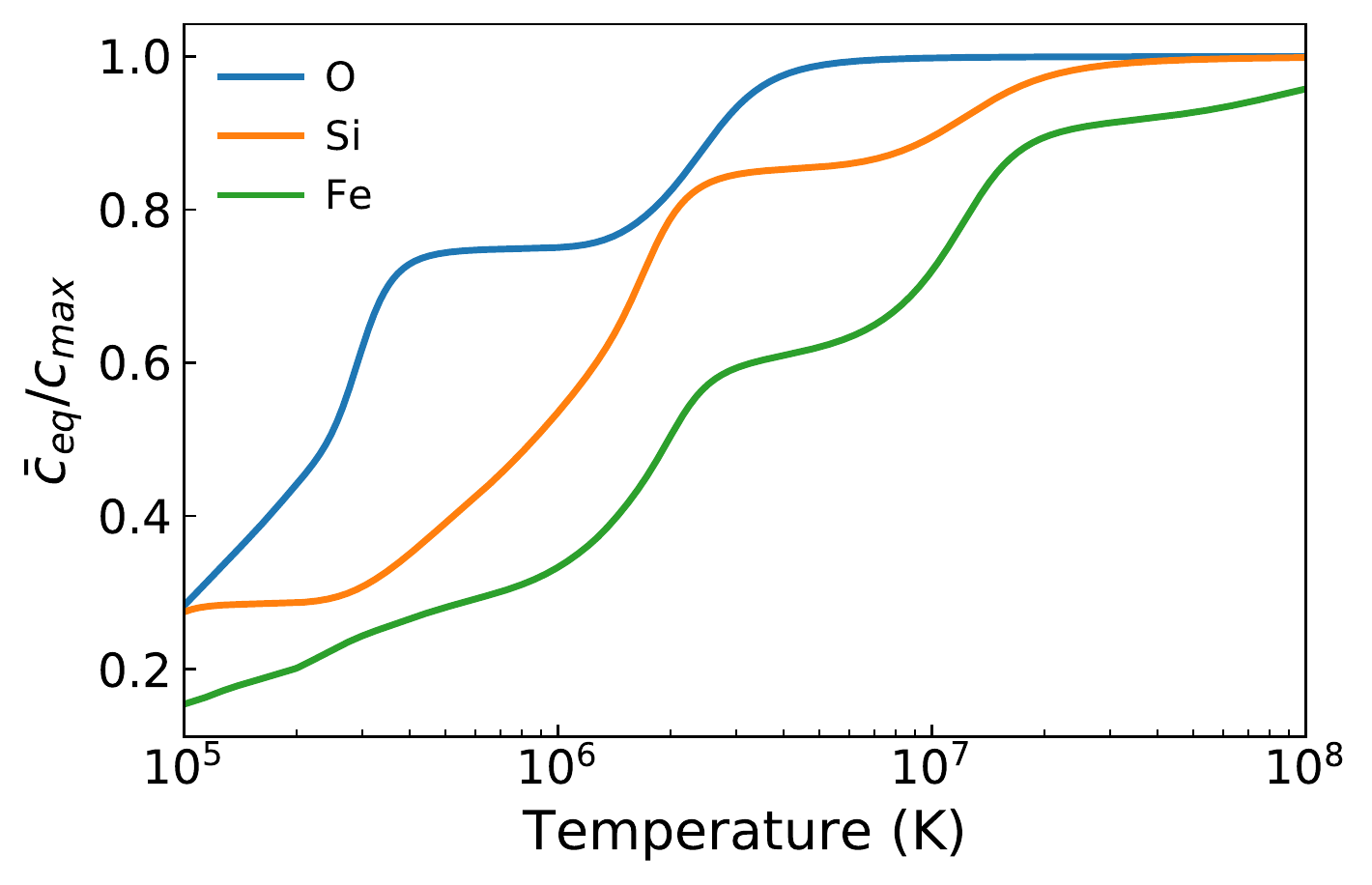}
  \caption{CIE average charge of different elements (O, Si, and Fe) as a function of temperature. 
  $C_{max}$ is the maximum charge for each elements (8, 14, and 26 respectively).
    \label{fig:avercharge}}
\end{figure}

{In the ISM the turbulent mixing of plasma is common \citep{Slavin1993},
which could be another source of over-ionization.}
When gas with different properties is advected into one of the highest resolution cells in a hydrodynamic simulation, the physical properties are effectively averaged.
The density, $\rho$, is averaged with volume weighting.
The temperature is calculated from the internal energy every time step
and is averaged with mass weighting,
$T = \frac{T_1m_1+T_2m_2}{m_1+m_2}$. The ion fraction for an element 
$f^{(i)}$ ($i=0,1,2...Z$, $Z$ is the atomic number) is also averaged with mass,
$f^{(i)}=\frac{f^{(i)}_1m_1+f^{(i)}_2m_2}{m_1+m_2}$. Therefore the average
charge can be expressed as
$\bar{c}=\sum\limits_{i=0}^{Z}f^{(i)}c_i=\frac{\bar{c}_1m_1+\bar{c}_2m_2}{m_1+m_2}$. 
From Fig.~\ref{fig:avercharge}, it can be seen that the equilibrium average charge is 
not linear with temperature. Even assuming the gas is in CIE before the mixing,
it can become effectively out of equilibrium as a result of mixing. However, since the ionization
is roughly linear over small ranges of temperature, mixing between 
materials with similar temperature will not cause substantial deviation from CIE.
Mixing is thus similar in its effects to thermal conduction.
Thermal conduction does not mix materials directly, but the internal energy 
is redistributed from the hotter to cooler gas. 
With enough 
elapsed time, this process also occurs when physically mixing plasmas. 
This becomes complex in a numerical simulation, however, because   
numerical viscosity and the effects of finite resolution, both of which
are connected to the simulation resolution, can create unphysical 
mixtures. 
By comparing different resolutions (See \S~\ref{sec:method}) of 2D simulations, 
we confirmed that the results shown above change very little.
In 3D simulations, with different resolutions, 
the slices of runtime variables look the same, and 
the mass of hot gas used in Fig.~\ref{fig:massratio} changes less
than about 1.5\%, which we deemed acceptable. 
More computation power is needed to find how much a better resolution could 
improve these conclusions in 3D simulations.

\subsection{Selection of elements and the ionization energy}
We use oxygen for most of the analysis in this work. Other elements can be used too.
As suggested by observations, some heavier elements, like Si, Mg, Fe, are easier to 
observe because of the high absorption commonly seen in MMSNR. However, the assumption of an 
equilibrium between the ion temperature and electron temperature should be carefully 
treated for those elements who are sensitive to electron temperature around $10^7\K$
(Fig.~\ref{fig:avercharge}). The electron temperature can be about $10^7\K$, while the ion
temperature is a factor of ten higher \citep[e.g.\ ][]{Patnaude2009}.

Although the ionization of some elements
are taken into account, the energy change from ionization and recombination 
is not calculated in current simulations. 
Considering the cosmic abundance of elements, heavy elements 
will not contribute much to the total energy. Both hydrogen and helium, however, will effectively `store' energy in their ionized states.
The energy will be released mainly into photons when they recombining back to lower states.
H and He have an ionization energy of 13.6\,eV and 79\,eV (fully ionized) respectively.
By assuming the mass ratio of He 0.79, and H 0.21, the ionization energy in a unit
mass is $E_{ion}=\frac{0.21}{m_H}\times13.6\eV+\frac{0.79}{m_{He}}\times79\eV$. If
100\,$M_{\odot}$ is swept by the SNR shock, $3\times10^{48}\erg$ (3\% of the explosion
energy) is needed for a full ionization. In fact, most ISM gas in our simulations is in a full ionization
or partially ionization state of H and He because of the fluctuation around temperature
$1\times10^4K$. Similarly, we can know the dissociation energy
for hydrogen molecules for the same swept mass is about $4\times10^{47}\erg$ (0.4\%
of the explosion energy).
It can then be assumed that the hydrogen and helium ionization does not 
affect the hydrodynamic evolution.

\subsection{Other possible effects}

Some observations have
found metal-enriched materials inside the SNRs \citep{Zhang2015}. 
The ejecta in SNR could affect the final results too, because it
has a larger metallicity than the ISM. After the shock front reaches about 10 pc from 
the explosion center the swept mass will be larger than the ejecta for ordinary SNRs
(ejecta mass is less than 8 $M_{\odot}$). 
We ignore the abundance from the ejecta as we focus on
the late stages of the SNR evolution. 
A further investigation of
simulations should be done with ejecta in the future for a more realistic model.

It is generally accepted that some of the high-energy cosmic ray (CR) particles are
accelerated in SNR, which could also be a source of the recombination.
The electrons in the ions are kicked out by the high-energy CR, appearing to be over-ionized
in the current temperature environment. The acceleration
process depends on the magnetic fields around the shock front. As no magnetic field is included in 
these simulations, we did not include this process.

\citet{Patnaude2005} have found that the interaction between a 
shock and a cloud with a smooth varying density changes the hydrodynamic instability 
from a sharp-edged cloud.
In addition, the ISM environment should be turbulent as simulated in \citet{ZhangD2018}, 
rather than a distribution of round clouds. 
To make a further comparison with observations it will be needed to consider
a turbulent environment with different cloud density gradients,
which we plan for a future project.


\section{Conclusions}

Including NEI calculations, several 2D and 3D SNR simulations are performed.
As expected, the shock front is always ionizing and the SNR in a homogeneous environment has
a good consistency with the Sedov-Taylor self-similarity solution. Both ionization
and recombination can be seen in the interior of the SNR with a cloudy environment. 
Both a direct analysis of the simulation results and a mimicked \xray\ observation
toward the SNR simulation are used. The thermal conduction is estimated for the 
contribution to the NEI. Especially for the cooling \xray\ emitting plasmas, thermal conduction
could contribute in a similar amount as the adiabatic expansion.

\section*{Acknowledgement}
We would like to thank Patrick Slane, John Raymond, Daniel Patnaude, and Hiroya Yamaguchi
for helpful comments and discussions.
This work is partially supported by the CSC, the Smithsonian Institution Scholarly Studies Program, the 973 Program grants 2017YFA0402600 and 2015CB857100, and NSFC grants 11773014, 11633007, 11851305, and 11503008.
G.Y.Z is also supported by the program A for Outstanding PhD candidate of Nanjing University 201802A019.
\software{FLASH code \citep{Fryxell2000}, yt \citep{Turk2011}, pyXSIM\citep{ZuHone2014}, SOXS\citep{ZuHone2014}, pyAtomDB\citep{Foster2012}}

\appendix

\section{Compare model ``B2'' with the theory}
\label{sec:theory}

\begin{figure}[htpb]
  \plotone{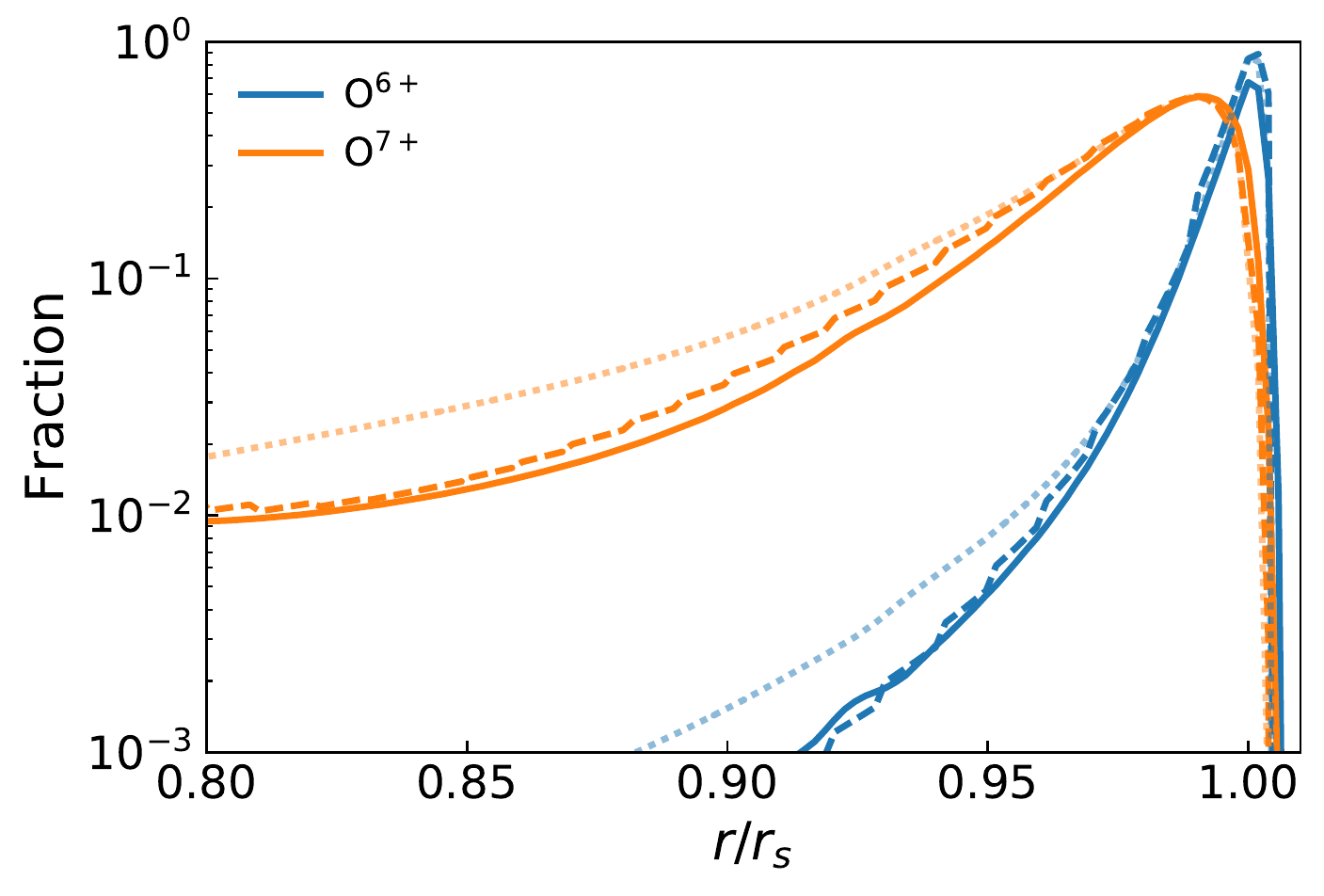}
  \caption{Ion fractions (O$^{6+}$ and O$^{7+}$) at the shock front of an SNR 
  in homogeneous environment (10000 yr). 
  The solid lines are from the simulation without 
  thermal conduction. The dotted lines are from the estimation of a theoretical
  Sedov-Taylor self-similarity solution of the blast wave.
  The dashed lines are the same theoretical model with the ionization history
  modified every 0.01
  in the horizontal axis.
  (See the text).
    \label{fig:env_sedov}}
\end{figure}

From the self-similarity solution, the shock radius is $r_s\propto t^{\frac{2}{5}}$,
and the shock velocity is $v_s\propto t^{-\frac{3}{5}}$.
To compare to the ionization state we set a strong shock model with shock 
front satisfying 
\begin{equation}
\frac{\rho_2}{\rho_1}=\frac{v_1}{v_2}=4
\end{equation}
where $\rho_1$, $v_1$ and $\rho_2$, $v_2$ are the density and velocity in the
upstream and downstream respectively.
At a later SNR age (e.g.\ after 10000 yr), 
shock front temperature, $T_s\propto v_s^2$, is higher than 10$^7\K$ where the 
ionizing rates of O$^{6+}$ and O$^{7+}$ change relatively slowly. By assuming
the shock front temperature does not change during a time interval, we can 
calculate
the ion fractions for several points in the post shock area (See dotted lines in
Fig.~\ref{fig:env_sedov}).
For each point, the initial ion population 
is adopted from the upstream values
(right side in the figure),
the temperature and the electron density are adopted from the theoretical value in situ.
The time interval is calculated by $\Delta t=\Delta r/v_2$, 
where $v_2$ is the downstream velocity
in the shock front reference frame. The maximum value of $v_2$ is 
$\frac{1}{4}v_{\rm s}$, and $v_s$ is the shock front velocity, changing with the 
shock front radius $r_s$ or time $t$ according
to the self-similarity solution.
The materials in the upstream are assumed to be equilibrium. The ion fractions in 
equilibrium and the ionization rates when it enter the shocked area
are calculated 
with values from AtomDB\footnote{\url{http://www.atomdb.org/}} \citep{Foster2012}. 
However, the time interval for a parcel of gas moving from the shock front $r/r_s=1$
to a downstream point at $r/r_s=0.8$ takes about 6000 yr.
The ionization history is not the same for the points in the downstream. To make 
a closer estimate from the theoretical model, for every 0.01 of $r/r_s$,
the ionization histories are corrected to a strong shock with physical parameters adopted
from the self-similar equations,
\begin{equation}
r_s\propto t^{2/5}
\end{equation}
\begin{equation}
v_s\propto t^{-3/5}
\end{equation}
\begin{equation}
T_s\propto t^{-6/5}
\end{equation}
The radial profiles for temperature, velocity are recalibrated for the self-similar solution,
where the density profile will not change because of the fixed compressing ratio.
Fig.~\ref{fig:env_sedov} shows that the
time corrected model gets closer to the simulation with some zig-zag features
caused by the sparse correction, although the real history is more complicated. 
Here, the model ``B2'' (without thermal conduction) is plotted as solid lines to compare
with the theoretical model. 

\section{Simulated \xray\ observations}
\label{sec:mimic}
\subsection{Spectrum generation method and the single point test}
\label{sec:singlepointtest}

Although we can determine the ionization state with the average charge as 
discussed in our previous work \citep{ZhangG2018}, a simulated spectrum remains the best way to compare directly with the SNR observations.
From the 3D simulation, we can generate a simulated spectrum with the physical
properties of the final results, using Python scripts SOXS\footnote{\url{http://hea-www.cfa.harvard.edu/~jzuhone/soxs/}} and pyXSIM\footnote{\url{http://hea-www.cfa.harvard.edu/~jzuhone/pyxsim/}}. 
It is assumed that the hydrogen in the simulation is fully ionized, which is 
true when the temperature is much larger than $1\times10^{4}\K$. Considering 
the X-ray emission temperature is above about $1\times10^{5}\K$, the exact hydrogen 
ion fraction will not significantly affect the spectrum.
The SOXS and pyXSIM codes generate a list of X-ray photons based on the non-equilibrium spectrum (including Doppler velocities) at every position in
the 3D simulation. By projecting the photons onto a simulated X-ray telescope, 
an event file is obtained that can be used to create images and spectra.
We modeled data from three \xray\ instruments: the Resolve Instrument on {\em XRISM}\footnote{\url{https://heasarc.gsfc.nasa.gov/docs/xarm/}}, the X-IFU on {\em Athena}\footnote{\url{http://www.the-athena-x-ray-observatory.eu/}}, and the LXM on {\em Lynx}\footnote{\url{https://www.lynxobservatory.com}}.

\begin{figure}[htpb]
\plottwo{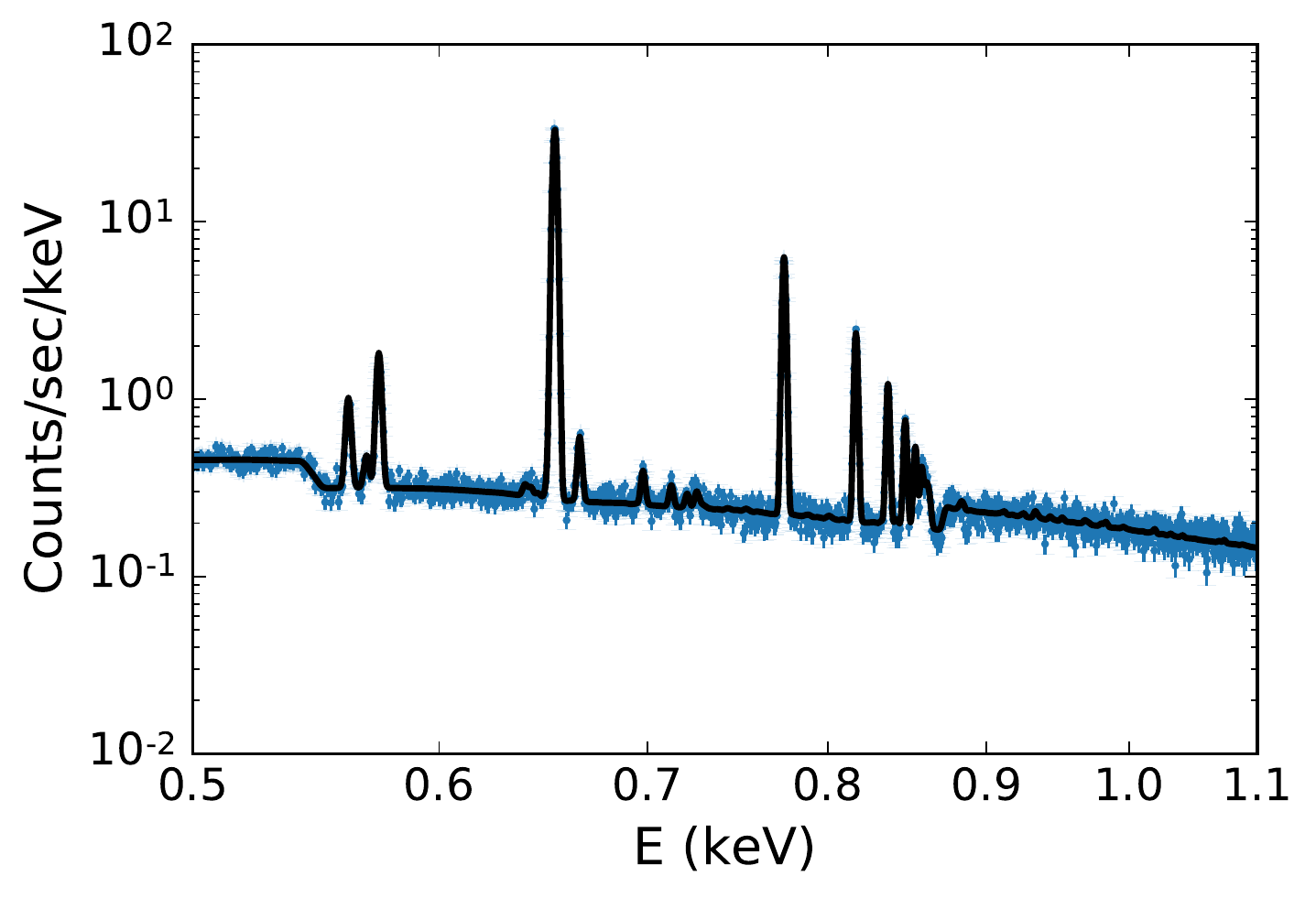}{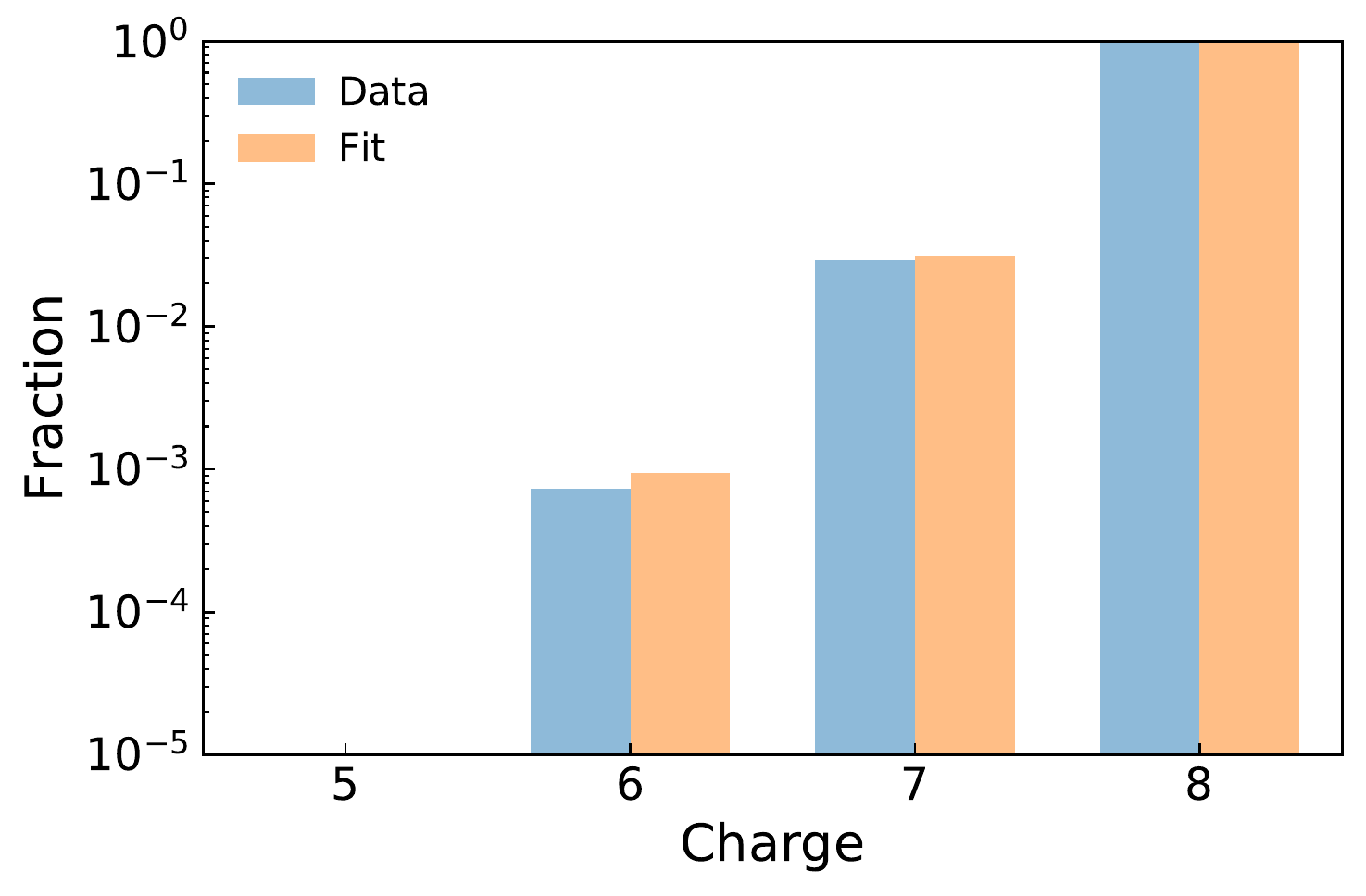}
\caption{A test spectrum from a single point in the simulation. The 
spectrum and the fit model are shown in blue and black in the left 
panel (See Table~\ref{tab:onepoint}). 
The ion fraction distribution (oxygen) that is obtained with the fit parameters
is shown in the right panel with the original ion fraction from data
as a reference.\label{fig:onepoint}}
\end{figure}

To test the spectral generation, both a single ionizing and a recombining pixel
in the simulation were selected. By fitting the spectra, we have compared them to the 
actual physical quantities in the pixels. the fit results from both the ionizing and 
recombining spectra are consistent with the simulation. Here is an example pixel
with an electron temperature 0.62 keV in the simulation.
Ignoring Doppler shifts for simplicity, the simulated spectrum from this one pixel, observed in a 100Ms {\em Athena} X-IFU exposure, is shown in Fig.~\ref{fig:onepoint}[Left]. 
We fit the spectrum using the {\em Xspec} {\em vrnei} model, which assumes a gas parcel in equilibrium at an initial temperature is then instantaneously heated (or cooled) and held at a new temperature. In the fit, the initial temperature was indeterminate and fixed to a maximum value of 10$\keV$. The best-fit electron temperature is $0.605\pm0.002$, only 3\%  from the simulation (See Table~\ref{tab:onepoint}). Nonetheless, this is a $7.5\sigma$\ difference; we expect this systematic offset is primarily due to using a simplified model ({\em vrnei}) that does not capture all of the changes this particlar cell in the 3D simulation has undergone.  
With the fit parameters and {\em pyAtomDB}, we can also compare the ion fractions predicted by the {\em vrnei} with those in the cell itself; these are shown in the right panel of Fig.~\ref{fig:onepoint}.
We conclude that this fit is largely consistent with the original data.

\begin{deluxetable}{ccc}
  \tablecaption{Single point spectrum fit ({\em vrnei})\label{tab:onepoint}}
  \tablehead{
  \colhead{Parameter} & \colhead{Unit} & \colhead{value} \\
  }
  \startdata
  kT & keV & 0.605$\pm0.002$ \\
  kT\_init\tablenotemark{$\dagger$} & keV & 10  \\
  Tau & s/cm$^{3}$ & $(1.92\pm0.09)\times10^{11}$ \\
  norm & & $(8.47\pm0.02)\times10^{-5}$\\
  $\chi^{2}\tablenotemark{$\ddagger$}$ & & 1.23 (3870) \\
  \enddata
  \tablenotetext{\dagger}{Initial temperature that has a maximum of 10$\keV$. }
  \tablenotetext{\ddagger}{Reduced $\chi^2$ with the degree of freedom in the bracket.}
  \tablenotetext{}{The abundances of H, He, and O are frozen to 1 solar abundance 
  \citep{Anders1989}.}
  \tablenotetext{}{The errors depict 90\% confidence ranges for each parameter.}
\end{deluxetable}


\subsection{Observations with Athena, XRISM, and Lynx}
\label{sec:proj}
\begin{figure}[htpb]
\plotone{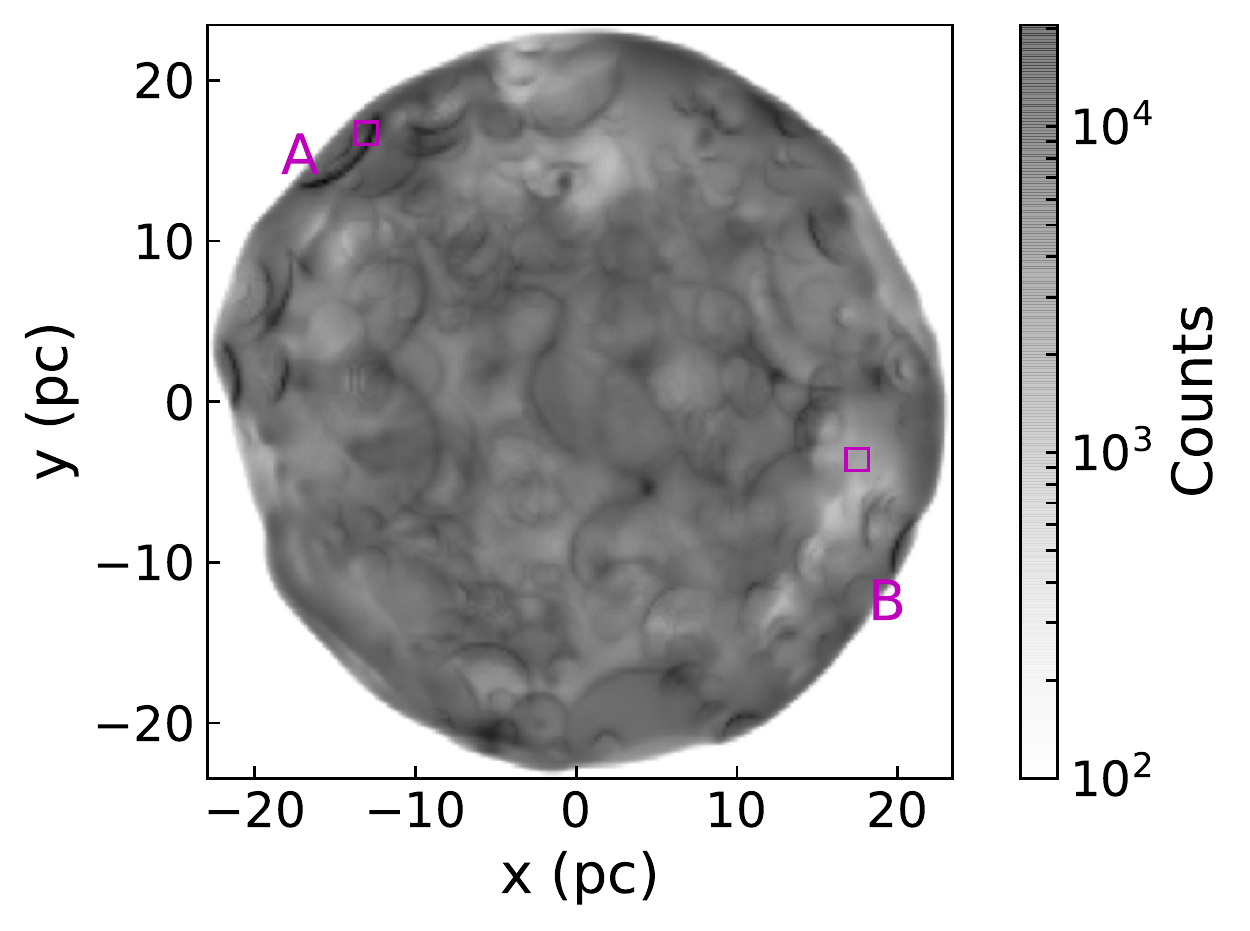}
\caption{Simulated {\em Athena} XIFU observation from a 3D simulation with projection
along $z$-axis. The gray scale depicts the distribution of generated events. 
The positions to extract Spectrum A and B have been labeled. \label{fig:observed_events}} 
\end{figure}

\begin{figure}[htpb]
\plottwo{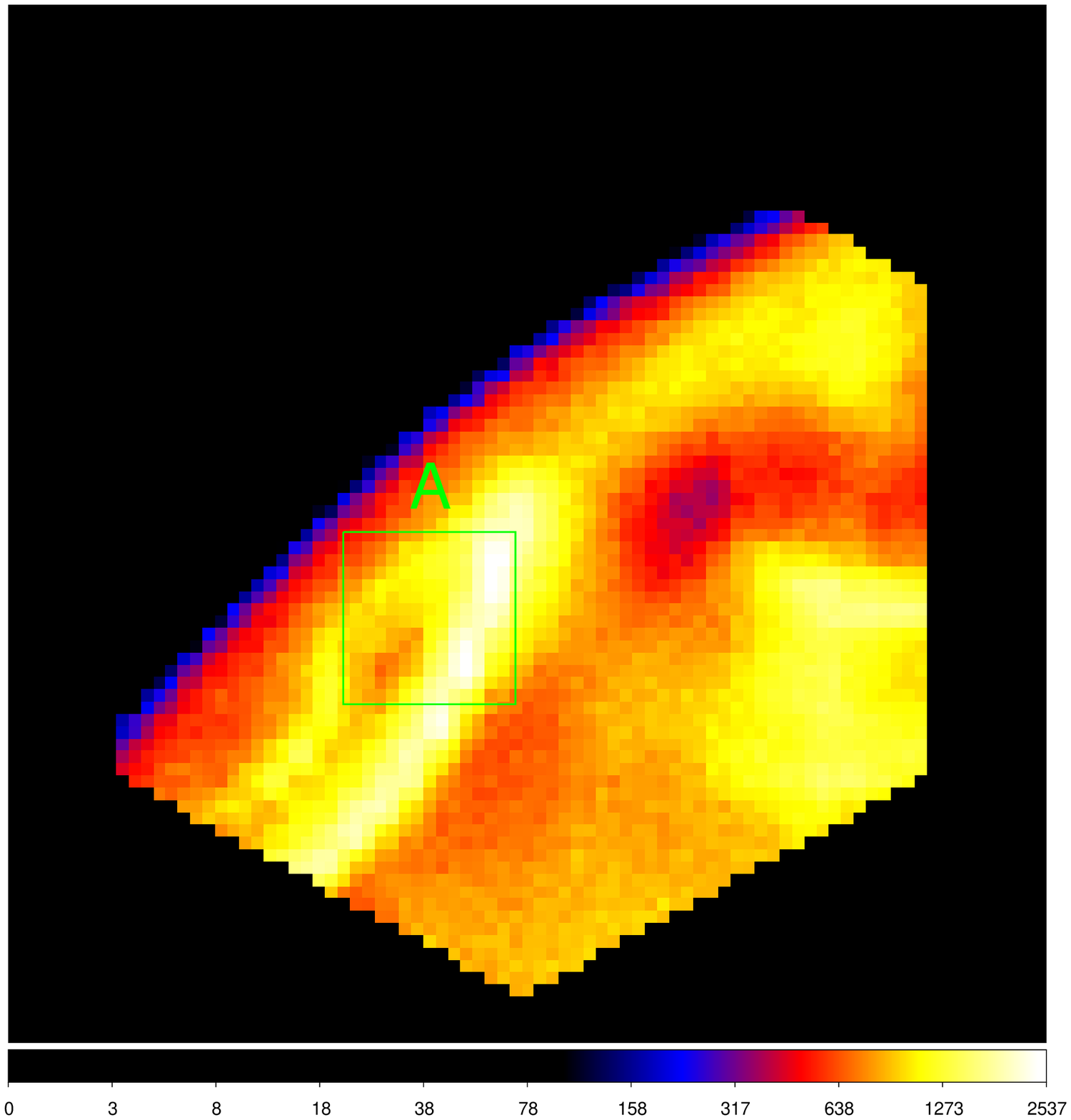}{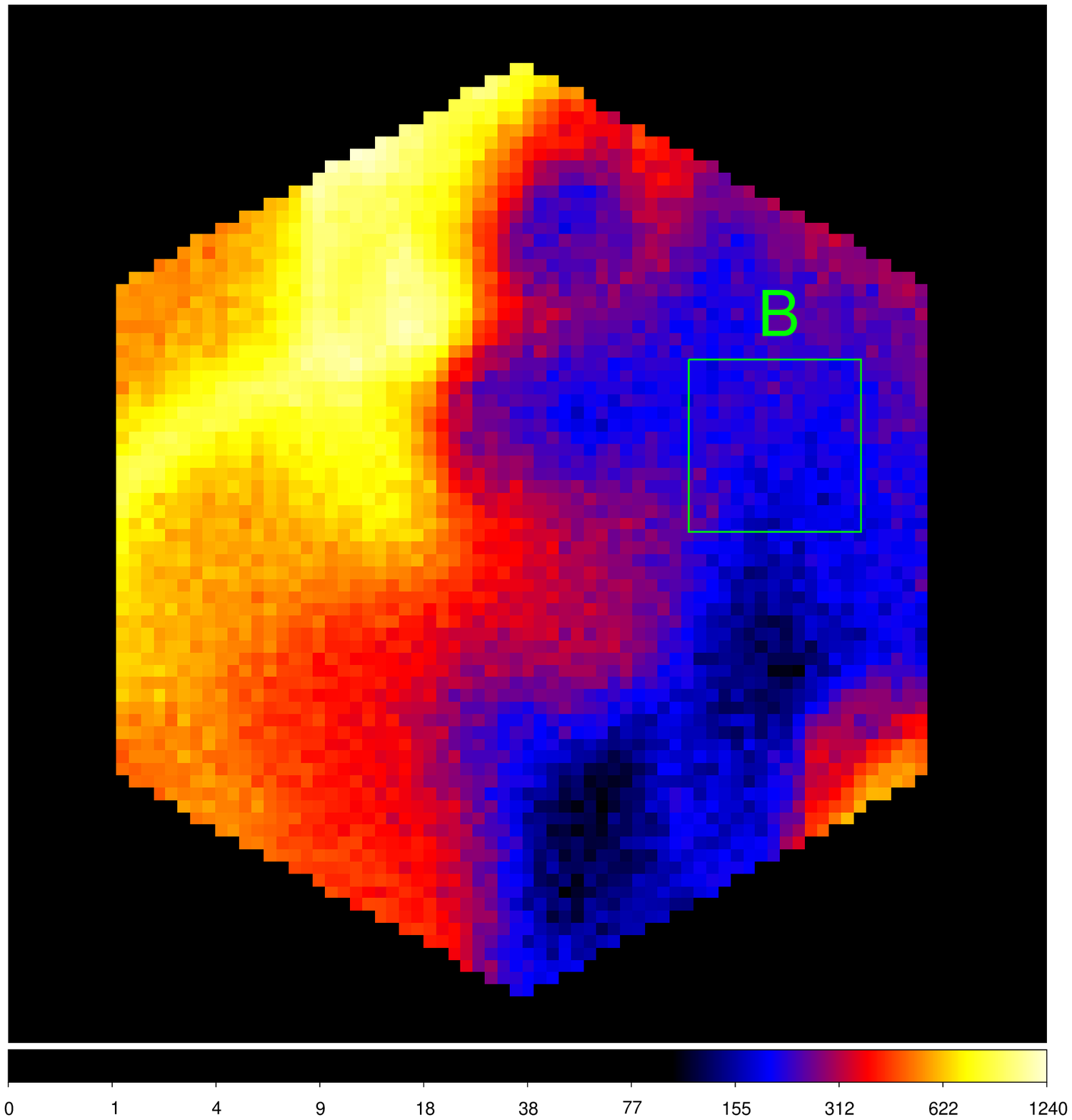}
\caption{Two simulated observations with {\em Athena}. 
Spectrum A and B are extracted from the corresponding boxes in these two 
observations.
\label{fig:real_observations}}
\end{figure}

\begin{figure}[htpb]
\plottwo{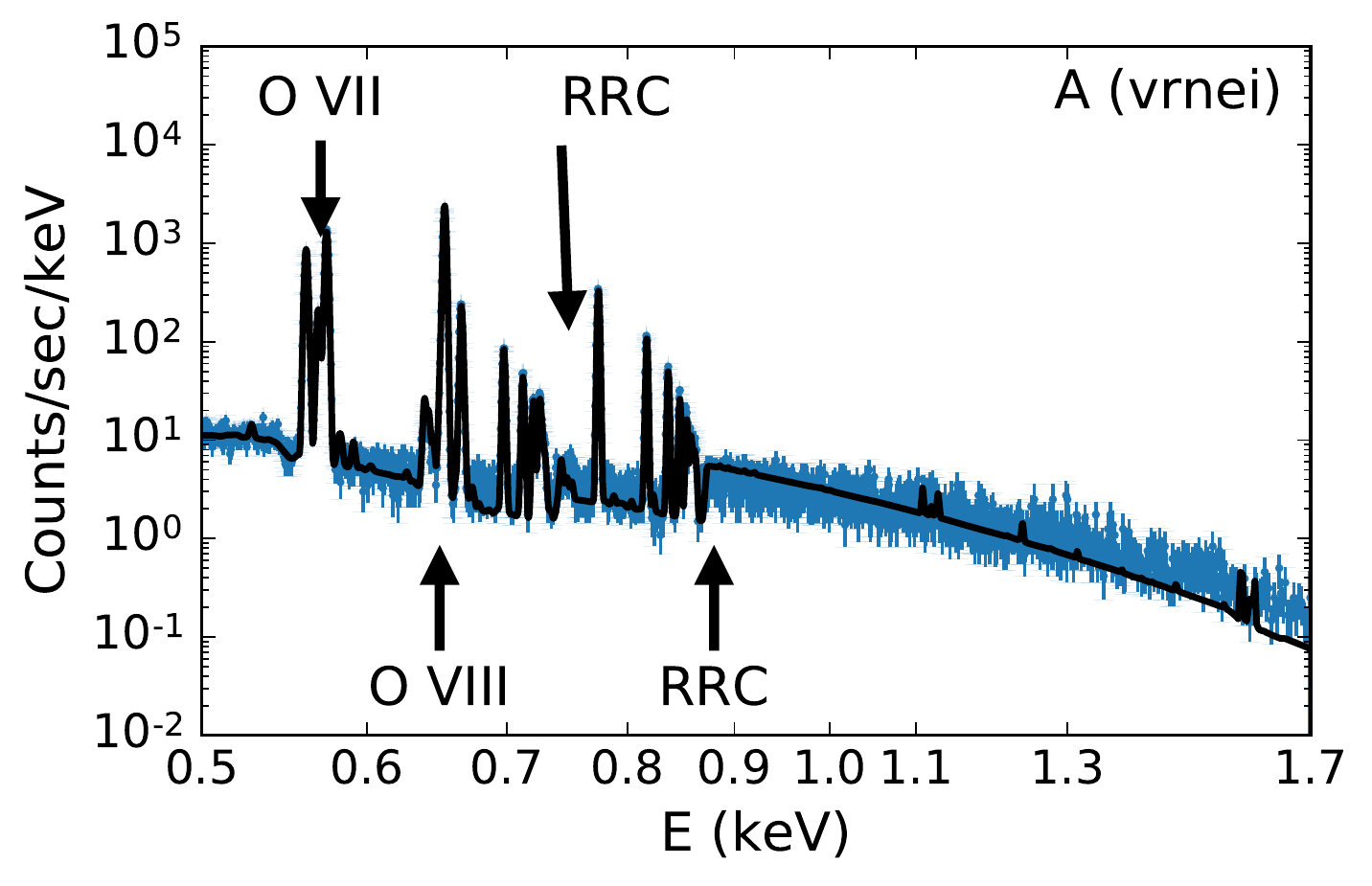}{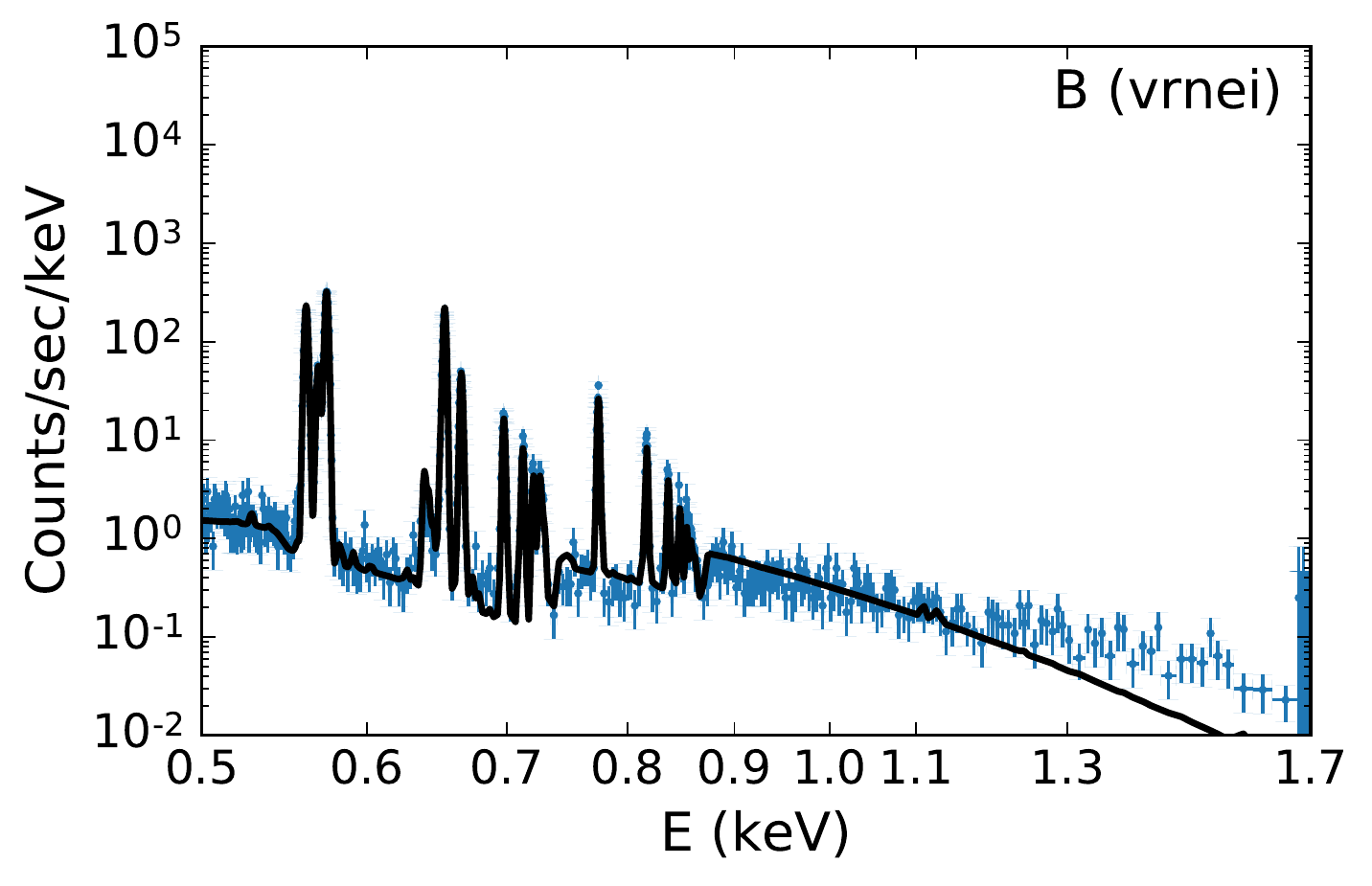}
\plottwo{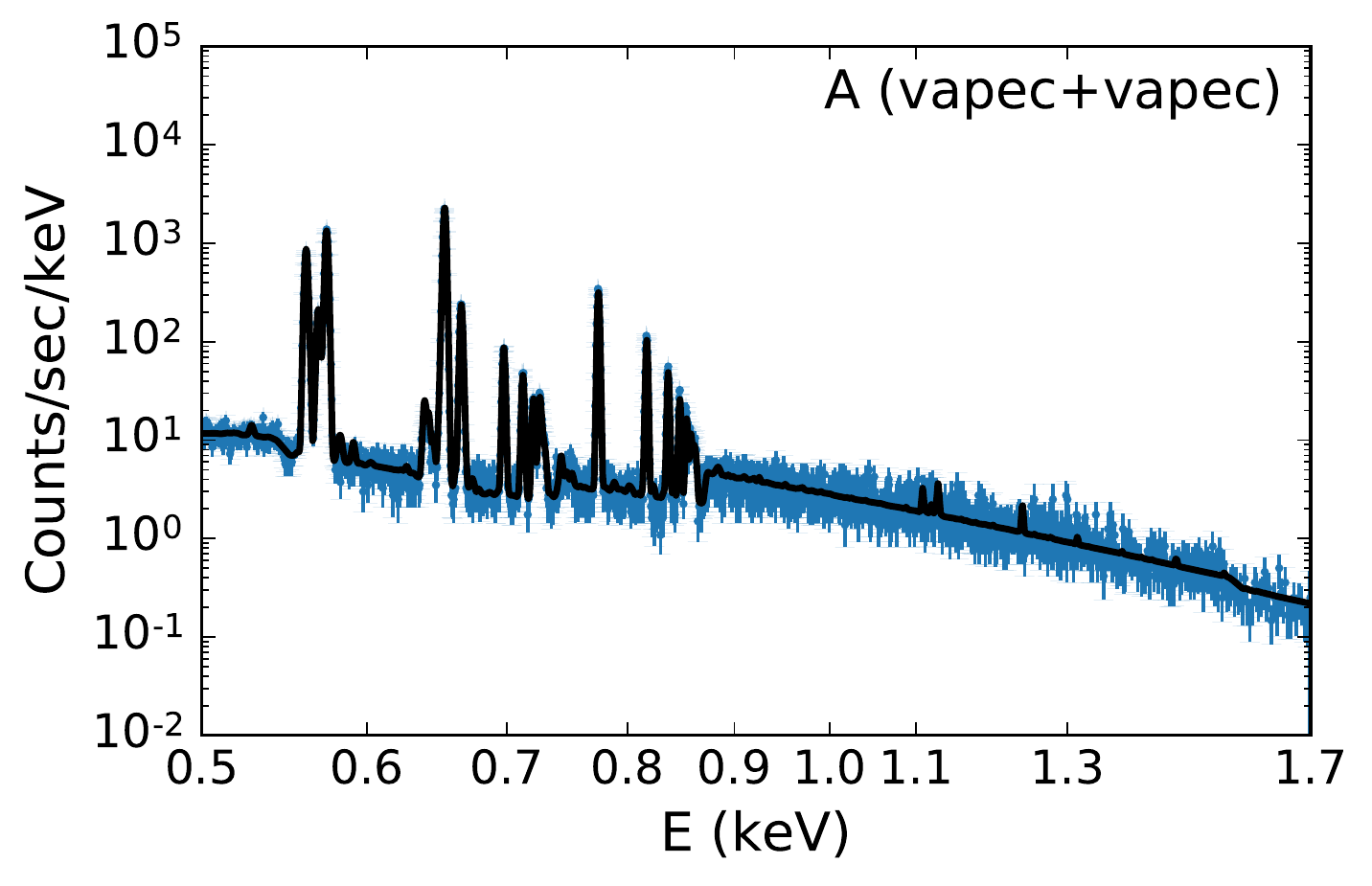}{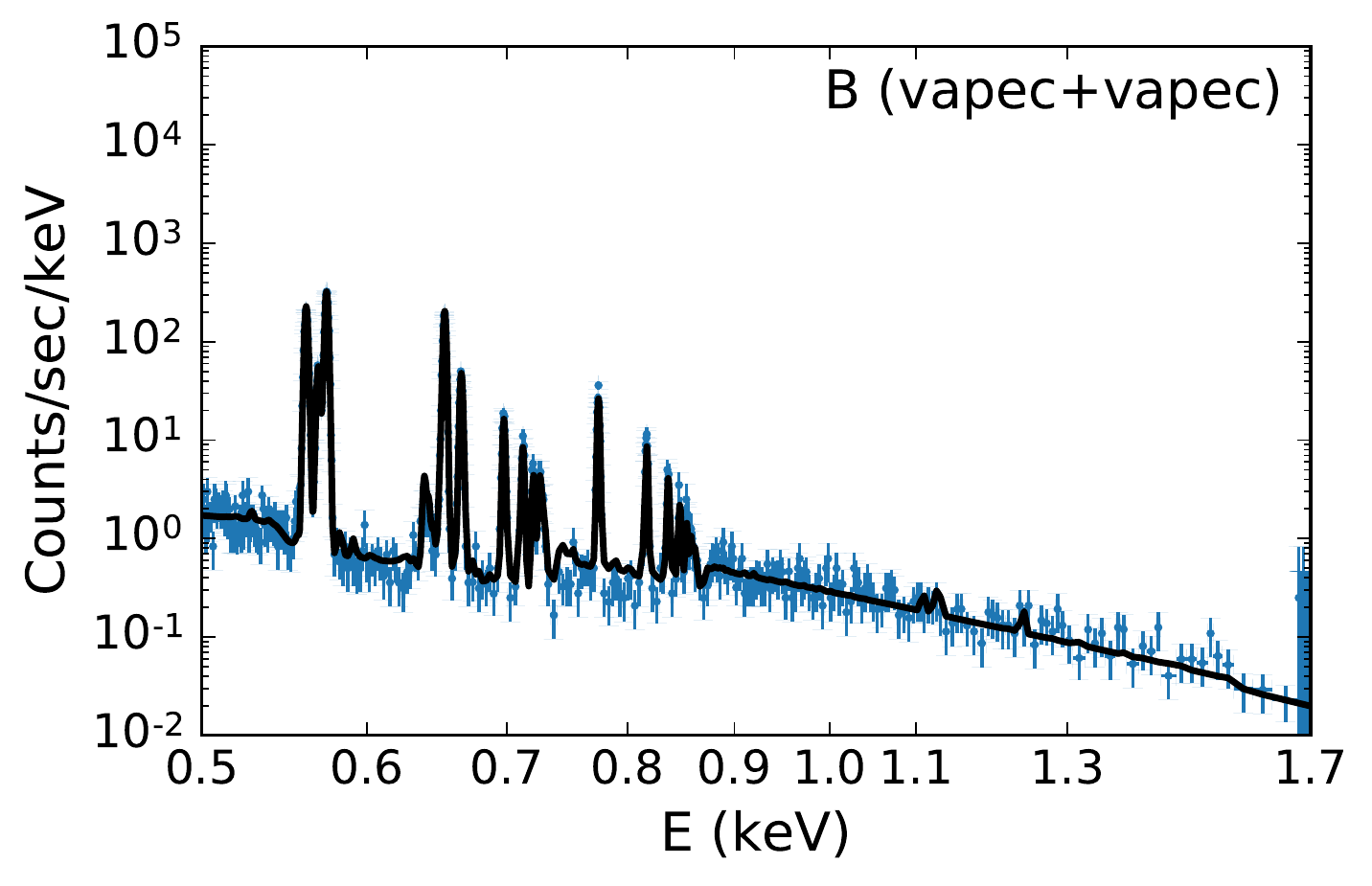}
\caption{Two example spectra (A and B) are shown below with the spatial regions depicted in 
the top panel.
Both the spectra A and B have an exposure time of 10 ks. They are fitted with a
``{\em vrnei}'' model and a two-component ``{\em vapec}'' model. The fit results are listed 
in Table~\ref{tab:fit} and Table~\ref{tab:fit_two}.
\label{fig:observed}}
\end{figure}

\begin{deluxetable}{cccc}
  \tablecaption{Fit parameters with {\em vrnei} model\label{tab:fit}}
  \tablehead{
    \colhead{Parameters} & \colhead{Unit} & \colhead{Spectrum A} & 
    \colhead{Spectrum B}  \\
  }
  \startdata
  kT & keV &0.2183 $\pm0.0006$ & 0.164 $\pm0.001$ \\
  kT\_init\tablenotemark{$\dagger$} & keV & 10  & 10 \\
  Tau & $\s/{\rm cm}^{3}$ & $(8.8\pm0.1)\times10^{11}$ &   $(1.34\pm0.03)\times10^{12}$\\
  norm & & $(2.53\pm0.02)\times10^{-3}$ & $(4.90\pm0.10)\times10^{-4}$ \\
  $\chi^{2}\tablenotemark{$\ddagger$}$ & & 3.21 (1704) & 1.75 (512) \\
  \enddata
  \tablenotetext{\dagger}{Initial temperature that has a maximum of 10$\keV$. When
  the fitted initial temperature exceed the maximum, it is set to the maximum value.}
  \tablenotetext{\ddagger}{Reduced $\chi^2$ with the degree of freedom in the bracket.}
  \tablenotetext{}{The abundances of H, He, and O are frozen to 1 solar abundance 
  \citep{Anders1989}.}
  \tablenotetext{}{The errors depict 90\% confidence ranges for each parameter.}
\end{deluxetable}

\begin{deluxetable}{cccc}
  \tablecaption{Fit parameters with two {\em vapec} models\label{tab:fit_two}}
  \tablehead{
    \colhead{Parameters} & \colhead{Unit} & \colhead{Spectrum A} & 
    \colhead{Spectrum B}  \\
  }
  \startdata
  kT\_c\tablenotemark{$\dagger$} & keV &  0.212$\pm0.002$ & 0.157$\pm0.003$ \\
  kT\_h & keV &  0.45$\pm0.02$ & 0.46$\pm0.03$ \\
  norm\_c & & $(1.66\pm0.02)\times10^{-3}$ & $(4.03\pm0.09)\times10^{-4}$ \\
  norm\_h & & $(8.06\pm0.47)\times10^{-4}$ & $(9.98\pm0.7)\times10^{-5}$ \\
  $\chi^{2}\tablenotemark{$\ddagger$}$ & & 2.38 (1703) & 1.30 (511) \\
  \enddata
  \tablenotetext{\dagger}{The subscripts ``c'' means the cold component; and ``h''
  means the hot component.}
  \tablenotetext{\ddagger}{Reduced $\chi^2$ with the degree of freedom in the bracket.}
  \tablenotetext{}{The abundances of H, He, and O are frozen to 1 solar abundance 
  \citep{Anders1989}.}
  \tablenotetext{}{The errors depict 90\% confidence ranges for each parameter.}
\end{deluxetable}

The simulated SNR model (with the age of 2$\times10^4\yr$) was assumed to be at
a distance of 5 kpc, with no absorption.
Using the response and field of view of the {\em Athena} X-IFU instrument {\citep{Nandra2013,Barret2018}}, 59 10 ksec observations
are adequate to cover the entire SNR.
In Fig.~\ref{fig:observed_events}, the \xray\ events distribution
image in energy range from 
0.35\,keV to 1.7\,keV shows the projected hot gas distribution inside the 
SNR. There are some ring or concave features 
in some areas, likely corresponding to the partially shocked clouds. 
Two example observations are shown in Fig.~\ref{fig:real_observations} with
the X-IFU field of view.
For each observation, twelve spectra are extracted in box regions 
(with a size of 1\arcmin$\times$1\arcmin each). 

With a {\em vrnei} model in {\em Xspec}\footnote{\url{https://heasarc.gsfc.nasa.gov/lheasoft/xanadu/xspec/}},
they can be fit to determine whether it is recombining.
Here we only use oxygen in solar abundance for lines 
with hydrogen and helium mainly contributing to the continuum; other elements are set to zero.
The electron number density can
be calculated from the fitted parameter ``norm'' in the model {\em vrnei}.
A spherical ball of X-ray emitting gas with a filling factor of 1 is assumed to get 
the volume of the observation area. The depth of the X-ray emitting gas 
is $L=\sqrt[]{R^2-r^2}$, where $R$ is the average radius of the remnant and 
$r$ is the projected radius of every observation.
Because there are cold materials in the simulated remnant,
this assumption will underestimate the density, a common problem in X-ray analysis of SNR.
Two spectra (A and B) are shown in the first row as examples (with the fit parameters 
in Table~\ref{tab:fit}). 
Because only ionized hydrogen and helium along with all oxygen ions are included in the simulated spectra, only the O\,\rom{7}, O\,\rom{8} lines and their radiative recombination continuum (RRC) features can be seen. 

However, both the $\chi^2$ in Table~\ref{tab:fit} and the high energy range in Fig.~\ref{fig:observed}
show that a single component of a vrnei model does not provide a good fit to the spectra.
We therefore tried two ``{\em vapec}'' models with different temperature to fit the same 
spectra. In the bottom panels in Fig.~\ref{fig:observed} and Table~\ref{tab:fit_two},
the high energy residuals are reduced when using a two-component model and the fitted $\chi^2$ 
is better than the one in a single recombining model. Of course, neither approach truly captures the underlying physics within the 3D simulation; we can only hope to capture some key components.

\begin{figure}[htpb]
\plottwo{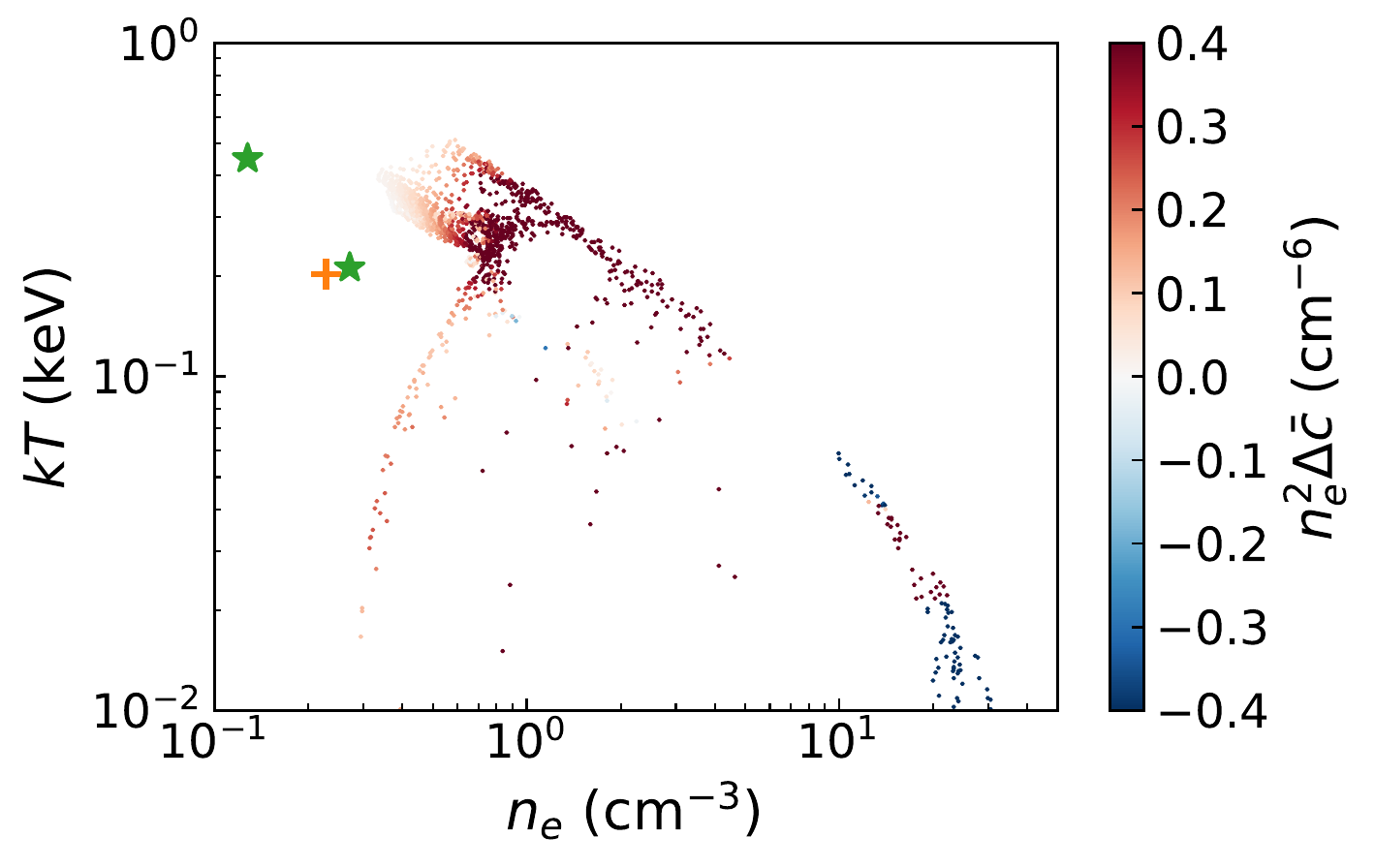}{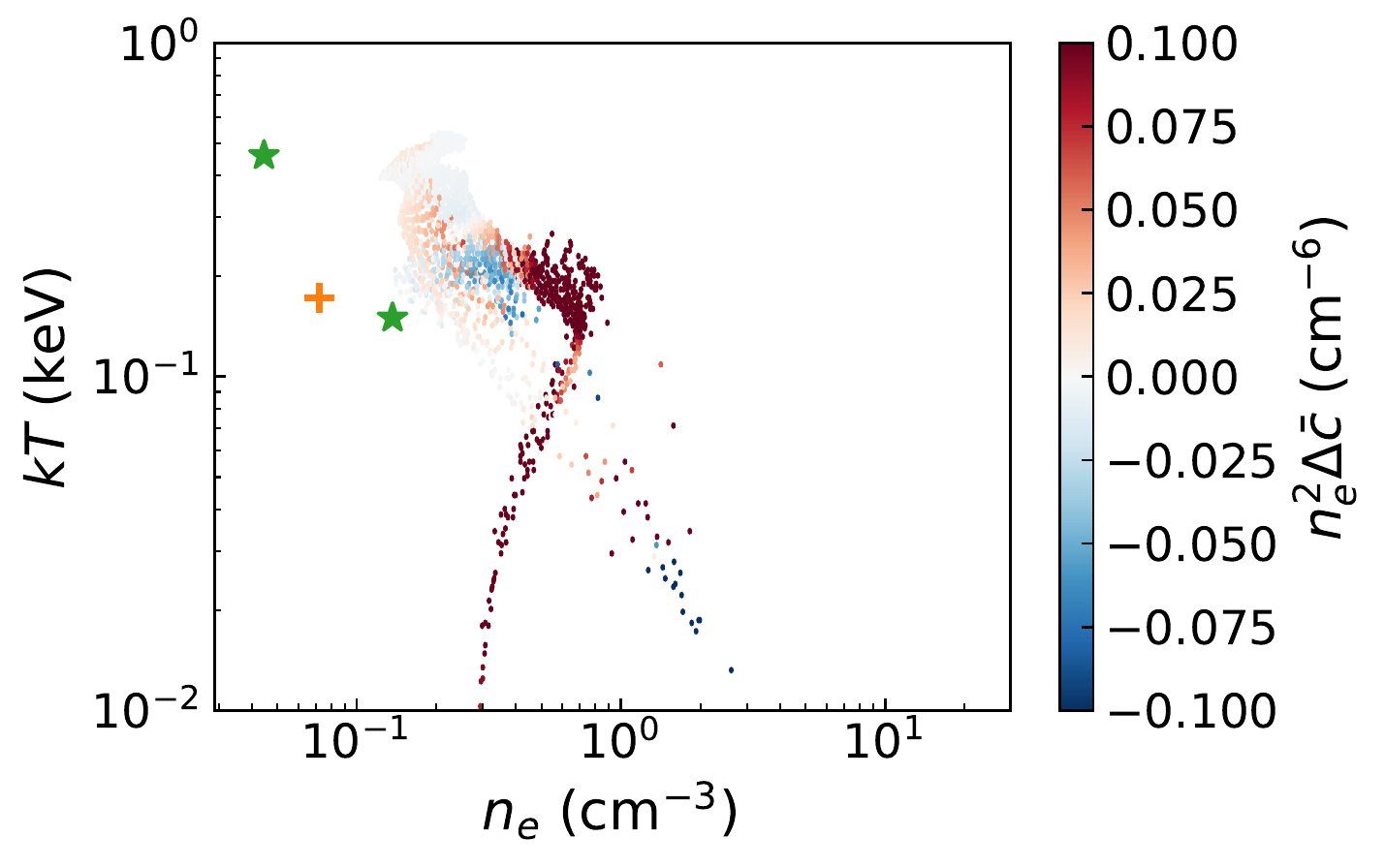}
\caption{The phase diagrams of the projected regions of spectrum A (left)
and spectrum B (right). The crosses show the fit parameters of 
the ``{\em vrnei}'' model; and the stars show the fit parameters of 
the two ``{\em vapec}'' model. The densities are underestimated 
as expected because of the assumption of filling factor (See the text).
\label{fig:phas_spec}}
\end{figure}

From the simulation data, all the physical quantities (e.g.\ temperature, 
density, and ion fractions) can be used to obtain the true parameters underlying the fitted spectra. In Fig.~\ref{fig:phas_spec}, the phase diagrams 
of the real data for spectrum A and B show each region's distinct evolution.
The average charge difference is weighted by $n_e^2$\ to more accurately capture each cell's effect on the total emission.
As these figures show, both the single- and two-temperature models have reasonable electron temperature values, given the breadth of the temperature distribution in the underlying region. 
The ionizing feature that is similar to the region ``a'' in Fig.~\ref{fig:dens_temp_delc_mass} appears in both diagrams, because the
ionizing shock front is always on the LOS. In the left panel, the phase
diagram of spectrum A shows a clear line from top left to bottom right, similar to the iosbaric lines in Fig.~\ref{fig:dens_temp_delc_mass}.
Both ionization and recombination appear in this line, which imply the 
thermal conduction could be the dominant process in these cells. 
In the right panel, the recombining gas is mainly situated in an area with a high temperature and low pressure, which is probably caused by adiabatic expansion. 

\begin{figure}[htpb]
\plottwo{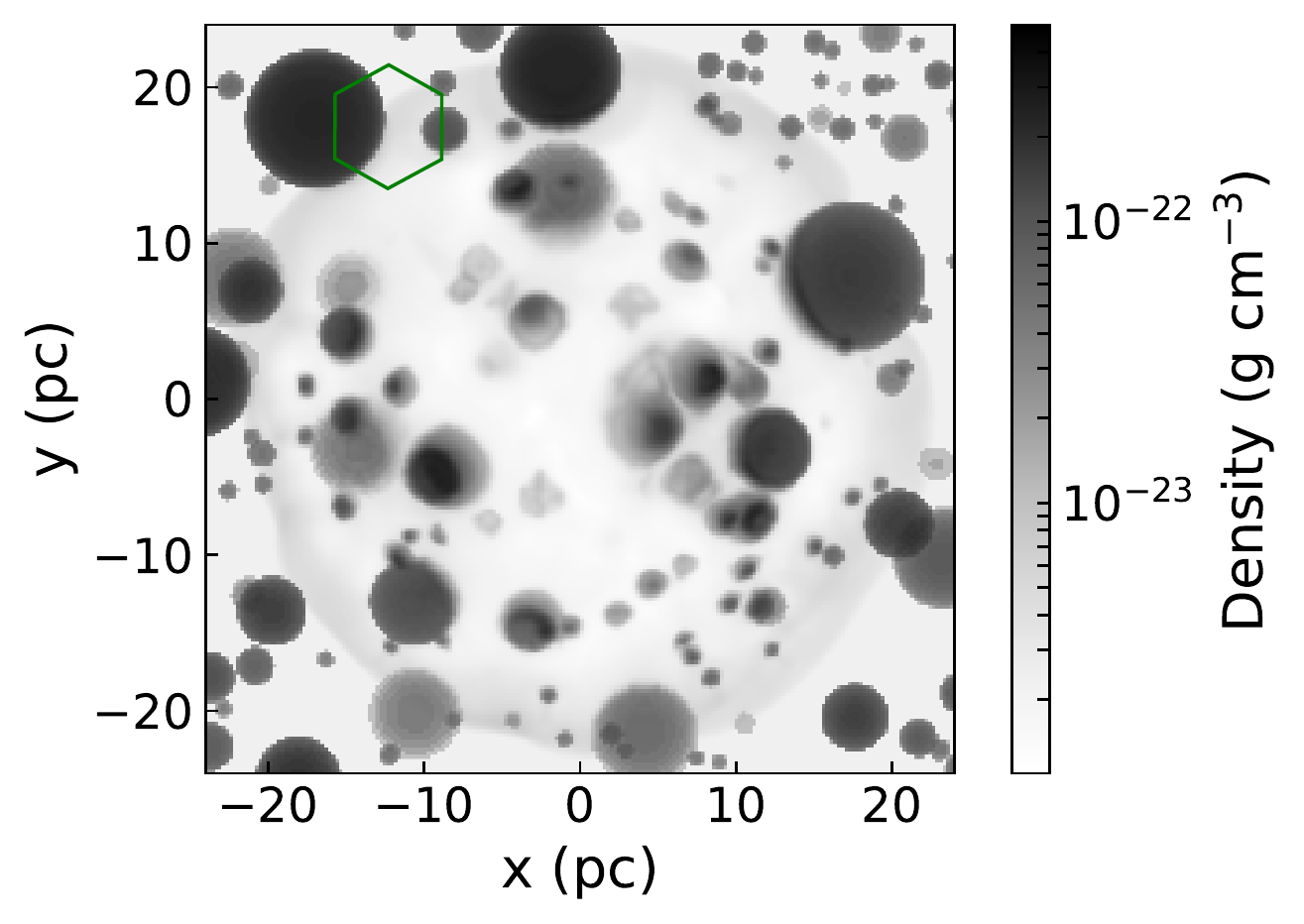}{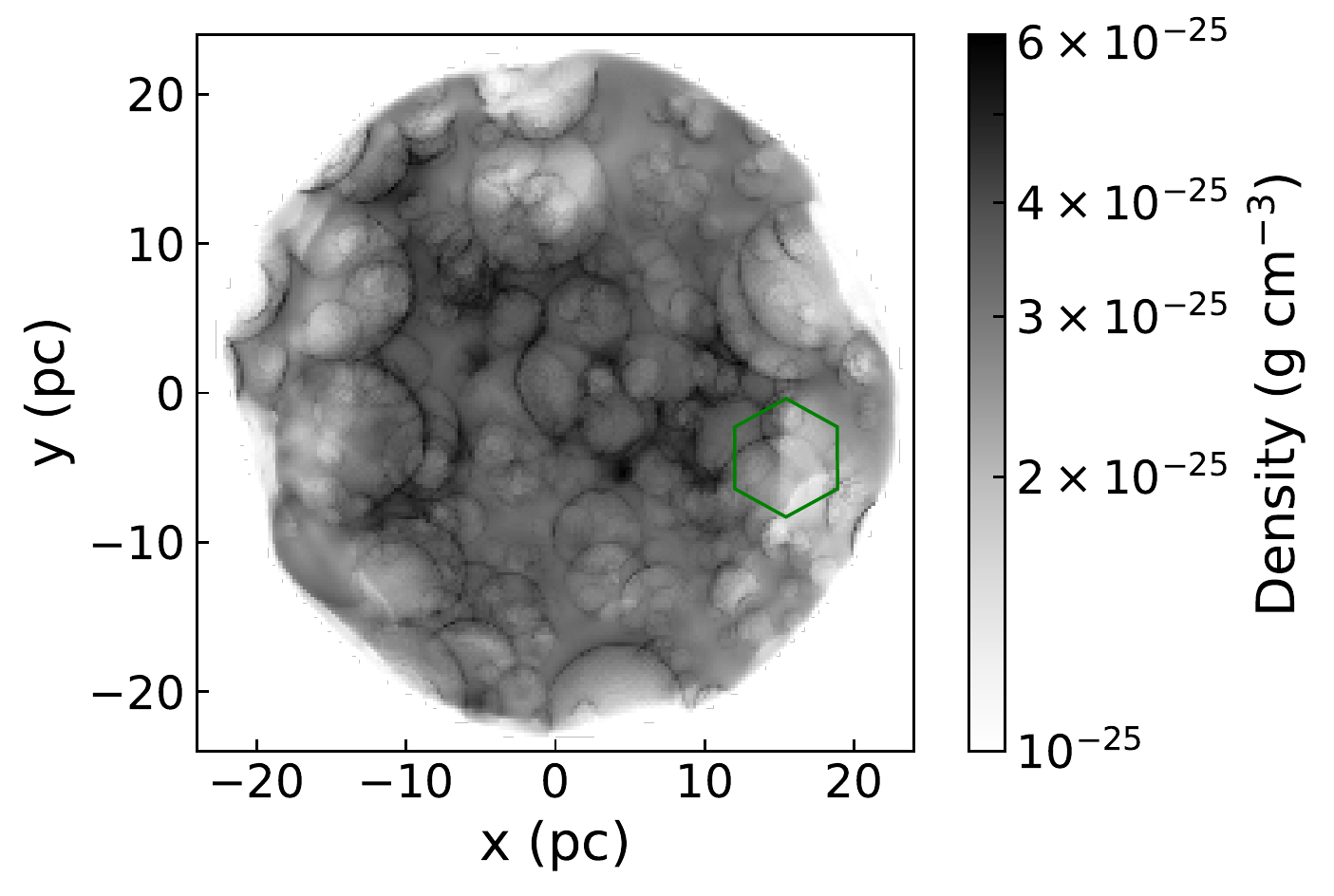}
\caption{Left panel: projected average density along $z$-axis from -9 pc to -2 pc.
Right panel: projected average density of \xray\ emitting gas ($T>1\times 10^6\K$) 
along $z$-axis. The observation regions for Spectrum A and B
in Fig.~\ref{fig:real_observations} are shown in green.
\label{fig:projected_dens}}
\end{figure}

Fig.~\ref{fig:phas_spec} also shows that the best-fit densities are underestimated (as expected), and are about one tenth of the actual values. Thus the  filling factor should be ~0.32 instead of the assumed 1.  In the left panel of Fig.~\ref{fig:projected_dens}, we plot the projected density within a range from $-9\parsec$ to $-2\parsec$ on the $z$-axis by averaging the density in this range along LOS. It shows the dense (non-X-ray-emitting) clouds in this thin slice of SNR. The spectrum A region is associated with a compressed cloud. In the right panel of Fig.~\ref{fig:projected_dens}, we plot the 
projected average density of hot gas ($T>1\times10^{6}\K$). It averages the density 
in a range from $-30\parsec$ to $30\parsec$ on the $z$-axis. The bottom right spectrum
B region is associated with a low density region.

Although both regions we selected for this example contain recombining plasma, the best-fit spectral models using diagnostics from oxygen alone only show that the gas has a broad temperature distribution. Real data obviously contrain lines from many more elements, which at the high spectral resolution provided by microcalorimeters will reveal both temperature and, we expect, ionization state information. We plan to do a next generation of simulations that will include all abundant elements and, we expect, indicate what analysis is needed to extract the time-dependent hydrodynamical parameters described in the main paper.

\begin{figure}[htpb]
\plottwo{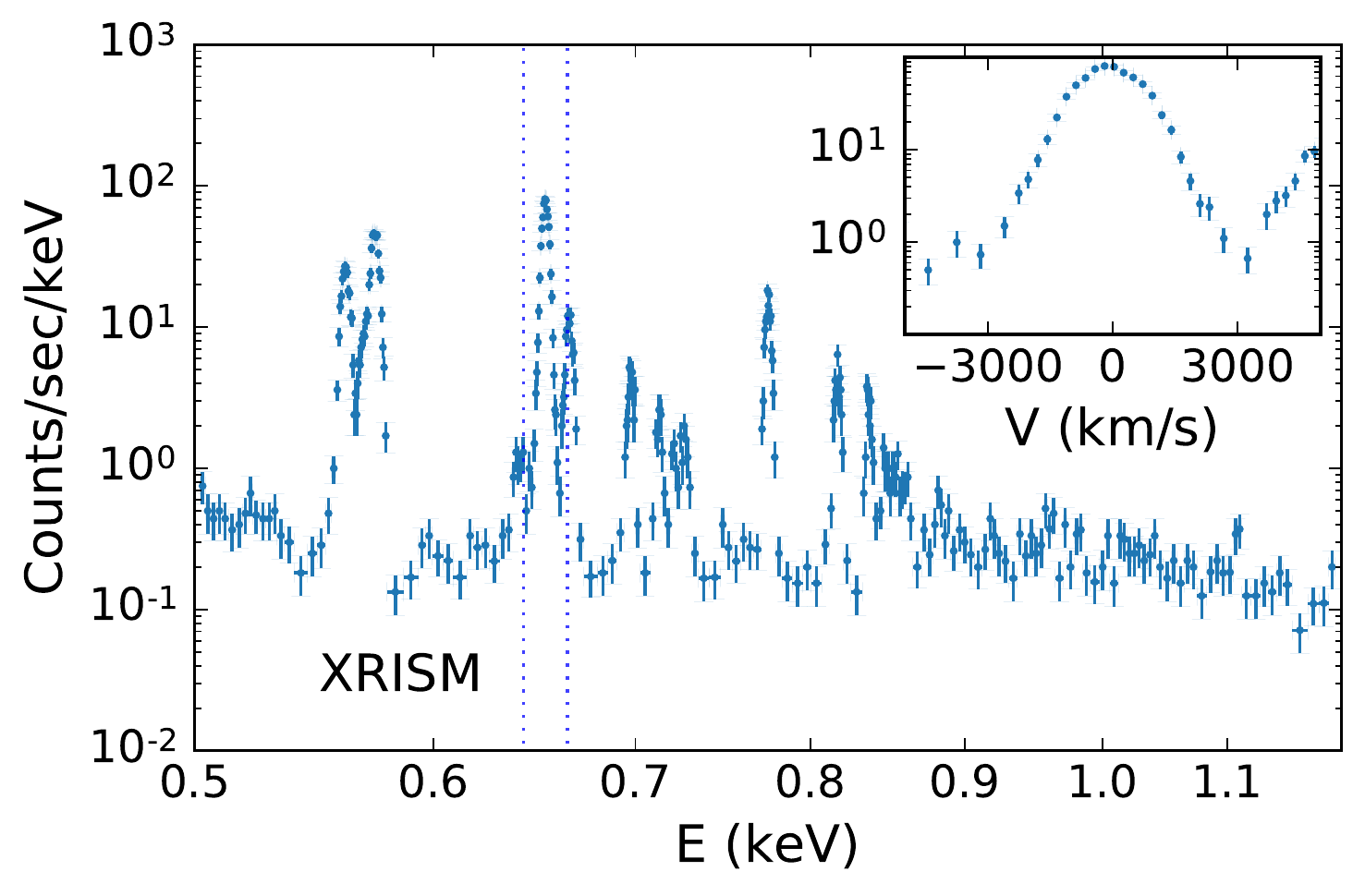}{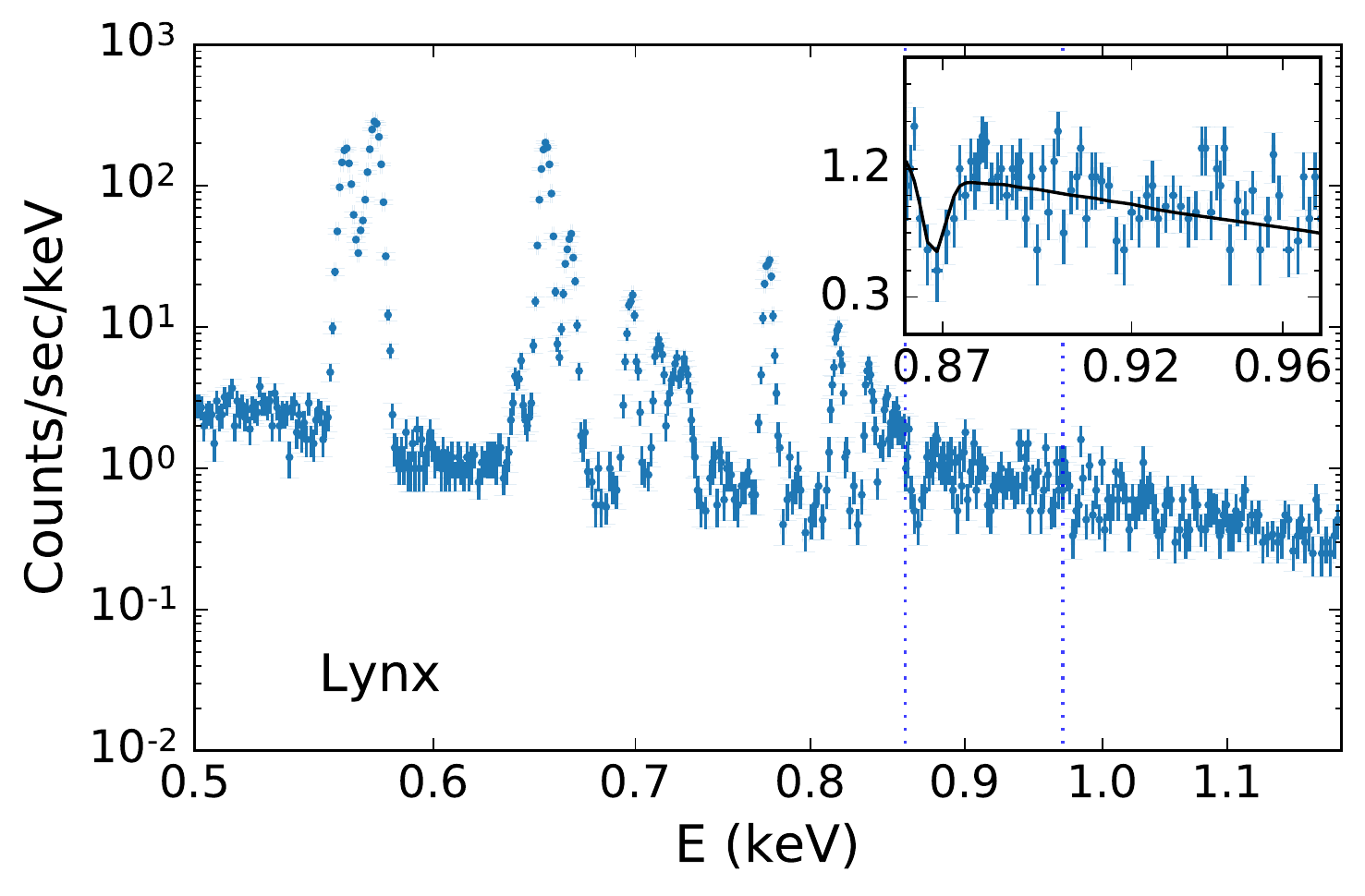}
\caption{The example spectra from {\em XRISM} and {\em Lynx}.
They are from regions similar 
to Spectrum A and B in Fig.~\ref{fig:observed} respectively. The {\em XRISM} spectrum
(left panel) is
extracted from a box of 3$\arcmin\times3\arcmin$ around the spectrum A area. 
The insert figure in this panel shows the velocity dispersion
of the strongest O\rom{8} line with the range depicted as dotted lines in the 
spectrum. The {\em Lynx} spectrum (right panel)
is extracted from a box of
1$\arcmin\times1\arcmin$ in the spectrum B area. The insert image in this panel
zooms in the RRC feature within the dotted lines. 
\label{fig:xrism_lynx}}
\end{figure}

With the instruments on {\em XRISM} and {\em Lynx}, 
we can also produce simulated observations. We use the same regions as Athena to simulate spectra for the {\em Lynx} LXM. As the spatial resolution is much lower, a larger region (3\arcmin$\times$3\arcmin\ box) was used for {\em XRISM} Resolve spectra. In Fig.~\ref{fig:xrism_lynx},
two example spectra are shown. They come from similar regions as the A and B spectra in
Fig.~\ref{fig:observed} and can be fit to similar temperatures. The exposure time 
of every observation is 10$\ks$ for both instruments. The future \xray\ instruments
have a higher spectral resolution which make it far easier to diagnose the RRC features
as shown in the insert figure in the right panel. 
If the observed hot gas has an bulk motion on the LOS, the emission line will show a
redshift, blueshift or broadening \citep[See][for an example]{Hitomi2016}.
The left panel shows the velocity dispersion of an O\rom{8} line in the insert figure.

\begin{figure}[htpb]
\plotone{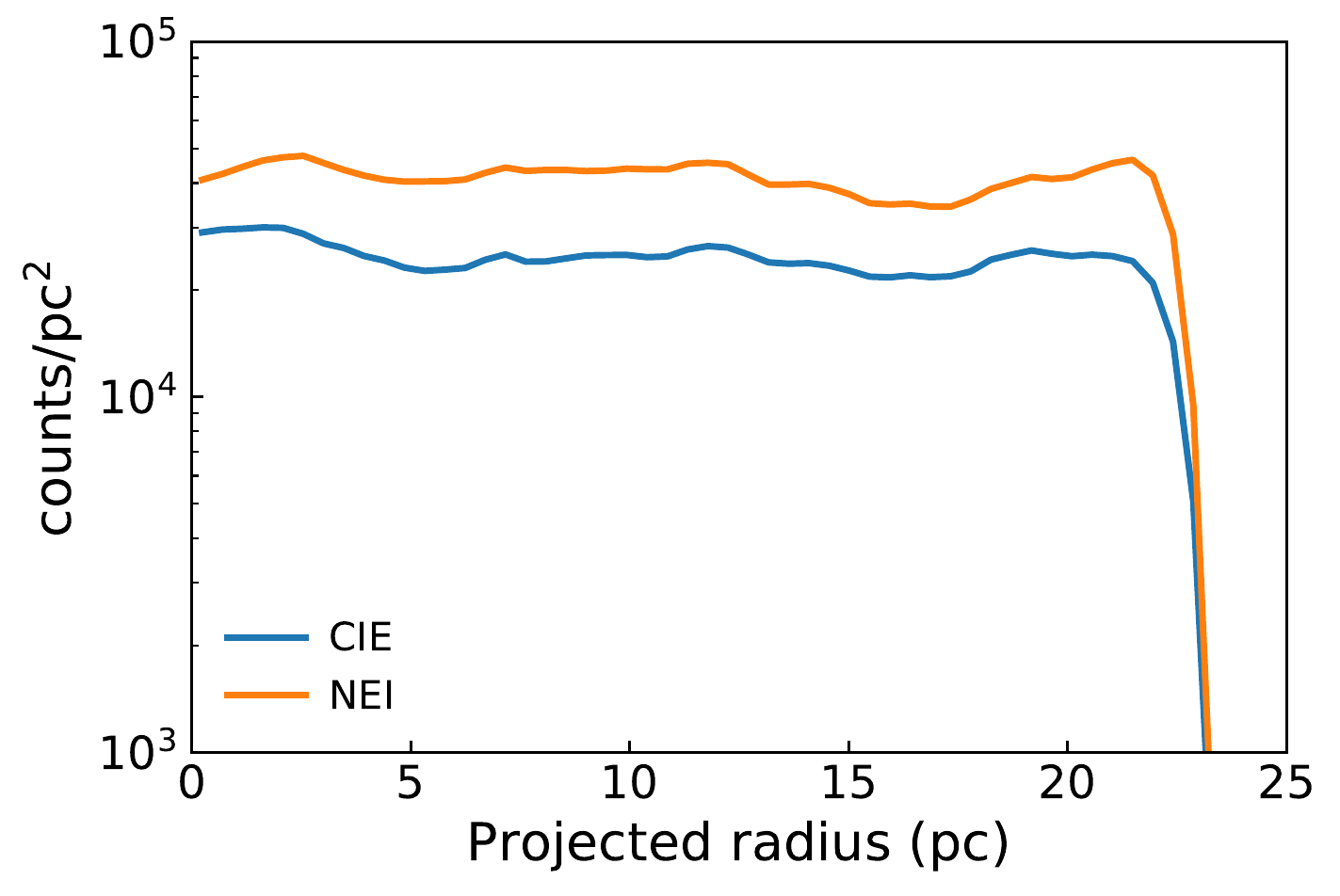}
  \caption{
  Surface brightness for the NEI and CIE for the same 3D simulation 
  result (at 2$\times10^4\yr$) with a shell thickness of 0.5 pc. 
  The ions of hydrogen, helium, and oxygen are taken into account for the
  emission, and the energy range is 0.35 keV to 1.2 keV. 
    \label{fig:sb}}
\end{figure}

If we calculate the emission from the 3D simulation by assuming that it is in CIE at each pixel, results can also be generated from the same 3D simulation result. In
Fig.~\ref{fig:sb}, the surface brightness of NEI model is compared to the CIE
assumption. The NEI model is about 30\% brighter than CIE in this band. 
Both of them have a flat surface brightness in the interior of the
SNR. In the shock front area, NEI model seems to have a brighter shell. 
When comparing to observations, other elements and energy ranges, omitted here, should of course also be included.

\subsection{Connection between the $\Delta\bar{c}$ and the observed parameters}
\label{sec:charge2tau}
\begin{figure}[htpb]
\plotone{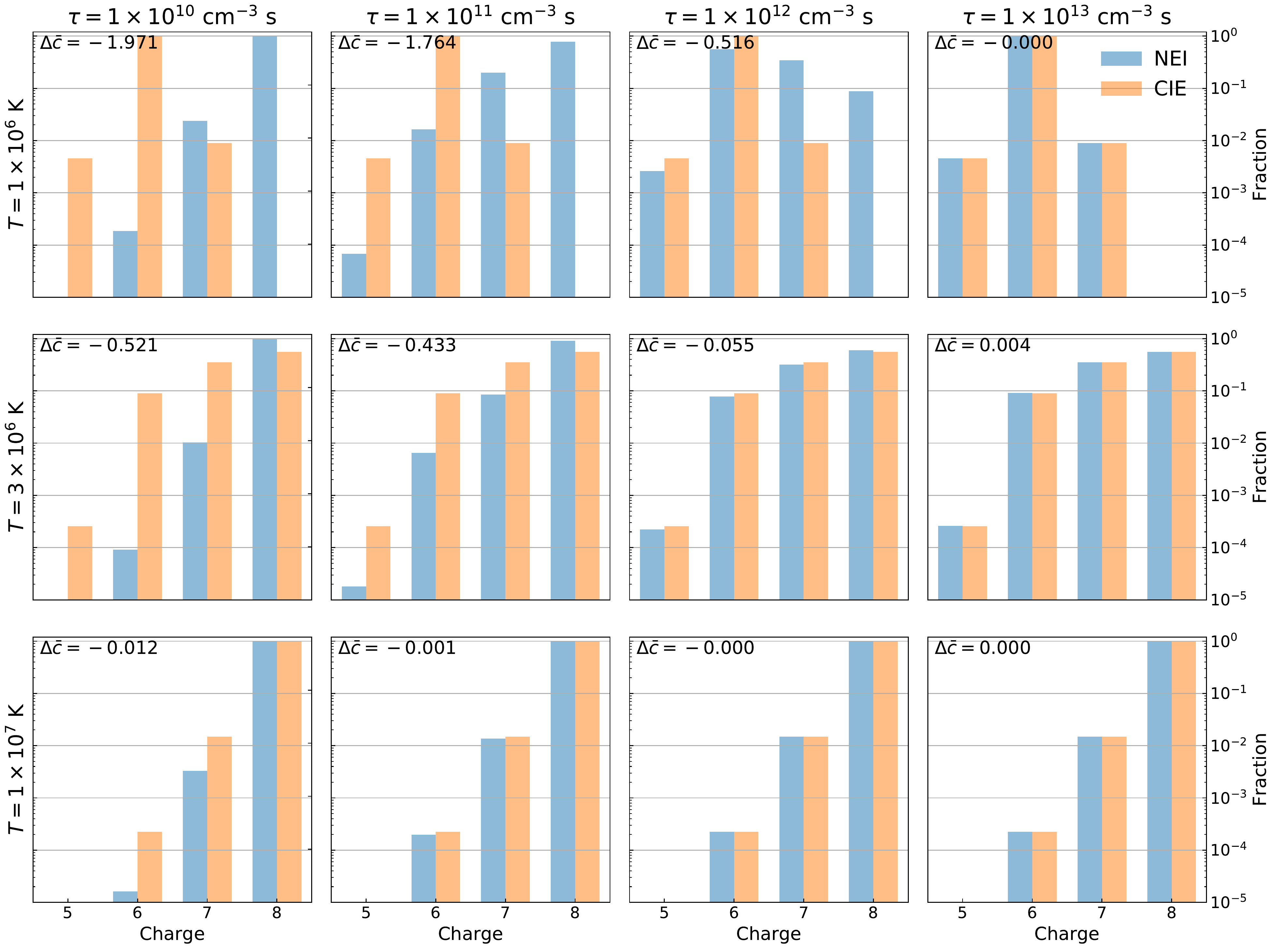}
\caption{Ion fraction of oxygen (recombining from 
$T_{init}=1\times10^{9}\K$) shown as a histogram. From top to bottom,
electron temperature changes from $1\times10^6\K$ to $1\times10^7\K$;
from left to right, the density-weighted timescale $\tau$ changes 
from $1\times10^{10}\cmmthree\s$ to $1\times10^{13}\cmmthree\s$. Only
the ion fractions of which the
charges are larger than 4 are shown in blue. The same ion fractions in
equilibrium at corresponding temperatures are shown in orange as reference.
The average charge differences defined in Equation \eqref{eq:aver} are
labeled in each panel with three significant digits.
\label{fig:charge2tau}}
\end{figure}

In our simulations, the average charge difference ($\Delta\bar{c}$) is used to show
whether a cell is in an ionizing or recombining state and how far the NEI state is away from the equilibrium. In observations, however, it is more typical to fit an electron temperature ($T$),
an initial temperature ($T_{init}$) and the density-weighted timescale ($\tau=n_e t$). To aid the observer, we show here the connection between $\Delta\bar{c}$ and $\tau$. 
Taking oxygen as an example,  in Fig.~\ref{fig:charge2tau}, an initial temperature of 1$\times10^{9}\K$
is used to show the recombination state with the temperature from $1\times10^{6}\K$
to $1\times10^{7}\K$ in three different ways. First, the ion fraction histograms
of the NEI state (blue) and the CIE state (orange) can be compared to each other. 
They are calculated with the {\em pyAtomDB} \citep{Foster2012}.
In an \xray\ observation, the O$^{6+}$ and O$^{7+}$\ ions that emit O\,\rom{7} and
O\,\rom{8} lines are the key ions, so we only show the high charge ions ($c>4$). We show the timescale $\tau$\ on the top of each column. In the right most column, $\tau=1\times 10^{13}\cmmthree\s$, 
where the NEI is very close to the equilibrium. We also calculate $\Delta\bar{c}$ by using the ion fraction for each panel. From the ion fractions, when $|\Delta\bar{c}|<0.001$, the difference between the NEI state and the CIE state is small. Therefore, we use $\Delta\bar{c}<-0.001$ and
$\Delta\bar{c}>0.001$\ to determine of ionization and recombination
respectively in our works to exclude the numerical fluctuation.

\bibliographystyle{aasjournal}
\bibliography{cite}



\end{CJK*}
\end{document}